\documentclass{pasj00}
\def\commenta{$^*$}
\def\commentb{$^\dagger$}
\def\commentc{$^\ddagger$}
\def\commentd{$^\|$}
\def\commente{$^\S$}

\def\vsnetalert#1{VSNET alert circular, #1}

\def\vsnetchat#1{VSNET chat circular, #1}
\def\vsnetid#1{VSNET ID circular, #1}

\def\vsnetcampaigndn#1{VSNET campaign circular (dwarf novae), #1}

\def\labelspace{}

\def\pasjcitesub#1#2{\authorcite{#1} \yearcite{#1}#2}
\def\pasjcitetsub#1#2{\authorcite{#1} (\yearcite{#1}#2)}
\def\pasjcitepsub#1#2{(\authorcite{#1} \yearcite{#1}#2)}
\def\pasjcite2sub#1#2{\authorcite{#1} \yearcite{#1},\yearcite{#2}}
\def\pasjcitet2sub#1#2{\authorcite{#1} (\yearcite{#1},\yearcite{#2})}
\def\pasjcitep2sub#1#2{(\authorcite{#1} \yearcite{#1},\yearcite{#2})}
\def\pasjcite2subyear#1#2#3{\authorcite{#1} \yearcite{#2}#3}
\def\pasjcitet2subyear#1#2#3{\authorcite{#1} (\yearcite{#2}#3)}
\def\pasjcitep2subyear#1#2#3{(\authorcite{#1} \yearcite{#2}#3)}

\DeclareAbbreviation\an{Astron. Nachr.}
\DeclareAbbreviation\AcA{Acta Astron.}
\DeclareAbbreviation\Ap{Astrophysics}
\DeclareAbbreviation\ARep{Astronomy Reports}
\DeclareAbbreviation\ATsir{Astron. Tsirk.}
\DeclareAbbreviation\BaltA{Baltic Astron.}
\DeclareAbbreviation\ibvs{Inf. Bull. Var. Stars}
\DeclareAbbreviation\JAVSO{Journal of the American. Assoc. Var. Star Observers}
\DeclareAbbreviation\JBAA{Journal of the  British Astron. Assoc.}
\DeclareAbbreviation\MitVS{Mitteil. \"{u}ber Ver\"{a}nder. Stern.}
\DeclareAbbreviation\MmSAI{Memorie della Soc. Astron. Italiana}
\DeclareAbbreviation\Msngr{The Messenger}
\DeclareAbbreviation\NewAR{New Astronomy Reviews}
\DeclareAbbreviation\Obs{Observatory}
\DeclareAbbreviation\PAZh{Pis'ma AZh}
\DeclareAbbreviation\PZ{Perem. Zvezdy}
\DeclareAbbreviation\PZP{Perem. Zvezdy Pril.}
\DeclareAbbreviation\RMxAA{Rev. Mexicana Astron. Astrof.}
\DeclareAbbreviation\VeSon{Ver\"{o}ff. der Sternwarte in Sonneberg}
\DeclareAbbreviation\VSOLJBul{VSOLJ Var. Star Bull.}
\DeclareAbbreviation\ZA{Z. Astrophys.}

\def\ASPConf#1#2{ASP Conf. Ser. #1, #2}
\def\IAUColloq#1#2{IAU Colloq. #1, #2}

\begin{document}
\SetRunningHead{T. Kato, Y. Sekine, and R. Hirata}{HV Virginis and WZ Sge-type Dwarf Novae}

\Received{}
\Accepted{}

\title{HV Virginis and WZ Sge-Type Dwarf Novae}

\author{Taichi \textsc{Kato}, Yoshiyuki \textsc{Sekine},
       Ryuko \textsc{Hirata}}
\affil{Department of Astronomy, Kyoto University,
       Sakyo-ku, Kyoto 606-8502}
\email{tkato@kusastro.kyoto-u.ac.jp, hirata@kusastro.kyoto-u.ac.jp}

\newcounter{hvref}
\newcounter{hvrem}



\maketitle

\begin{abstract}
   A dwarf nova HV Vir was observed photometrically for eight nights
during the outburst in 1992 April --- May.  The star showed two distinct
types of periodic variation: (1) 82.20-min (0.05708 d) double-humped
variation with decaying amplitudes during the early stage of the outburst,
and (2) 83.80-min (0.05820 d) superhumps in later stages.  We attributed
the former to ``early superhumps", which are only seen in the earliest
stage of WZ Sge-type outbursts.  The superhump period and evolution of the
superhumps, together with general characteristics of the light curve,
make HV Vir a typical WZ Sge-type dwarf nova.  HV Vir also showed
a large increase of the superhump period during the superoutburst.
Upon the recognition of the WZ Sge-type nature of an object previously
considered as a nova, we present a comprehensive list of candidates
for WZ Sge-type dwarf novae, and related systems.
\end{abstract}

\section{Introduction}

\subsection{WZ Sge and Related Stars}

   Although presently classified as a dwarf nova, WZ Sge is peculiar in
many respects.  The star exhibited three historical outbursts (1913, 1946,
1978) separated by 32.5 years, and most recently experienced its fourth
outburst in 2001 (\pasjcitesub{ish01wzsgeiauc7669}{b};
\pasjcitesub{mat01wzsgeiauc7669}{b}).
The light curves of its outbursts resemble that of a fast nova in
that they show 1) a rapid initial decline, 2) a subsequent slower decline
lasting 60 to 100 d\footnote{
  The most straightforward evidence can be found in
  \pasjcitetsub{tar79wzsgeiauc3344}{c}, who observed WZ Sge at $V$=14.4 on
  1979 April 8, 128 d after the initial outburst detection.  Further
  observational evidence for this phenomenon is rather sparse
  (cf. \cite{ort80wzsge}; \cite{ric92wzsgedip}), than is generally believed.
}, and 3) large ($\sim$8 mag) outburst amplitudes.
From these features, the object had been long believed to be a recurrent
nova (\cite{may46wzsge}; \cite{cam47wzsge}; see also a historical review
by \citet{mat80wzsge}).
Spectroscopic and photometric observations during the 1978 outburst,
however, confirmed that WZ Sge is a dwarf nova, and not
a classical nova\footnote{
  Although the early outburst spectra (\cite{nat78wzsgeiauc3311};
  \cite{bro78wzsgeiauc3313}; \cite{wal78wzsgeiauc3315}) were already
  indicative of a dwarf nova, rather than a recurrent nova, the first
  correct classification as a dwarf nova was suggested by \citet{boh79wzsge}
  from photometric observations.  \citet{bai79wzsge}, \pasjcitetsub{pat79SH}{b}
  and \citet{pap79SHmodel} also supported this interpretation.
  \citet{cra79wzsgespec} presented the spectra in outburst, and noted
  the difference from an ordinary nova.  Later literature supporting
  the dwarf nova classifications include \citet{gil80wzsgeSH},
  \citet{wal80wzsgespec} and \citet{pat81wzsge}.
}.
The most striking discovery was the presence of superhumps
by \citet{boh79wzsge}\footnote{
  The emergence of a slightly longer period than the orbital period
  was first noted by \citet{gui79wzsgeiauc3319}.
  \pasjcitetsub{tar79wzsgeiauc3320}{b}
  yielded the first estimate of this period.  However, the period increase
  was initially considered to be a consequence of the `nova' outburst.
  \citet{boh79wzsge} first correctly described this period as superhumps,
  and identified WZ Sge to be an SU UMa-type dwarf nova.  The first
  indication of the emergence of superhumps in reported photometry can
  be found in \citet{hei79wzsge} and \pasjcitetsub{tar79wzsgeiauc3320}{b}.
}
(for a review of superhumps, see \cite{war85suuma}),
which are characteristic to
SU UMa-type dwarf novae [see \citet{osa96review} for a review of
dwarf novae; see also \citet{war85suuma} for basic observational properties,
and \pasjcitetsub{war95suuma}{b} for a more recent review of SU UMa-type
stars].

   There had been several competing theories to explain the peculiar
nature of WZ Sge.  One is the mass-transfer burst model, which was used
to explain the large-amplitude periodic humps early in its 1978 outburst
\citep{pat81wzsge}.
The period of the humps was equal to the orbital period ($P_{\rm orb}$),
and they were considered to reflect the hot spot enhanced by the
mass-transfer burst, although the hump maxima occurred 0.17 orbital
phase prior to the orbital humps in quiescence.

   The other is the extension of thermal and tidal instability theory
of SU UMa-stars (\cite{osa89suuma}; \cite{osa93unifiedmodel}) towards
the lowest mass-transfer rate \citep{osa95wzsge}.
Numerical simulations have shown that thermal instability of the
accretion disk occurs rarely in such conditions, and it always leads
to the tidal instability to trigger a superoutburst \citep{osa95wzsge}.

   \citet{las95wzsge} assumed the evaporation of the inner disk, and argued
that rare outbursts can be caused by a small increase of the mass-transfer
rate, without an assumption of the very low quiescent viscosity.  This
scenario may be considered as a combination of the mass-transfer burst
model and the disk instability model (see, however, the arguments
by \citet{mey98wzsge} and \citet{min98wzsge} on the scenario of
\citet{las95wzsge}).

   Some WZ Sge-type dwarf novae are known to show complex post-superoutburst
rebrightening(s), which are rarely seen in other SU UMa-type dwarf novae.
The most remarkable objects are EG Cnc (\pasjcitesub{kat97egcnc}{a};
\pasjcitesub{mat98egcnc}{b}; \pasjcitesub{pat98egcnc}{a}), and WZ Sge
itself (the 2001 superoutburst: Ishioka et al. in preparation).
In order to explain this unique feature,
\citet{osa97egcnc} proposed a working model, in which the the disk viscosity
in post-superoutburst WZ Sge-type stars are somehow maintained higher than
that the pre-superoutburst level.  Although the underlying physical mechanism
was not clear ar the time of the initial proposition, \citet{osa01egcnc}
further succeeded to explain, by a combination of the schemes of
\citet{osa95wzsge}, \citet{mey98wzsge} and the decay of magnetic
turbulence in the quiescent dwarf nova disk \citep{gam98}, the variety of
phenomena in WZ Sge-type dwarf novae.

   From the observational side, discrimination of the theories has been
difficult owing to the rarity of outbursts of WZ Sge.  Searches for
more ``WZ Sge-type'' objects (\cite{bai79wzsge}; \cite{dow81wzsge};
\cite{pat81wzsge}; \cite{dow90wxcet}; \cite{odo91wzsge}) among dwarf
novae have been a natural consequence of the motivation to confirm the
universality of the features found in WZ Sge itself.
In spite of the efforts, a number of the candidates have turned out to be
rather normal SU UMa-stars exhibiting normal outbursts (e.g. WX Cet
(\cite{odo91wzsge}; \cite{rog01wxcet}; \pasjcitesub{kat01wxcet}{c}),
and VY Aqr (\cite{dellaval90vyaqr}; \cite{aug94vyaqr}; \cite{pat93vyaqr})
see also \cite{odo91wzsge} and references therein), and the attempt
remained rather unsuccessful.

\subsection{HV Virginis as a WZ Sge-Type Candidate}

   HV Vir was discovered by \citet{sch31hvvir} in outburst on 1929
February 11.  \authorcite{due84hvvir} (\yearcite{due84hvvir}b,
\yearcite{due87novaatlas})
identified the object as a classical nova based on its large outburst
amplitude and its light curve constructed from his examination of
archival plates.  However, we suspected it to be a WZ Sge-type dwarf
nova because of its high galactic latitude ($l$=\timeform{319.9D},
$b$=+\timeform{63.8D}, which makes HV Vir a very distant ($\sim$100 kpc),
halo or even an extragalactic object based on the usual absolute
maximum magnitude vs rate-of-decline (MMRD) relation of galactic
novae (originally proposed by \citet{sch57novaabsmag} and formulated by
\citet{pfa76novaabsmag}.  See \citet{coh85novaabsmag};
\citet{vandenver86novaabsmag}; \citet{san88virgonovaabsmag};
\pasjcitet2subyear{cap90LMCnovaabsmag}{cap90comanovaabsmag}{a,b};
\citet{liv92novaabsmag} for recent applications, formulations, and
discussions, and for a recent review, see \cite{dellaval95novaabsmag}).
This suggestion was strengthened by our
unpublished CCD photometry in quiescence (1990) which showed possible
humps separated by 84$\pm$4 m, a period very close to that of WZ Sge
(81.6 min).

   HV Vir was again caught in outburst at visual magnitude 12.0 on 1992
April 20.928 UT, when the object was still brightening
\citep{sch92hvviriauc}.
The outburst was subsequently confirmed photoelectrically with
a conspicuous ultraviolet excess, and the maximum was probably reached
on April 21.4 at $V$=11.5 \citep{kil92hvviriauc}.
We started photometric observations within a day of the maximum,
on April 22, using the multicolor photoelectric polarimeter at the Dodaira
Station, National Astronomical Observatory, Japan, and later using a CCD
camera at Ouda Station, Kyoto University.  This early time-resolved
photometry eventually enabled us to elucidate the nature of the object.

\section{Observations}

\subsection{Photoelectric Photometry at Dodaira}

   The observations were performed with the multi-channel photoelectric
polarimeter \citep{kik88Dodaira}, attached to the Cassegrain focus of
the 0.91-m reflector at Dodaira Station, National Astronomical
Observatory.  The journal of the observations is summarized in table
\ref{tab:log}.  The time resolution was 35 s, interrupted by measuring
the local standard (SAO 119899 = BD +2$^{\circ}$ 2664,
\timeform{13h21m41.6s}, +\timeform{02D05'14''} (J2000.0), $V$=5.69,
$B-V$=+0.06, A2 V) in every 18 observations of the variable.
The aperture size was \timeform{18.0''}, which was confirmed on the
image-intensified TV monitor to large enough to include several times
the seeing size.  The typical error of a single measurement was 0.02 mag.
A very gradual change of the extinction of the sky was calibrated using
this local standard, and was interpolated to estimate the magnitude of
the variable.  Because the observed color was almost constant throughout
the runs, we added the counts of the seven color bands covering
360 -- 800 nm, after correcting for atmospheric extinction in each band,
and obtained ``white-light'' magnitudes to achieve the best
signal-to-noise ratio for detecting small amplitude variations.

\subsection{CCD Photometry at Ouda}

   The observations were performed with the CCD camera (Thomson TH7882
chip, 576 $\times$ 384 pixels) attached to the 0.6-m reflector at Ouda
Station, Kyoto University \citep{Ouda} for successive six nights from 
1992 May 1 through May 6.  An on-chip summation of 3$\times$3 pixels
(the unbinned 1 pixel corresponds to \timeform{1.0''} and the typical
seeing size was \timeform{6''}) were adopted to minimize the readout time
and noise.  The interference filter which is designed to reproduce the
Johnson $V$-band was adopted.  The exposure was set to 60 s.
The readout time between exposures was 8 s.  Bias frames were taken every
ten frames.  The typical error of a single measurement was 0.03 mag.
The total number of useful frames was 1341.

\begin{table}
\caption{Log of observations}\label{tab:log}
\begin{center}
\begin{tabular}{lcccrc}
\hline\hline
Date & Start\commenta & End\commenta & Exp\commentb & $N$\commentc &
       Site\commentd \\
\hline
1992    & \phantom{0}$h$ \phantom{0}$m$ &
                  \phantom{0}$h$ \phantom{0}$m$ & & & \\
Apr. 22 & 11 44 & 15 42 & 35 & 234 & D \\
Apr. 23 & 10 33 & 14 07 & 35 & 216 & D \\
May   1 & 10 58 & 15 58 & 60 &  94 & O \\
May   2 & 15 17 & 16 11 & 60 &   7 & O \\
May   3 & 10 36 & 17 33 & 60 & 329 & O \\
May   4 & 10 28 & 17 49 & 60 & 360 & O \\
May   5 & 10 44 & 17 47 & 60 & 339 & O \\
May   6 & 10 41 & 16 35 & 60 & 212 & O \\
\hline
 \multicolumn{6}{l}{\commenta Time in UT. Apr. 0 UT= JD 2448712.5,} \\
 \multicolumn{6}{l}{\phantom{\commenta} and May 0 UT= JD 2448742.5.} \\
 \multicolumn{6}{l}{\commentb Exposure time (s).} \\
 \multicolumn{6}{l}{\commentc Number of useful measurements or frames.} \\
 \multicolumn{6}{l}{\commentd D (Dodaira), O (Ouda).} \\
\end{tabular}
\end{center}
\end{table}

   The summary of observation is listed in table \ref{tab:log}.
The CCD frames were, after standard de-biasing and flat-fielding,
analyzed using automatic microcomputer-based aperture photometry package
developed by one of the authors (TK).  The magnitudes of the object were
determined relatively using a local standard star GSC 300.603 located at
\timeform{13h20m54.80s}, +\timeform{01D52'06.8''} (J2000.0),
$V$=14.11, $B-V$=+0.56.  The synthetic aperture size was \timeform{12.0''}.
The constancy of the comparison during the observations were not confirmed
very satisfactorily ($\sim$0.05 mag), because of lack of suitable check
stars in the same field.
A long-term variation larger than 0.01 mag was, however, reasonably
ruled out by averaging the frames on each night, and comparing with
fainter anonymous stars.

\section{Results}

\subsection{The course of the outburst}\label{sec:lc}

   HV Vir showed a rapid magnitude increase by at least 1.2 mag in
1.011 d prior to the detection of the outburst \citep{sch92hvviriauc}.
The visual maximum was reached on 1992 April 21.4, at $V$=11.5
\citep{kil92hvviriauc}.  The object started to fade instantaneously at
a rate of 0.3 mag d$^{-1}$, down to $V$=12.4 on April 24
(\cite{kil92hvviriauc}; \cite{bru92hvvir}).  The light curve obtained at
Dodaira Station (figure \ref{fig:april}) clearly shows this rapid
fading.  A similar rapid initial fade by 2 mag in 6 d was observed during
the 1929 outburst \pasjcitepsub{due84hvvir}{b}.
Taking the poor calibration of old photographic photometry into account,
the behavior of the two outbursts is strikingly similar.  Such a rapid
initial fade is quite similar to that observed in previous outbursts
of WZ Sge itself (e.g. \cite{ort80wzsge}; \cite{pat81wzsge}\footnote{
  There exists an argument against the sharp, initial peak recorded
  in the past observations.  The brightest estimates during the 1913 and 1946
  outbursts were photographic ones, which are shown to be systematically
  brighter than the modern visual scale by more than 0.5--1.0 mag (Kuulkers,
  private communication).  Such an effect may have skewed the past
  representative light curves of WZ Sge.  The most recent (2001) outburst
  of WZ Sge (see $\langle$
  http://www.kusastro.kyoto-u.ac.jp/vsnet/DNe/wzsge01.html $\rangle$)
  did not show such a sharp peak as in the past.
})
and WZ Sge-type candidate V592 Her (\cite{due98v592her}; Kato et al.,
in preparation, see also the VSNET page\footnote{
  $\langle$http://www.kusastro.kyoto-u.ac.jp/vsnet/DNe/\\
  v592her9808.html$\rangle$
}

\begin{figure}
  \begin{center}
    \FigureFile(88mm,60mm){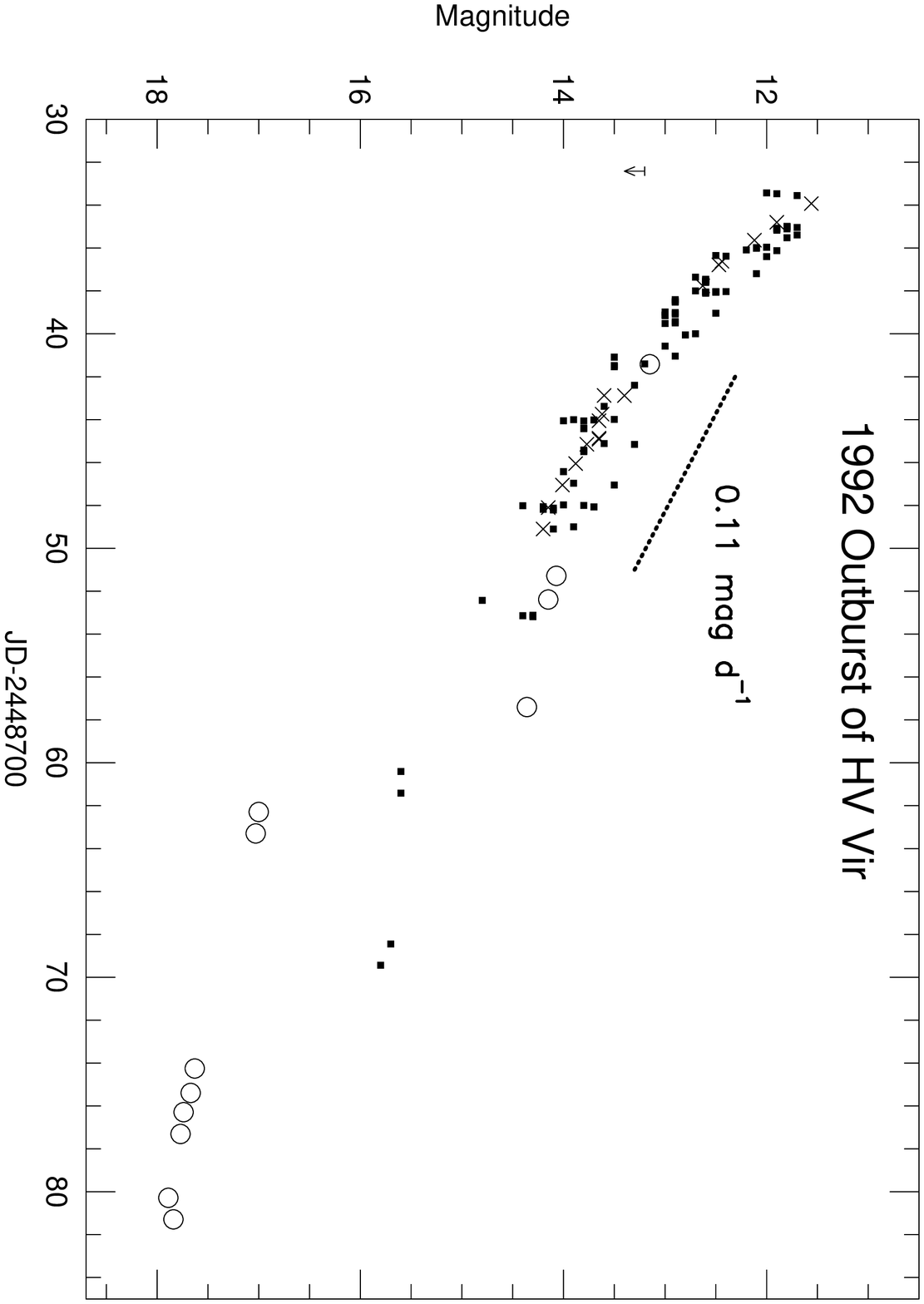}
  \end{center}
  \caption{The overall behavior of the 1992 outburst of HV Vir,
  constructed from photoelectric $V$-band (crosses), CCD $R$ measurements
  and visual (points) observations taken from the literature, and visual
  variable star databases (see text).  The arrow at (32.417, 13.2)
  represents the upper limit.  The dotted line marks the linearly declining
  portion of the superoutburst.  The general behavior of the light curve
  closely resembles that of the 1946 outburst of WZ Sge.}
  \label{fig:lc}
\end{figure}

\begin{figure}
  \begin{center}
    \FigureFile(88mm,60mm){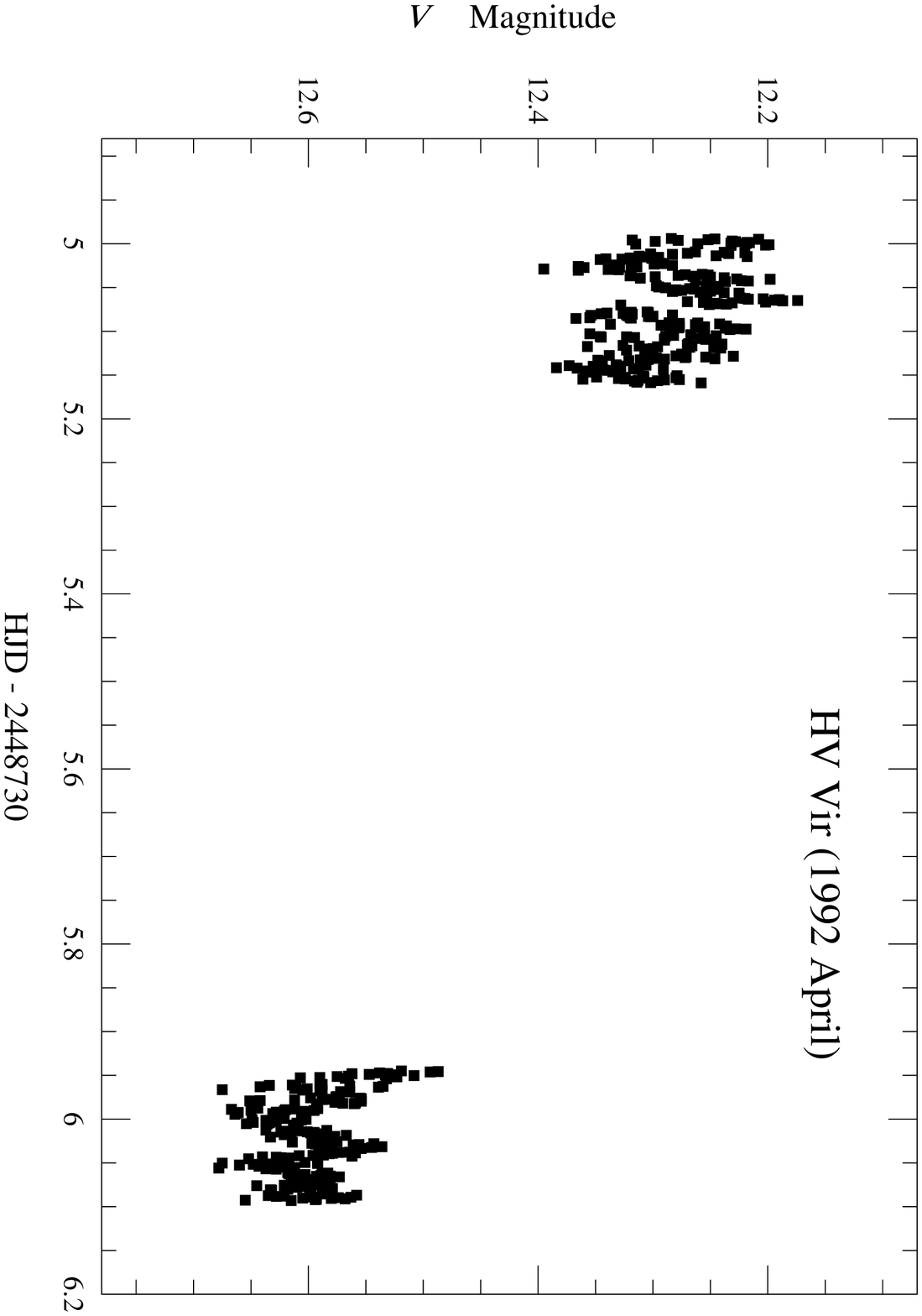}
  \end{center}
  \caption{`White-light' light curve of HV Vir on April 22 and 23
  taken at Dodaira Station.  A rapid, linear decline at a mean rate of
  0.33 mag d$^{-1}$ is clearly seen.  Small fluctuations superimposed on
  the general fading are early superhumps (see subsection \ref{sec:earlyhump}
  for more details).}
  \label{fig:april}
\end{figure}

   Following this fade, the star entered a stage of slower decline
of 0.15 mag d$^{-1}$.  The visual light curve constructed from
observations from IAU Circulars, $R$-band CCD observations from
\citet{lei94hvvir}, visual variable star database supplied from Variable
Star Observers League in Japan (VSOLJ) and the data supplied by G. Poyner,
and B. Worraker (British Astronomical Association; BAA) clearly
demonstrates this feature (figure \ref{fig:lc}).  The main superoutburst
(superoutburst plateau) lasted for 23 d, and followed by a rapid decline
by 2.6 mag to $R$=17.0 within 4.8 d \citep{lei94hvvir}.  Visual observations
suggest that the object brightened again to $\sim$15.5 mag (May 25--26),
34 d after the outburst maximum.
This may have been a post-outburst rebrightening as is frequently
observed in WZ Sge-type dwarf novae
(cf. \pasjcitesub{kat98super}{a})\footnote{
  The presence of various post-superoutburst phenomena in rarely outbursting
  systems (VY Aqr, WZ Sge, AL Com, WX Cet, V592 Her, DV Dra and UZ Boo)
  was first pointed out by \citet{ric92wzsgedip}.
  More recent descriptions can be found in \pasjcitetsub{how95TOAD}{a},
  \citet{kuu96TOAD} and \citet{kuu00wzsgeSXT}.
}.
The object then entered a stage of a prolonged fading tail, as observed in
a WZ Sge-type dwarf nova, EG Cnc (\pasjcitesub{kat97egcnc}{a};
\pasjcitesub{pat98egcnc}{a}).
The outburst is similar to that of the 1946 outburst of WZ Sge\footnote{
  The apparent absence of recorded post-superoutburst rebrightening(s)
  in 1946 may have been a result of the sparse coverage after the main
  superoutburst.  On the occasion of the 2001 superoutburst of WZ Sge,
  very short ($\sim$2.5 to 1.5 d) recurrence time of short ($\sim$1 d)
  rebrightenings was observed (e.g. \pasjcitesub{kat01wzsgealert6345}{d};
  \pasjcitesub{kat01wzsgealert6432}{e}).  Events with such short time-scales
  may have been easily missed by a limited density of observations.
}
\citep{ort80wzsge} before the rapid fading, and is different from usual
superoutbursts of SU UMa stars in its rapid initial decline, long
duration of the superoutburst plateau, and prolonged fading tail.

\begin{figure}
  \begin{center}
    \FigureFile(88mm,60mm){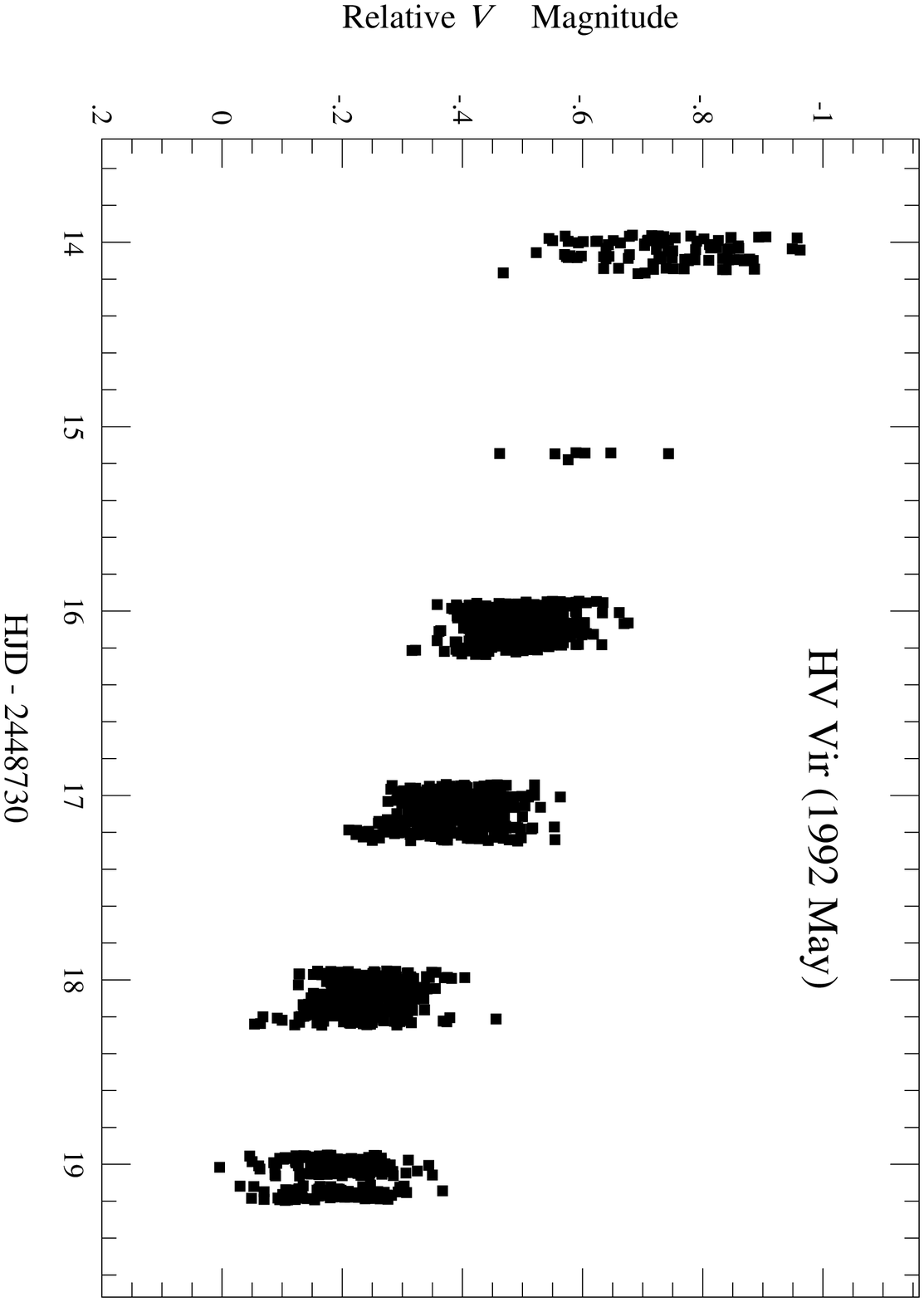}
  \end{center}
  \caption{CCD $V$-band light curve of HV Vir on May 1--6, taken at
  Ouda Station.  The linear decline at a slower rate of 0.11 mag
  d$^{-1}$ is characteristic of an SU UMa-type superoutburst.
  Small fluctuations superimposed on the general fading are well-developed
  superhumps (see subsection \ref{sec:sh}).}
  \label{fig:may}
\end{figure}

\subsection{Hump features in the early stage of the outburst}
\label{sec:earlyhump}

\begin{figure}
  \begin{center}
    \FigureFile(88mm,60mm){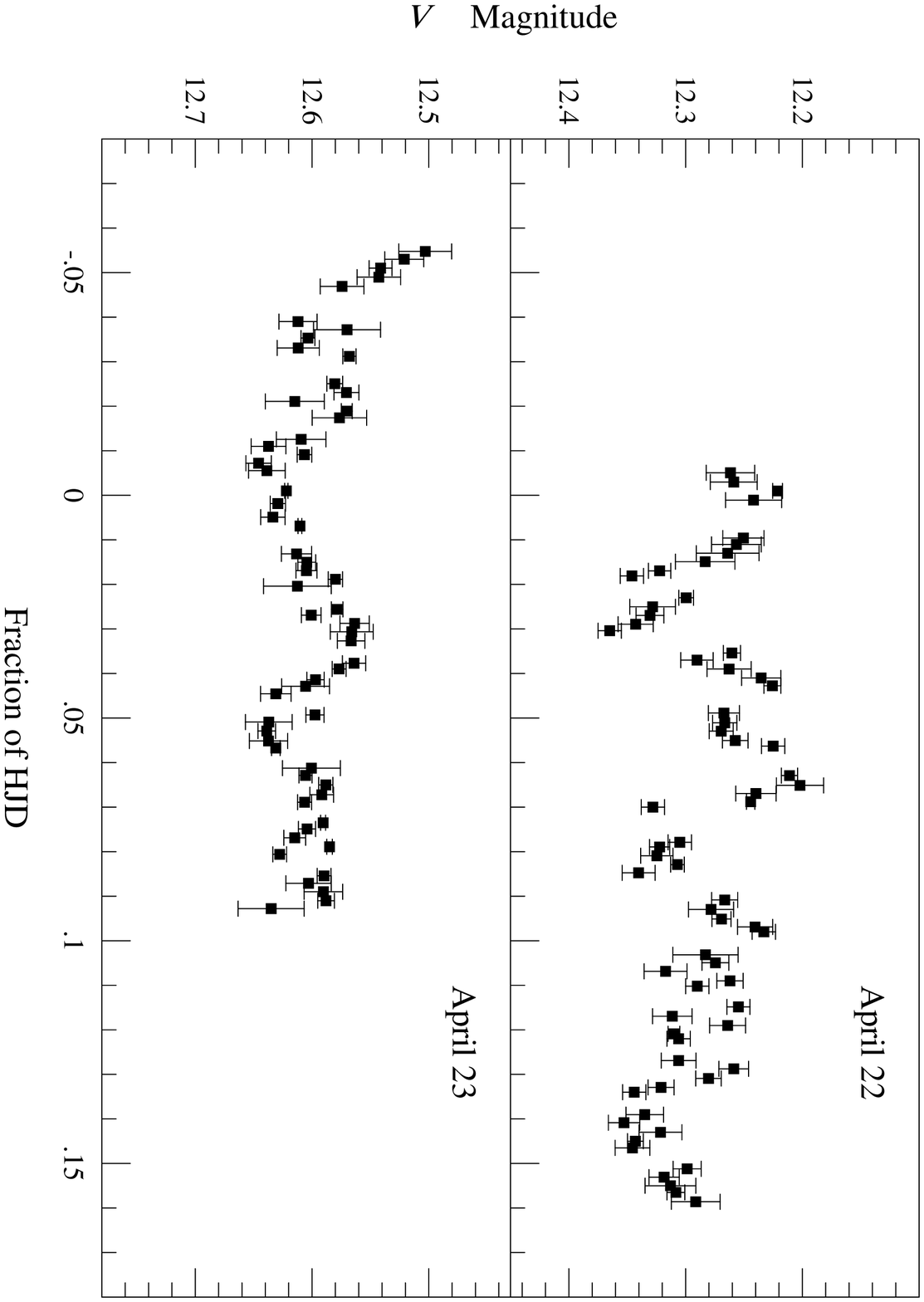}
  \end{center}
  \caption{`White-light' light curve of HV Vir on April 22 and 23
  taken at Dodaira, within two days of the visual maximum.
  Each point and error bar represent the average of 0.002 d bin and
  the standard error of the average, respectively.
  Low-amplitude ($\sim$0.10 mag) modulations existed on both nights.
  The light curve is dominated by rather sharp minima and broader maxima,
  recurring every 0.057 d (82 min), which we interpret as
  ``early superhumps" (cf. subsection \ref{sec:earlyhump}),
  characteristic to a WZ Sge-type superoutburst.  The initial part of the
  observations on April 23 were obtained at a low altitude, and was
  relatively poorly photometrically calibrated.}
  \label{fig:dodaira}
\end{figure}

   We analyzed the photoelectric observations at Dodaira on April 22 and
23.  The resultant light curve is shown in figure \ref{fig:dodaira}.
The magnitude scale is approximately adjusted to the $V$-band.

\begin{figure}
  \begin{center}
    \FigureFile(88mm,120mm){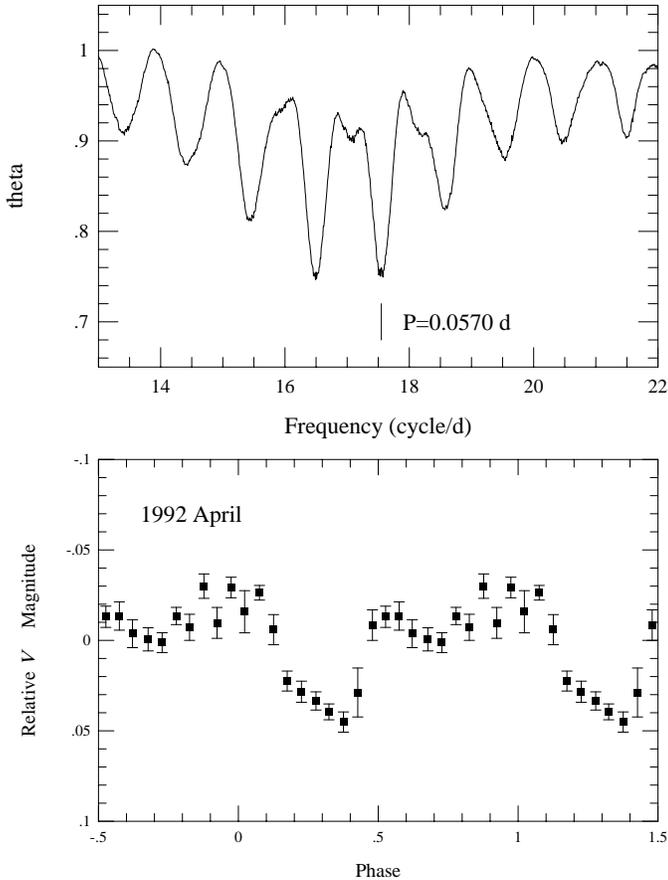}
  \end{center}
  \caption{(Upper) Period analysis of April 22 and 23 data, by using
  the Phase Dispersion Minimization (PDM) method \citep{PDM},
  after removing the linear trend of the steady decline.
  The signal at 17.55(2) d$^{-1}$ corresponds to the period of
  0.05698(8) d.  A stronger signal around 16.55 d$^{-1}$ is its
  one-day alias, which can be safely rejected using other maxima times
  from the literature.
  (Lower) Phase-averaged light curve with a period of 0.057085 d
  (the best-determined period using all data, see text).
  The complex, doubly humped profile is characteristic to
  ``early superhumps" observed in other WZ Sge-type dwarf novae.}
  \label{fig:earlysh}
\end{figure}

   Recurring hump features with an amplitude of 0.1 mag are evident.
We used the Phase Dispersion Minimization (PDM) method \citep{PDM},
after removing the linear trend of the steady decline.  The resultant
period of 0.05698(8) d well expresses the light variation, which is
characterized by a rather complex, doubly humped profile, with a narrower
minimum (figure \ref{fig:earlysh}).  The signal in the upper panel of
figure \ref{fig:earlysh} at 17.55(2) d$^{-1}$ corresponds to the period
of 0.05698(8) d.  A stronger signal around 16.55 d$^{-1}$ is its one-day
alias, which can be safely rejected using other maxima times (table
\ref{tab:earlyhump}) taken from the literature.

   Similar light variations were independently reported by
\citet{bar92hvvir} and \citet{men92hvviriauc} for the intervals April
24.86---26.56 UT and April 28.77---29.06 UT, respectively\footnote{
  Other observers (\cite{bru92hvvir}; \cite{how92hvviriauc};
  \cite{ing92hvvir}; \cite{man92hvviriauc}; \cite{szk92hvviriauc})
  also reported preliminary results, which were either summarized in
  \citet{lei94hvvir} or discussed in subsection \ref{sec:sh}.
}.
Although their observations were not long enough to precisely determine
the period, our period is found to express excellently all the
``pronounced humps'' listed by \citet{bar92hvvir}.  Together with the
hump maxima obtained from our photometry,
we refined the period as 0.057085(23) d (82.20$\pm$0.03 min) expressed by
the following linear ephemeris.

\begin{equation}
\rm{HJD_{max}} = 2448735.0069(12) + 0.057085(23) E. \label{equ:reg1}
\end{equation}

   This period corresponds to P$_{\rm 1}$ by \citet{bar92hvvir}.
The almost zero residuals from this period (table \ref{tab:earlyhump})
precludes alternative periods.  Since this period is based on a longer
baseline observations than one with Dodaira data, we consider this value
as the best-determined period.
[N.B. The epochs of hump maxima in table \ref{tab:earlyhump} are taken
from \citet{lei94hvvir}, which provides a table of the summary of
\citet{bar92hvvir} and \citet{men92hvviriauc}].

\begin{table}
\caption{Timings of early superhumps}\label{tab:earlyhump}
\begin{center}
\begin{tabular}{lrrc}
\hline\hline
HJD\commenta & $E$\commentb & $O-C$\commentc & Ref.\commentd \\
\hline
35.005  &  0 & $-$0.002\phantom{0} & 1 \\
35.066  &  1 &  0.002\phantom{0} & 1 \\
35.977  & 17 & $-$0.000\phantom{0} & 1 \\
36.033  & 18 & $-$0.001\phantom{0} & 1 \\
36.091  & 19 & $-$0.001\phantom{0} & 1 \\
37.4060 & 42 &  0.0015 & 2 \\
37.4607 & 43 & $-$0.0009 & 2 \\
37.5171 & 44 & $-$0.0016 & 2 \\
37.5810 & 45 &  0.0053 & 2 \\
39.3467 & 76 &  0.0013 & 2 \\
39.4002 & 77 & $-$0.0023 & 2 \\
39.4568 & 78 & $-$0.0028 & 2 \\
39.5185 & 79 &  0.0019 & 2 \\
\hline
 \multicolumn{4}{l}{\commenta HJD$-$2448700.} \\
 \multicolumn{4}{l}{\commentb Cycle count.} \\
 \multicolumn{4}{l}{\commentc Against equation \ref{equ:reg1}.} \\
 \multicolumn{4}{l}{\commentd 1: this paper,} \\
 \multicolumn{4}{l}{\phantom{\commentd} 2: \citet{lei94hvvir}.} \\
\end{tabular}
\end{center}
\end{table}

   The amplitude of the hump features decayed from 0.10 mag on April 22
to 0.06 mag for April 24---26 \citep{bar92hvvir}.  The inferred amplitude
of 0.05 mag by \citet{men92hvviriauc} on Apr. 28 might follow this
sequence, but its identity is uncertain.  It may be safe to say that
the amplitude of 82.20-min periodicity was no larger than 0.05 mag on
April 28.  This hump feature (``early superhumps", which are humps only
seen during the earliest stage of WZ Sge-type outbursts, having
periods close to $P_{\rm orb}$) will be discussed in subsection
\ref{sec:dis:earlysh}.

\subsection{Superhumps}\label{sec:sh}

   After the 82.20-min periodicity decayed, another distinct type
of variability appeared in late April, which was first described by
\citet{szk92hvviriauc}.  Their observations starting on April 30 showed
definite superhumps with an amplitude of 0.2 mag and a period of
84.1$\pm$0.4 m.  Although observations by \citet{men92hvviriauc} suggest
superhumps had appeared on April 28, the large differences in amplitude
and period from the present observation make identification inconclusive.
The superhumps may have just started to develop at the time of the
observation by \citet{men92hvviriauc}.

\begin{figure}
  \begin{center}
    \FigureFile(88mm,60mm){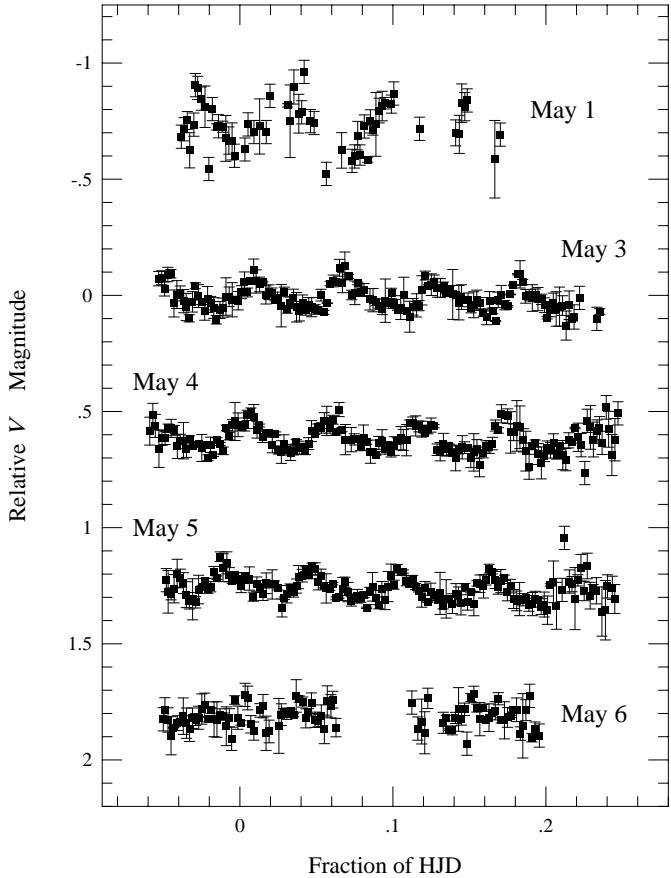}
  \end{center}
  \caption{Enlarged light curve of HV Vir in early May.
  Each point and error bar represent the average of 0.002 d bin and
  the standard error of the average, respectively.
  Periodic humps with a period of 0.05820 d (=83.80 min) and an amplitude
  of 0.16 mag are clearly seen.
  The humps show steeper rise and slower decline, characteristic of
  superhumps of SU UMa stars, are distinct from those observed in April
  (figure \ref{fig:dodaira}.}
  \label{fig:ouda}
\end{figure}

\begin{figure}
  \begin{center}
    \FigureFile(88mm,120mm){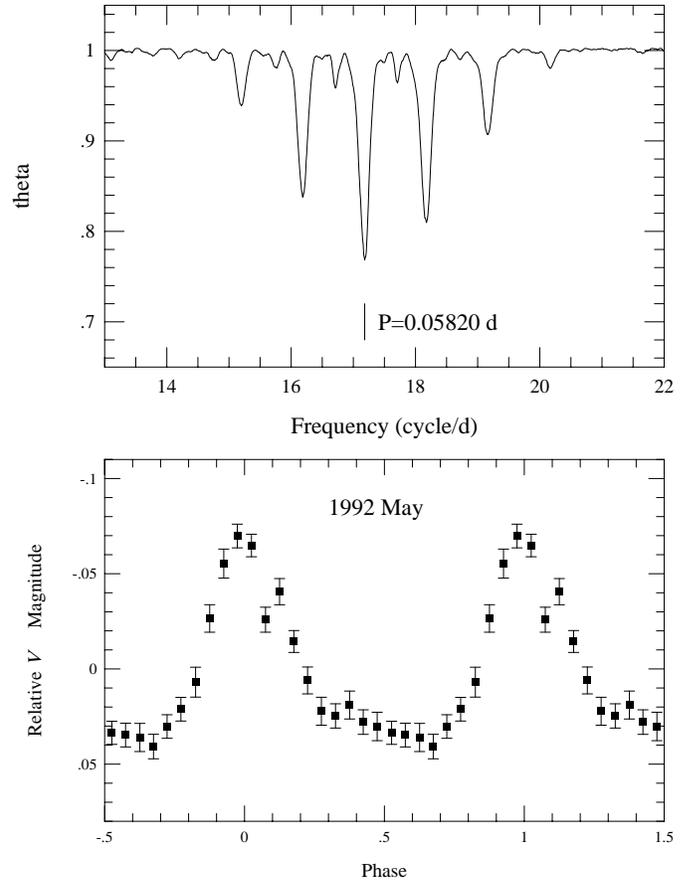}
  \end{center}
  \caption{(Upper) Period analysis of the May 1--6 data, by using
  the Phase Dispersion Minimization (PDM) method \citep{PDM},
  after removing the linear trend of the steady decline.
  (Lower) Phase-averaged light curve with a period of 0.05820 d.
  The superhumps were quite regular in shape, showing a steeper
  rise and gradual decline.}
  \label{fig:sh}
\end{figure}

   The same features were independently discovered by us from the
Ouda data (figure \ref{fig:ouda}).
The superhumps were quite regular in shape, showing a steeper
rise and gradual decline throughout our observations (figure \ref{fig:sh}).
The amplitude of superhumps slightly decreased from 0.30 mag on May 1 to
0.16 mag on May 6.  The period analysis using PDM yields a period of
0.05820(4) d (83.80$\pm$0.03 min)
  \footnote{
  As shown in subsection \ref{sec:psh}, the superhump period showed a
  systematic variation through the outburst.  We refer to this 83.80 m
  as the representative period of superhumps in the following sections.
  }.
The value is in agreement with a period of 84.1$\pm$0.4 min by
\citet{szk92hvviriauc}.
The appearance of superhumps confirms the SU UMa-type nature of HV Vir.

\section{Discussion}

\subsection{Frequency of Superoutbursts}

   \citet{lei94hvvir} examined Sonneberg plates, and found another bright
outburst (reaching 11.2 mag) occurring in 1970.  \citet{lei94hvvir} also
reported possible positive detections in 1939 and 1981, and implied that
the outburst cycle length is $\sim$10 yrs or shorter.  Their possible
detections in 1939 and 1981 being much fainter (around 13.5 -- 14.0 mag,
close to the plate limit) than the 1992 outburst, these detections,
even if they were real outbursts, should be more properly treated as
a different kind of outbursts.  The only known {\it major} outbursts
(likely superoutbursts) are in 1929, 1970 and 1992.  HV Vir has been
intensively monitored by the members of
VSNET Collaboration
  \footnote{
  $\langle$http://www.kusastro.kyoto-u.ac.jp/vsnet/$\rangle$
  }
since the 1992 outburst.  The lack of outburst detection up to 2001
suggests that the cycle length of major outbursts is an order of
$\sim$10 yrs, or even longer.  The observed interval of major outbursts
is close to that of AL Com, a ``twin" system to WZ Sge
(see \pasjcitesub{kat96alcom}{b}, \pasjcitesub{nog97alcom}{a}
and \cite{pat96alcom} for extensive discussions).

\subsection{Superhump Period}\label{sec:psh}

   The presence of two superhump periods (i.e. those of early superhumps
and ordinary superhumps) in HV Vir, which were not properly
recognized at the time of \citet{lei94hvvir}, together with a large
observational gap in \citet{lei94hvvir}, caused an ambiguity in cycle
count identifications.
All superhump timings between HJD 2448742.8728 and 2448752.4897 were
first collected from our observations and from \citet{lei94hvvir}.  The
longest gap of observations being 2.1 d in the present analysis has
greatly improved the quality of the analysis compared to
\citet{lei94hvvir}, which greatly suffered from the long gap of 7.3 d
in the middle of the superoutburst.  Our superhump period
(subsection \ref{sec:sh}) was accurate enough to unambiguously determine
the cycle counts of all available observations, which are summarized
in table \ref{tab:sh}.  This result indicates that the cycle count
in \citet{lei94hvvir} was in error by 1 cycle, leading to an erroneous
interpretation of the period and its change.

   \citet{lei94hvvir} further identified the period as the possible orbital
period later in their observations.  This period, however, is more
likely identified with late superhumps (\cite{hae79lateSH};
\cite{vog83lateSH}; \cite{vanderwoe88lateSH}; \cite{hes92lateSH}),
which are known to have similar periods with ordinary superhumps
(i.e. a few percent longer than $P_{\rm orb}$), but have phases $\sim$0.5
different from those of ordinary superhumps.
The late superhumps are known to persist for a long time (several tens of
days) in WZ Sge-type outbursts (\pasjcitesub{kat97egcnc}{a};
\pasjcitesub{pat98egcnc}{a}).
The reported period of 0.05799(3) d by \citet{lei94hvvir} within the
range of variation of the superhump period (cf. equation \ref{equ:reg2b})
also supports this identification.

\begin{table}
\caption{Timings of superhumps}\label{tab:sh}
\begin{center}
\begin{tabular}{lrrrc}
\hline\hline
HJD\commenta & $E$\commentb & $O-C_1$\commentc & $O-C_2$\commentd
             & Ref.\commente \\
\hline
42.8728 &   0 &  0.0064 & $-$0.0023 & 2 \\
43.7448 &  15 &  0.0041 & $-$0.0006 & 2 \\
43.8653 &  17 &  0.0081 &  0.0038 & 2 \\
43.9198 &  18 &  0.0043 &  0.0002 & 2 \\
43.9780 &  19 &  0.0042 &  0.0004 & 1 \\
44.0362 &  20 &  0.0041 &  0.0005 & 1 \\
44.0978 &  21 &  0.0074 &  0.0041 & 1 \\
44.1471 &  22 & $-$0.0016 & $-$0.0047 & 1 \\
45.9520 &  53 & $-$0.0035 & $-$0.0015 & 1 \\
46.0104 &  54 & $-$0.0034 & $-$0.0013 & 1 \\
46.0689 &  55 & $-$0.0032 & $-$0.0010 & 1 \\
46.1272 &  56 & $-$0.0032 & $-$0.0009 & 1 \\
46.1840 &  57 & $-$0.0046 & $-$0.0022 & 1 \\
46.9430 &  70 & $-$0.0033 &  0.0001 & 1 \\
47.0066 &  71 &  0.0020 &  0.0055 & 1 \\
47.0573 &  72 & $-$0.0056 & $-$0.0021 & 1 \\
47.1178 &  73 & $-$0.0034 &  0.0002 & 1 \\
47.1766 &  74 & $-$0.0029 &  0.0008 & 1 \\
47.2396 &  75 &  0.0018 &  0.0055 & 1 \\
47.9893 &  88 & $-$0.0062 & $-$0.0022 & 1 \\
48.0480 &  89 & $-$0.0058 & $-$0.0018 & 1 \\
48.1069 &  90 & $-$0.0051 & $-$0.0012 & 1 \\
48.1656 &  91 & $-$0.0047 & $-$0.0008 & 1 \\
48.2261 &  92 & $-$0.0025 &  0.0014 & 1 \\
48.9847 & 105 & $-$0.0016 &  0.0018 & 1 \\
49.0420 & 106 & $-$0.0026 &  0.0008 & 1 \\
49.1552 & 108 & $-$0.0060 & $-$0.0027 & 1 \\
51.2592 & 144 & $-$0.0002 & $-$0.0015 & 2 \\
52.2552 & 161 &  0.0049 & $-$0.0000 & 2 \\
52.3162 & 162 &  0.0076 &  0.0024 & 2 \\
52.3710 & 163 &  0.0041 & $-$0.0013 & 2 \\
52.4311 & 164 &  0.0060 &  0.0003 & 2 \\
52.4897 & 165 &  0.0063 &  0.0003 & 2 \\
\hline
 \multicolumn{5}{l}{\commenta HJD-2448700.} \\
 \multicolumn{5}{l}{\commentb Cycle count.} \\
 \multicolumn{5}{l}{\commentc Against equation \ref{equ:reg2a}.} \\
 \multicolumn{5}{l}{\commentd Against equation \ref{equ:reg2b}.} \\
 \multicolumn{5}{l}{\commente 1: this paper, 2: \citep{lei94hvvir}} \\
\end{tabular}
\end{center}
\end{table}

   A linear regression to the superhump times gives the following
ephemeris.

\begin{equation}
\rm{HJD_{max}} = 2448742.8664(16) + 0.058285(17) E. \label{equ:reg2a}
\end{equation}

\begin{figure}
  \begin{center}
    \FigureFile(88mm,60mm){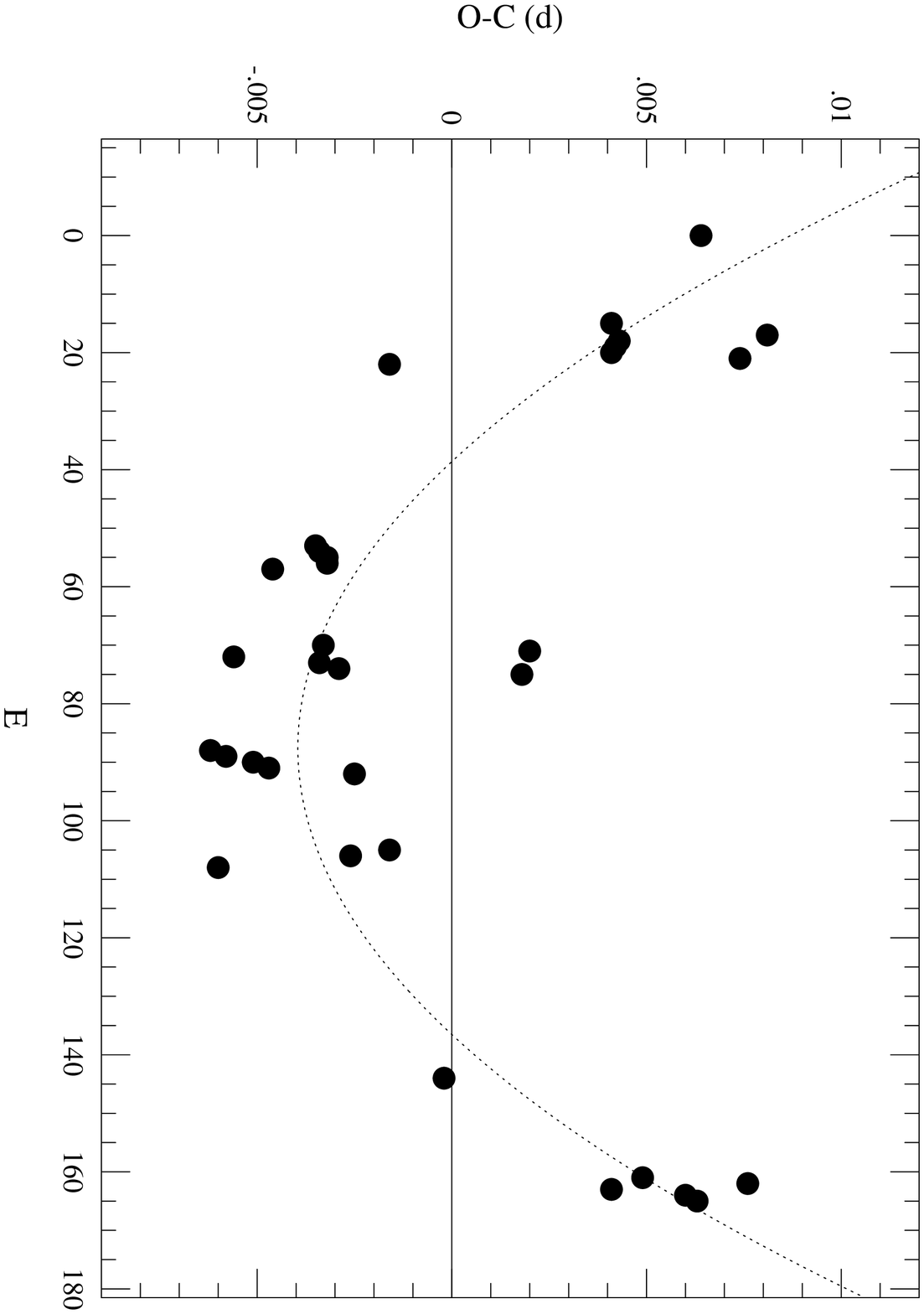}
  \end{center}
  \caption{$O-C$ diagram of superhump maxima.
  The parabolic fit corresponds to equation \ref{equ:reg2b}.}
  \label{fig:ocsh}
\end{figure}

   The $O-C$'s ($O-C_1$ in table \ref{tab:sh}; figure \ref{fig:ocsh})
against this ephemeris indicates that the superhump period ($P_{\rm SH}$)
became longer, in contrast to the conclusion by \citet{lei94hvvir}.
As shown in $O-C_2$ in table \ref{tab:sh}, the times of superhump
maxima in this interval can be well represented by the following
quadratic equation.

\begin{eqnarray}
\rm{HJD_{max}} & = 2448742.8751(12) + 0.057996(31) E \nonumber \\
    & + 1.65(17) \times 10^{-6} E^2. \label{equ:reg2b}
\end{eqnarray}

   The quadratic term corresponds to $\dot{P}/P$=+5.7(0.6) $\times$
10$^{-5}$.

\begin{figure}
  \begin{center}
    \FigureFile(88mm,60mm){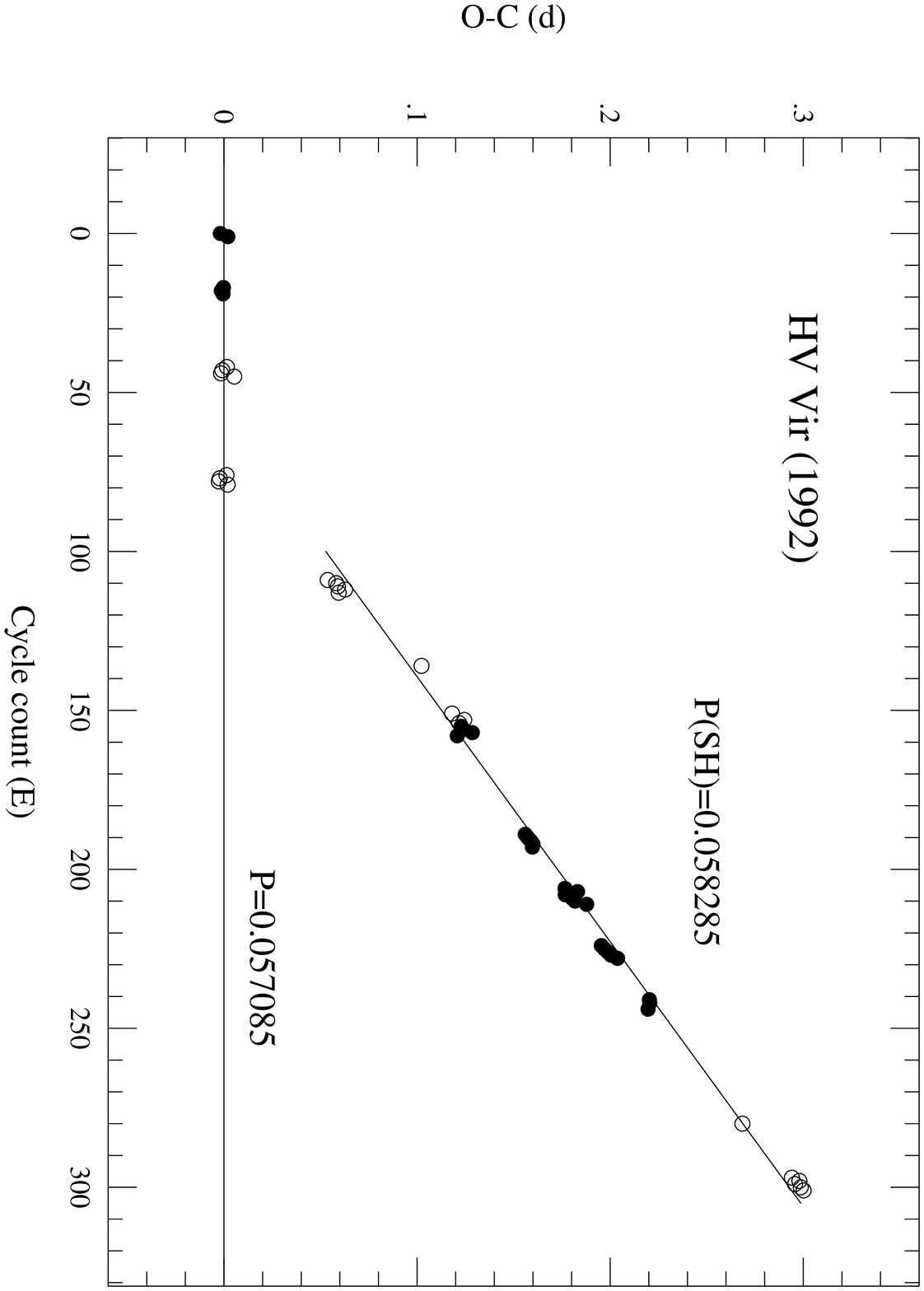}
  \end{center}
  \caption{$O-C$ diagram of against the 0.057085 d period (the period of
  early superhumps) throughout the superoutburst.  $E$ is the cycle
  count since HJD 2448735.0069 (cf. equation \ref{equ:reg1}).  Filled and
  open circles correspond to maxima timings by this work and by
  \citet{lei94hvvir}, respectively.  This diagram corresponds to figure 1
  of \citet{boh79wzsge} and figure 4 of \citet{pat81wzsge} for the 1978
  outburst of WZ Sge.}
  \label{fig:ocall}
\end{figure}

   Figure \ref{fig:ocall} represents the $O-C$ diagram of superhump
maximum of against the period of 0.057085 d (the period of early superhumps)
throughout the superoutburst.  $E$ is the cycle count since HJD 2448735.0069
(cf. equation \ref{equ:reg1}).  This diagram corresponds to figure 4 of
\citet{pat81wzsge} for WZ Sge.  The marked resemblance of the overall
$O-C$ variation to that of WZ Sge also strengthens the similarity of
HV Vir to WZ Sge.

\subsection{Early Superhumps}\label{sec:dis:earlysh}

   As described in subsection \ref{sec:earlyhump} and \ref{sec:sh}, two
distinct types of periodic humps appeared in different stages of the
outburst. The longer period (83.80 min) is identified by many authors as
genuine superhumps from their characteristics of the hump light curves.
The similarity of the general light curve in this stage (see subsection
\ref{sec:lc}) with those of superoutbursts in usual SU UMa stars also
supports this identification.

   As described in subsection \ref{sec:earlyhump}, the 82.20-min periodicity
showed a rather complex, doubly humped profile, with a narrower minimum.
All of superoutbursts of well-observed WZ Sge-type dwarf novae are
recently recognized to commonly show semi-periodic modulations at the
earliest stage of superoutburst (AL Com: \pasjcitetsub{kat96alcom}{b};
EG Cnc: \pasjcitetsub{mat98egcnc}{b}; RZ Leo: \citet{ish00rzleoiauc};
\pasjcitetsub{ish01rzleo}{a}; WZ Sge in 2001:
\pasjcitetsub{ish01wzsgeiauc7669}{b};
\pasjcitet2subyear{kat01wzsgeiauc7672}{kat01wzsgeiauc7678}{b,f}).
These modulations are called ``early superhumps"
(e.g. \pasjcitetsub{kat98super}{a}
  \footnote{
  This feature is also referred to as ``orbital" superhumps
  (\pasjcite2subyear{kat96alcomproc}{kat96alcom}{a,b} or
  {\it outburst orbital hump} \pasjcitepsub{pat98egcnc}{a}.
  }).
Since the early-stage modulations in HV Vir bear all characteristics
common to other WZ Sge-type dwarf novae, we regard the 82.20-min
periodicity as early superhumps.

   Since early superhumps are known to have the same or extremely close
periods to orbital periods in all well-observed cases, we propose this
period to be the orbital period of HV Vir, although the identification
should be confirmed by future radial velocity and photometric observations.
The proposed orbital period 0.05708 d (82.20 min) is one of the shortest
among well-established dwarf novae.

   The resultant fractional superhump excess
($\epsilon=P_{\rm SH}/P_{\rm orb}-1$) is 2.0\%.  This excess corresponds
to an extreme mass-ratio ($q$=$M_2/M_1$) of 0.1 \citep{mol92SHexcess}.

\subsection{Delay of the Superhump Development}

   The superhumps are now widely believed to be caused by the
tidal instability of the accretion disks (\cite{whi88tidal};
\cite{hir90SHexcess}; 
\authorcite{lub91SHa} (\yearcite{lub91SHb}a,b, \yearcite{lub92SH})),
and the delay of the superhump development offers a suitable measure of
the growth rate of the tidal instability.

   The full development of superhumps in HV Vir took 7 to 9 d following
the visual maximum, in contrast to 2 to 4 d in usual SU UMa stars.
This delay is similar to
WZ Sge (10 d for the 1978 outburst:
          \cite{boh79wzsge}; \cite{pat81wzsge};
        12 d for the 2001 outburst:
          \pasjcitesub{kat01wzsgeiauc7678}{f}),
AL Com ($\sim$7 d: \pasjcitesub{nog97alcom}{a};
        \cite{pat96alcom}),
EG Cnc ($\sim$8 d: \pasjcitesub{kat97egcnc}{a}; \pasjcitesub{pat98egcnc}{a})
and WX Cet (4 -- 7 d: \cite{odo91wzsge};
more recent observations of WX Cet by \pasjcitetsub{kat01wxcet}{c},
however, showed that the delay is less than 2 d, suggesting that this
star is more related to a usual SU UMa-type dwarf nova).
This finding is compatible with the slow growth rate, which has been shown
to be proportional to the square of the mass ratio $q = M_{2}/M_{1}$
(\pasjcite2subyear{lub91SHa}{lub91SHb}{a,b}), expected from the estimated
small mass-ratio of HV Vir, and also strengthens the classification of
HV Vir as a WZ Sge-type object.

\subsection{Superhump Period Change}

   The periods of ``textbook" superhumps in usual SU UMa-type dwarf
novae are known to decrease during superoutburst (e.g. \cite{war85suuma};
\cite{pat93vyaqr}).
The period derivative ($P_{\rm dot} = \dot{P}/P$) has a rather common
negative value ($\sim -5 \times 10^{-5}$), which has been generally
attributed to the decreasing apsidal motion due to the decreasing disk
radius.  However, recent observations (including the present work)
have shown a number of systems which show positive period derivatives.
Since many of them are associated with either WZ Sge-like systems,
or systems with short $P_{\rm orb}$ \pasjcitepsub{kat98super}{a},
we have made a systematic survey of observed
$P_{\rm dot}$ in SU UMa-type dwarf novae.  The result is summarized in
table \ref{tab:pdot}.

\begin{table}
\caption{Superhump period change}\label{tab:pdot}
\setcounter{hvref}{0}
\setcounter{hvrem}{0}
\begin{center}
\begin{tabular}{llcc}
\hline\hline
Object\commenta & $P_{\rm SH}$\commentb & $P_{\rm dot}$\commentc & Ref. \\
\hline
V485 Cen      & 0.04216 & 28(3)       & \ref{count:pdot:v485cen-1} \\
WZ Sge (1978) & 0.05722$^{\ref{rem:pdot:wzsge-1}}$ & $-$1(4)
                                      & \ref{count:pdot:wzsge-1} \\
WZ Sge (2001) & 0.05719 & 0.1(0.8)$^{\ref{rem:pdot:wzsge-2}}$
                                      & \ref{count:pdot:wzsge-2} \\
AL Com (1995) & 0.0572  & 2.1(0.3)    & \ref{count:pdot:alcom-1} \\
HV Vir        & 0.05820 & 5.7(0.6)    & \ref{count:pdot:hvvir-1} \\
SW UMa (1991) & 0.0583  &   6(4)      & \ref{count:pdot:swuma-1} \\
SW UMa (1996) & 0.0583  & 4.4(0.4)    & \ref{count:pdot:swuma-2} \\
WX Cet (1996) & 0.0593  &   4(2)      & \ref{count:pdot:wxcet-1} \\
WX Cet (1998) & 0.05949 & 8.5(1.0)    & \ref{count:pdot:wxcet-2} \\
T Leo         & 0.0602  & $-$0.5(0.3)$^{\ref{rem:pdot:tleo-1}}$
                                      & \ref{count:pdot:tleo-1} \\
EG Cnc        & 0.06038 & 2.0(0.4)    & \ref{count:pdot:egcnc-1} \\
V1028 Cyg     & 0.06154 & 8.7(0.9)    & \ref{count:pdot:v1028cyg-1} \\
V1159 Ori     & 0.0642  & $-$3.2(1)   & \ref{count:pdot:v1159ori-1} \\
VY Aqr        & 0.0644  & $-$8(2)     & \ref{count:pdot:vyaqr-1} \\
OY Car        & 0.06443 & $-$5(2)     & \ref{count:pdot:oycar-1} \\
UV Per        & 0.06641 & $-$2.0(1)   & \ref{count:pdot:uvper-1} \\
CT Hya        & 0.06643 & $-$2(8)     & \ref{count:pdot:cthya-1} \\
SX LMi        & 0.0685  & $-$8(2)     & \ref{count:pdot:sxlmi-1} \\
RZ Sge (1994) & 0.07042 & $-$10(2)    & \ref{count:pdot:rzsge-1} \\
RZ Sge (1996) & 0.07039 & $-$11.5(1)  & \ref{count:pdot:rzsge-2} \\
CY UMa        & 0.0724  & $-$5.8(1.4) & \ref{count:pdot:cyuma-1} \\
V1251 Cyg     & 0.07604 & $-$12(4)    & \ref{count:pdot:v1251cyg-1} \\
VW Hyi        & 0.07714 & $-$6.5(0.6) & \ref{count:pdot:vwhyi-1} \\
Z Cha         & 0.07740 & $-$4(2)     & \ref{count:pdot:zcha-1} \\
SU UMa        & 0.0788  & $-10$(3)    & \ref{count:pdot:suuma-1} \\
HS Vir        & 0.08077 & $-$4(1)     & \ref{count:pdot:hsvir-1} \\
EF Peg (1991) & 0.0871  & $-$2.2(1)$^{\ref{rem:pdot:efpeg-1}}$
                                      & \ref{count:pdot:efpeg-1} \\
V344 Lyr      & 0.09145 & $-$0.8(0.4) & \ref{count:pdot:v344lyr-1} \\
YZ Cnc        & 0.09204 & $-$7(2)     & \ref{count:pdot:yzcnc-1} \\
TU Men        & 0.1262  & $-$9(2)     & \ref{count:pdot:tumen-1} \\
\hline
 \multicolumn{4}{l}{\commenta Year of the outburst in parentheses.} \\
 \multicolumn{4}{l}{\commentb Typical superhump period (d).} \\
 \multicolumn{4}{l}{\commentc $\dot{P}/P$ unit in 10$^{-5}$.} \\
\end{tabular}
\end{center}
{\footnotesize
  {\bf Remarks:}
   \refstepcounter{hvrem}\label{rem:pdot:wzsge-1}
    \arabic{hvrem}. -- \citet{pat81wzsge} originally gave a period of
    0.05714 d.  However, more precise identifications of superhump maxima
    based on the more accurately determined superhump period on the
    2001 superoutburst have refined the period as given in the table
    \pasjcitepsub{kat01wzsgealert6285}{c}.
   \refstepcounter{hvrem}\label{rem:pdot:wzsge-2}
    \arabic{hvrem}. -- Period and period derivative determined from the
    observations of the VSNET Collaboration team
    (\pasjcitesub{ish01wzsgeiauc7669}{b};
    \authorcite{kat01wzsgeiauc7672} \yearcite{kat01wzsgeiauc7678}b,e,f).
   \refstepcounter{hvrem}\label{rem:pdot:tleo-1}
    \arabic{hvrem}. -- $P_{\rm dot}$ through the entire superoutburst
    in 1993.  \citet{lem93tleo} gave $-$6 $\times$ 10$^{-5}$ for the initial
    part of the same superoutburst.  See \pasjcitetsub{kat97tleo}{a}
    for a complete discussion.
   \refstepcounter{hvrem}\label{rem:pdot:efpeg-1}
    \arabic{hvrem}. -- The value is from \pasjcitetsub{kat01efpeg}{a}.
    Matsumoto et al. (in preparation) gave a virtually zero $P_{\rm dot}$
    for the 1997 superoutburst.

  {\bf References:}
   \refstepcounter{hvref}\label{count:pdot:v485cen-1}
    \arabic{hvref}. \citet{ole97v485cen} ;
   \refstepcounter{hvref}\label{count:pdot:wzsge-1}
                         \label{count:pdot:vyaqr-1}
                         \label{count:pdot:oycar-1}
                         \label{count:pdot:vwhyi-1}
                         \label{count:pdot:zcha-1}
                         \label{count:pdot:suuma-1}
                         \label{count:pdot:yzcnc-1}
                         \label{count:pdot:tumen-1}
    \arabic{hvref}. \citet{pat93vyaqr} ;
   \refstepcounter{hvref}\label{count:pdot:wzsge-2}
    \arabic{hvref}. Kato et al., in preparation ;
   \refstepcounter{hvref}\label{count:pdot:alcom-1}
    \arabic{hvref}. \pasjcitetsub{nog97alcom}{a} ;
   \refstepcounter{hvref}\label{count:pdot:hvvir-1}
    \arabic{hvref}. this work ;
   \refstepcounter{hvref}\label{count:pdot:swuma-1}
    \arabic{hvref}. Kato, in preparation ;
   \refstepcounter{hvref}\label{count:pdot:swuma-2}
    \arabic{hvref}. \citet{nog98swuma} ;
   \refstepcounter{hvref}\label{count:pdot:wxcet-1}
    \arabic{hvref}. Nogami et al., in preparation ;
   \refstepcounter{hvref}\label{count:pdot:wxcet-2}
    \arabic{hvref}. \pasjcitetsub{kat01wxcet}{c} ;
   \refstepcounter{hvref}\label{count:pdot:tleo-1}
    \arabic{hvref}. \pasjcitetsub{kat97tleo}{a} ;
   \refstepcounter{hvref}\label{count:pdot:egcnc-1}
    \arabic{hvref}. \pasjcitetsub{kat97egcnc}{a}, reanalyzed together with
    the data by \pasjcitesub{pat98egcnc}{a} ;
   \refstepcounter{hvref}\label{count:pdot:v1028cyg-1}
    \arabic{hvref}. \citet{bab00v1028cyg} ;
   \refstepcounter{hvref}\label{count:pdot:v1159ori-1}
    \arabic{hvref}. \pasjcitetsub{pat95v1159ori}{b} ;
   \refstepcounter{hvref}\label{count:pdot:uvper-1}
    \arabic{hvref}. Kato, in preparation ;
   \refstepcounter{hvref}\label{count:pdot:cthya-1}
    \arabic{hvref}. \pasjcitetsub{kat99cthya}{a} ;
   \refstepcounter{hvref}\label{count:pdot:sxlmi-1}
    \arabic{hvref}. \pasjcitetsub{nog97sxlmi}{b} ;
   \refstepcounter{hvref}\label{count:pdot:rzsge-1}
    \arabic{hvref}. \citet{kat96rzsge} ;
   \refstepcounter{hvref}\label{count:pdot:rzsge-2}
    \arabic{hvref}. \pasjcitetsub{sem97rzsge}{a} ;
   \refstepcounter{hvref}\label{count:pdot:cyuma-1}
    \arabic{hvref}. \citet{har95cyuma} ;
   \refstepcounter{hvref}\label{count:pdot:v1251cyg-1}
    \arabic{hvref}. \pasjcitetsub{kat95v1251cyg}{a}, reanalyzed ;
   \refstepcounter{hvref}\label{count:pdot:hsvir-1}
    \arabic{hvref}. \pasjcitetsub{kat98hsvir}{b} ;
   \refstepcounter{hvref}\label{count:pdot:efpeg-1}
    \arabic{hvref}. \pasjcitetsub{kat01efpeg}{a};
   \refstepcounter{hvref}\label{count:pdot:v344lyr-1}
    \arabic{hvref}. \citet{kat93v344lyr} ;
}
\end{table}

\begin{figure}
  \begin{center}
    \FigureFile(88mm,60mm){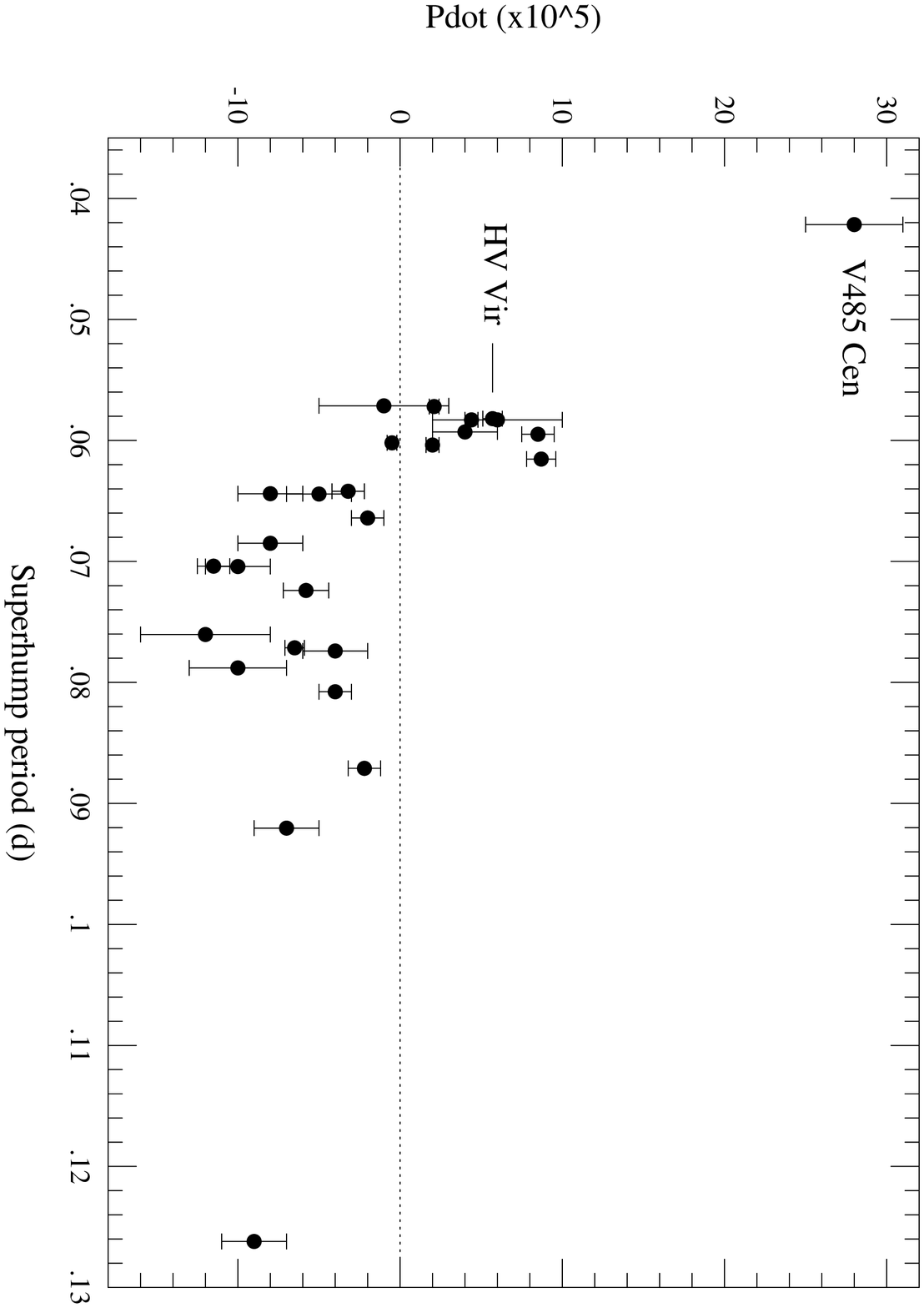}
  \end{center}
  \caption{$\dot{P}/P$ versus $P_{\rm SH}$ drawn from table
  \ref{tab:pdot}.  While most of long-period systems show a decrease of
  superhump periods, short-period systems or infrequently outbursting
  SU UMa-type systems tend to show an increase of superhump periods.
  Although $\dot{P}/P$ seems to comprise a continuum against $P_{\rm SH}$,
  systems with $P_{\rm SH}$ shorter than 0.062 d show a strong tendency
  of positive $\dot{P}/P$.}
  \label{fig:pdot}
\end{figure}

   Figure \ref{fig:pdot} presents the relation between $\dot{P}/P$ and
$P_{\rm SH}$.  While most of long-period systems show the ``textbook"
{\it decrease} of superhump periods, short-period systems or infrequently
outbursting SU UMa-type systems (also listed as ``large-amplitude
SU UMa-type dwarf novae" in table \ref{tab:candidate} predominantly
show an {\it increase} of superhump periods.  This concentration of
positive $P_{\rm dot}$ systems among short-period systems may be either
a consequence of low $q$ and/or low mass-transfer rate ($\dot{M}$).
Recent discoveries of zero or marginally positive $\dot{P}$ systems
(V725 Aql: \citet{uem01v725aql}; EF Peg: Matsumoto et al, in preparation)
in long $P_{\rm orb}$ systems may be a suggestion that low $\dot{M}$
is more responsible, but the origin of this phenomenon is still not
well understood (cf. \pasjcitesub{kat01wxcet}{c}).

\subsection{Guideline to a Future Search for WZ Sge Stars}

   Previous searches for WZ Sge stars (e.g. \cite{odo91wzsge})
have been rather unsuccessful because because the search has been confined
to known dwarf novae with two or more outbursts.  As in the case of HV Vir
and WZ Sge, the cycle length of WZ Sge-type stars are very long, and they
may have experienced only one outburst in recent years.  Although light
curves of WZ Sge stars resemble that of a fast nova, making it difficult
to discriminate such object from galactic novae, a number of WZ Sge-type
dwarf novae are expected among historical records of galactic novae.

   A criterion for identifying WZ Sge type objects amongst galactic novae
is the rate-of-decline vs outburst amplitude of novae.  Fast novae usually
show amplitudes greater than 10 mag,
which is in agreement with the current picture of galactic novae,
having a quiescent absolute magnitude of M$_{\rm V}$=+4
(\cite{war87CVabsmag}; \cite{can98DNabsmag}; \cite{sma00DNabsmag}).
The method is easily applied to fast novae without spectroscopic
confirmation, but with measured quiescent magnitudes.

   Based on these criteria, and based on a comprehensive study in
the past observations of dwarf novae and candidates, we present a new set
of promising candidates of WZ Sge-type stars which need to be carefully
studied.  Table \ref{tab:wzsgemember} lists properties of the confirmed
WZ Sge-type dwarf novae.  Other probable WZ Sge-type dwarf novae are
listed in table \ref{tab:probablewzsge}.  Candidates are given in
table \ref{tab:candidate}.  A table of large-amplitude SU UMa-type
dwarf novae (table \ref{tab:largeampsuuma}) is also given, which have
some common properties with WZ Sge-type dwarf novae, or which were listed
as WZ Sge-type candidates in the past literature.  The existence of
superhumps in these systems have been established (i.e. established
SU UMa-type dwarf novae).  None of these systems, however, were observed
to show definite early superhumps.

   Some of the objects selected from possible old novae have been
observed in quiescence, and are shown to have cataclysmic nature
(for example, see \cite{muk90faintCV}).  Further spectroscopic observations
of these objects will allow determination of quiescent mass transfer
rate which is expected to be extremely low in WZ Sge stars
\citep{osa95wzsge},
in good contrast to galactic novae with high mass-transfer rates.
Furthermore, photometric observations in quiescence may provide
information about their orbital periods.  Since all well-studied
WZ Sge-type dwarf novae have orbital periods below 2 hr, and may be
photometrically selected against galactic novae, which usually have
orbital periods longer than 3 hr.

\section{Conclusion}

(1) HV Vir, previously recognized as a galactic nova, was 
observed in outburst in 1992 April --- May, first time since 1929.
The general features of the light curve, together with its
large amplitude (7.7 mag) and long recurrence time (typically longer
than $\sim$10 yrs) closely resemble to those of WZ Sge, an enigmatic
dwarf nova, in all respects.

(2) Photoelectric and CCD photometry in April 22 --- 23 and May 1 -- 6
revealed two distinct types of periodic modulation: 82.20-min period
and 83.80-min superhumps.  The detection of superhumps confirms the
SU UMa-type nature of the object.

(3) The 82.20-min periodicity was detected only in the earliest stage
of the outburst, and soon decayed.  Based on striking similarity with
``early superhumps" observed in recently established WZ Sge-type dwarf
novae, we identified the periodicity as ``early superhumps", which
likely have an identical period with the orbital period.

(4) The small fractional superhump excess
($\epsilon=P_{\rm SH}/P_{\rm orb}-1$) of 2.0\% implies an extreme 
($q$=$M_2/M_1$$\sim$0.1) binary mass ratio.
This binary parameter, combined with theoretical models, is compatible
with the slow development of superhumps, and satisfies the required
condition for the WZ Sge-type phenomenon.

(5) The present success in identifying new WZ Sge-type objects
encouraged a further search for analogs in the literature.
A new comprehensive list of candidates for WZ Sge-type objects is
presented.

\vskip 3mm

The authors are grateful to the staffs of the Dodaira Station for helping
with the observations. We are also grateful to VSNET members, to Gary Poyner
and Bill Worraker (BAA), and the members of Variable Stars Observers
League in Japan (VSOLJ) for supplying visual observations.

\newpage

\begin{table*}
\caption{Confirmed WZ Sge-type dwarf novae.}\label{tab:wzsgemember}
\setcounter{hvref}{0}
\setcounter{hvrem}{0}
\begin{center}
\begin{tabular}{lcccllcl}
\hline\hline
Object    & Max  & Min & Amplitude\commenta & P$_{\rm orb}$\commentb
            & P$_{\rm SH}$\commentc & Remakes & References \\
\hline
WZ Sge    &  7.0 & 15.5 &  8.5 & 0.05667 & 0.05722
                                   & -
                                     & \ref{count:confirmed:wzsge-1},
                                       \ref{count:confirmed:wzsge-2},
                                       \ref{count:confirmed:wzsge-3},
                                       \ref{count:confirmed:wzsge-4},
                                       \ref{count:confirmed:wzsge-5},
                                       \ref{count:confirmed:wzsge-6},
                                       \ref{count:confirmed:wzsge-7},
                                       \ref{count:confirmed:wzsge-8},
                                       \ref{count:confirmed:wzsge-10},
                                       \ref{count:confirmed:wzsge-11},
                                       \ref{count:confirmed:wzsge-12},
                                       \ref{count:confirmed:wzsge-13},
                                       \ref{count:confirmed:wzsge-14}, \\
          &      &      &       &  &&& \ref{count:confirmed:wzsge-15},
                                       \ref{count:confirmed:wzsge-16},
                                       \ref{count:confirmed:wzsge-17},
                                       \ref{count:confirmed:wzsge-18},
                                       \ref{count:confirmed:wzsge-19},
                                       \ref{count:confirmed:wzsge-20},
                                       \ref{count:confirmed:wzsge-21},
                                       \ref{count:confirmed:wzsge-22},
                                       \ref{count:confirmed:wzsge-23},
                                       \ref{count:confirmed:wzsge-24},
                                       \ref{count:confirmed:wzsge-25}, \\
          &      &      &       &  &&& \ref{count:confirmed:wzsge-26},
                                       \ref{count:confirmed:wzsge-27},
                                       \ref{count:confirmed:wzsge-28},
                                       \ref{count:confirmed:wzsge-29},
                                       \ref{count:confirmed:wzsge-30},
                                       \ref{count:confirmed:wzsge-31},
                                       \ref{count:confirmed:wzsge-32},
                                       \ref{count:confirmed:wzsge-33},
                                       \ref{count:confirmed:wzsge-34},
                                       \ref{count:confirmed:wzsge-35},
                                       \ref{count:confirmed:wzsge-36}, \\
          &      &      &       &  &&& \ref{count:confirmed:wzsge-37},
                                       \ref{count:confirmed:wzsge-38},
                                       \ref{count:confirmed:wzsge-39},
                                       \ref{count:confirmed:wzsge-40},
                                       \ref{count:confirmed:wzsge-41},
                                       \ref{count:confirmed:wzsge-42},
                                       \ref{count:confirmed:wzsge-43},
                                       \ref{count:confirmed:wzsge-44},
                                       \ref{count:confirmed:wzsge-45},
                                       \ref{count:confirmed:wzsge-46},
                                       \ref{count:confirmed:wzsge-47}, \\
          &      &      &       &  &&& \ref{count:confirmed:wzsge-48},
                                       \ref{count:confirmed:wzsge-49},
                                       \ref{count:confirmed:wzsge-50},
                                       \ref{count:confirmed:wzsge-51},
                                       \ref{count:confirmed:wzsge-52},
                                       \ref{count:confirmed:wzsge-53},
                                       \ref{count:confirmed:wzsge-54},
                                       \ref{count:confirmed:wzsge-55},
                                       \ref{count:confirmed:wzsge-56},
                                       \ref{count:confirmed:wzsge-57},
                                       \ref{count:confirmed:wzsge-58}, \\
          &      &      &       &  &&& \ref{count:confirmed:wzsge-59},
                                       \ref{count:confirmed:wzsge-60},
                                       \ref{count:confirmed:wzsge-61},
                                       \ref{count:confirmed:wzsge-62},
                                       \ref{count:confirmed:wzsge-63},
                                       \ref{count:confirmed:wzsge-64} \\
AL Com    & 12.8 & 20.5 &  7.7 & 0.05666 & 0.05722
                                   & \ref{rem:confirmed:earlysh}
                                     & \ref{count:confirmed:alcom-1},
                                       \ref{count:confirmed:alcom-2},
                                       \ref{count:confirmed:alcom-3},
                                       \ref{count:confirmed:alcom-4},
                                       \ref{count:confirmed:alcom-5},
                                       \ref{count:confirmed:alcom-6},
                                       \ref{count:confirmed:alcom-7},
                                       \ref{count:confirmed:alcom-8},
                                       \ref{count:confirmed:alcom-9},
                                       \ref{count:confirmed:alcom-10},
                                       \ref{count:confirmed:alcom-11}, \\
          &      &      &       &  &&& \ref{count:confirmed:alcom-12},
                                       \ref{count:confirmed:alcom-13},
                                       \ref{count:confirmed:alcom-14},
                                       \ref{count:confirmed:alcom-15},
                                       \ref{count:confirmed:alcom-16},
                                       \ref{count:confirmed:alcom-17},
                                       \ref{count:confirmed:alcom-18},
                                       \ref{count:confirmed:alcom-19},
                                       \ref{count:confirmed:alcom-20},
                                       \ref{count:confirmed:alcom-21},
                                       \ref{count:confirmed:alcom-22}, \\
          &      &      &       &  &&& \ref{count:confirmed:alcom-23},
                                       \ref{count:confirmed:alcom-24},
                                       \ref{count:confirmed:alcom-25},
                                       \ref{count:confirmed:alcom-26},
                                       \ref{count:confirmed:alcom-27},
                                       \ref{count:confirmed:alcom-28},
                                       \ref{count:confirmed:alcom-29},
                                       \ref{count:confirmed:alcom-30},
                                       \ref{count:confirmed:alcom-31} \\
EG Cnc    & 11.9 & 18.0 &  6.1 & 0.05877 & 0.06038
                                   & \ref{rem:confirmed:earlysh},
                                     \ref{rem:confirmed:egcnc-1}
                                     & \ref{count:confirmed:egcnc-1},
                                       \ref{count:confirmed:egcnc-2},
                                       \ref{count:confirmed:egcnc-3},
                                       \ref{count:confirmed:egcnc-4},
                                       \ref{count:confirmed:egcnc-5},
                                       \ref{count:confirmed:egcnc-6},
                                       \ref{count:confirmed:egcnc-7},
                                       \ref{count:confirmed:egcnc-8},
                                       \ref{count:confirmed:egcnc-9} \\
V2176 Cyg & 13.3 & 19.9 &  6.6 & - & 0.0561
                                   & \ref{rem:confirmed:v2176cyg-1}
                                     & \ref{count:confirmed:v2176cyg-1},
                                       \ref{count:confirmed:v2176cyg-2},
                                       \ref{count:confirmed:v2176cyg-3},
                                       \ref{count:confirmed:v2176cyg-4},
                                       \ref{count:confirmed:v2176cyg-5} \\
HV Vir    & 11.5 & 19.2 &  7.7 & 0.05708 & 0.05820
                                   & \ref{rem:confirmed:earlysh}
                                     & \ref{count:confirmed:hvvir-1},
                                       \ref{count:confirmed:hvvir-2},
                                       \ref{count:confirmed:hvvir-3} \\
RZ Leo    & 12.3 & 19.3 &  7.0 & 0.07616\commentd & 0.07853
                                   & \ref{rem:confirmed:earlysh}
                                     & \ref{count:confirmed:rzleo-1},
                                       \ref{count:confirmed:rzleo-2},
                                       \ref{count:confirmed:rzleo-3},
                                       \ref{count:confirmed:rzleo-4},
                                       \ref{count:confirmed:rzleo-5},
                                       \ref{count:confirmed:rzleo-6},
                                       \ref{count:confirmed:rzleo-7},
                                       \ref{count:confirmed:rzleo-8}, \\
          &      &      &       &  &&& \ref{count:confirmed:rzleo-9},
                                       \ref{count:confirmed:rzleo-10},
                                       \ref{count:confirmed:rzleo-11},
                                       \ref{count:confirmed:rzleo-12},
                                       \ref{count:confirmed:rzleo-13},
                                       \ref{count:confirmed:rzleo-14},
                                       \ref{count:confirmed:rzleo-15},
                                       \ref{count:confirmed:rzleo-16}, \\
          &      &      &       &  &&& \ref{count:confirmed:rzleo-17},
                                       \ref{count:confirmed:rzleo-18} \\
\hline
 \multicolumn{7}{l}{\commenta Amplitude of outburst (mag)} \\
 \multicolumn{7}{l}{\commentb Orbital period (d).} \\
 \multicolumn{7}{l}{\commentc Superhump period (d).} \\
\end{tabular}
\end{center}
{\footnotesize
  {\bf Remarks:}
   \refstepcounter{hvrem}\label{rem:confirmed:earlysh}
    \arabic{hvrem}. -- P$_{\rm orb}$ is taken from the best estimate
    of the period of early superhumps.
   \refstepcounter{hvrem}\label{rem:confirmed:egcnc-1}
    \arabic{hvrem}. -- Period of early superhumps by
    \pasjcitetsub{mat98egcnc}{b} and \pasjcitetsub{kat97egcnc}{a},
    which is in agreement with the period by quiescent photometry by
    \pasjcitetsub{mat98egcncqui}{a}.
    \pasjcitetsub{pat98egcnc}{a} proposed a different (0.05997 d) period.
   \refstepcounter{hvrem}\label{rem:confirmed:v2176cyg-1}
    \arabic{hvrem}. -- The first indication of a WZ Sge-type dwarf nova
    was proposed by \pasjcitetsub{kat97v2176cyg}{b}.  The lack of further
    outbursts (VSNET) also supports this identification.
   \\
}
{\footnotesize
  {\bf References:}
   \refstepcounter{hvref}\label{count:confirmed:wzsge-1}
    \arabic{hvref}. \citet{mac19wzsge} ;
   \refstepcounter{hvref}\label{count:confirmed:wzsge-2}
    \arabic{hvref}. \citet{hum38} ;
   \refstepcounter{hvref}\label{count:confirmed:wzsge-3}
    \arabic{hvref}. \citet{him46wzsge} ;
   \refstepcounter{hvref}\label{count:confirmed:wzsge-4}
    \arabic{hvref}. \citet{bey51wzsge} ;
   \refstepcounter{hvref}\label{count:confirmed:wzsge-5}
    \arabic{hvref}. \citet{gre57wzsge} ;
   \refstepcounter{hvref}\label{count:confirmed:wzsge-6}
    \arabic{hvref}. \citet{kra62wzsge} ;
   \refstepcounter{hvref}\label{count:confirmed:wzsge-7}
    \arabic{hvref}. \citet{krz62wzsge} ;
   \refstepcounter{hvref}\label{count:confirmed:wzsge-8}
    \arabic{hvref}. \citet{esk63wzsge} ;
   \refstepcounter{hvref}\label{count:confirmed:wzsge-9}
    \arabic{hvref}. \citet{krz64wzsge} ;
   \refstepcounter{hvref}\label{count:confirmed:wzsge-10}
    \arabic{hvref}. \citet{war72wzsge} ;
   \refstepcounter{hvref}\label{count:confirmed:wzsge-11}
    \arabic{hvref}. \citet{rob78wzsge} ;
   \refstepcounter{hvref}\label{count:confirmed:wzsge-12}
    \arabic{hvref}. \citet{fab78wzsgeparameter} ;
   \refstepcounter{hvref}\label{count:confirmed:wzsge-13}
    \arabic{hvref}. \citet{pat78wzsgeiauc3311} ; 
   \refstepcounter{hvref}\label{count:confirmed:wzsge-14}
    \arabic{hvref}. \citet{nat78wzsgeiauc3311} ; 
   \refstepcounter{hvref}\label{count:confirmed:wzsge-15}
    \arabic{hvref}. \citet{kru78wzsgeiauc3312} ; 
   \refstepcounter{hvref}\label{count:confirmed:wzsge-16}
    \arabic{hvref}. \citet{bro78wzsgeiauc3313} ; 
   \refstepcounter{hvref}\label{count:confirmed:wzsge-17}
    \arabic{hvref}. \citet{wal78wzsgeiauc3315} ; 
   \refstepcounter{hvref}\label{count:confirmed:wzsge-18}
    \arabic{hvref}. \citet{kru79wzsgeiauc3318} ; 
   \refstepcounter{hvref}\label{count:confirmed:wzsge-19}
    \arabic{hvref}. \citet{gui79wzsgeiauc3319} ; 
   \refstepcounter{hvref}\label{count:confirmed:wzsge-20}
    \arabic{hvref}. \pasjcitetsub{tar79wzsgeiauc3320}{b} ; 
   \refstepcounter{hvref}\label{count:confirmed:wzsge-21}
    \arabic{hvref}. \pasjcitetsub{tar79wzsgeiauc3344}{c} ; 
   \refstepcounter{hvref}\label{count:confirmed:wzsge-22}
    \arabic{hvref}. \pasjcitetsub{tar79wzsge}{a} ; 
   \refstepcounter{hvref}\label{count:confirmed:wzsge-23}
    \arabic{hvref}. \citet{boh79wzsge} ; 
   \refstepcounter{hvref}\label{count:confirmed:wzsge-24}
    \arabic{hvref}. \citet{bro79wzsgeatsir} ;
   \refstepcounter{hvref}\label{count:confirmed:wzsge-25}
    \arabic{hvref}. \citet{hei79wzsge} ; 
   \refstepcounter{hvref}\label{count:confirmed:wzsge-26}
    \arabic{hvref}. \citet{bro79wzsge} ; 
   \refstepcounter{hvref}\label{count:confirmed:wzsge-27}
    \arabic{hvref}. \citet{cra79wzsgespec} ;
   \refstepcounter{hvref}\label{count:confirmed:wzsge-28}
    \arabic{hvref}. \citet{rit79wzsge} ;
   \refstepcounter{hvref}\label{count:confirmed:wzsge-29}
    \arabic{hvref}. \citet{sma79wzsge} ;
   \refstepcounter{hvref}\label{count:confirmed:wzsge-30}
    \arabic{hvref}. \citet{ort80wzsge} ;
   \refstepcounter{hvref}\label{count:confirmed:wzsge-31}
    \arabic{hvref}. \citet{bro80wzsgespec} ;
   \refstepcounter{hvref}\label{count:confirmed:wzsge-32}
    \arabic{hvref}. \citet{fab80wzsgeUV} ;
   \refstepcounter{hvref}\label{count:confirmed:wzsge-33}
    \arabic{hvref}. \citet{gil80wzsgeSH} ;
   \refstepcounter{hvref}\label{count:confirmed:wzsge-34}
    \arabic{hvref}. \citet{pat80wzsge} ;
   \refstepcounter{hvref}\label{count:confirmed:wzsge-35}
    \arabic{hvref}. \citet{wal80wzsgespec} ;
   \refstepcounter{hvref}\label{count:confirmed:wzsge-36}
    \arabic{hvref}. \citet{pat81wzsge} ;
   \refstepcounter{hvref}\label{count:confirmed:wzsge-37}
    \arabic{hvref}. \citet{fri81wzsge} ;
   \refstepcounter{hvref}\label{count:confirmed:wzsge-38}
    \arabic{hvref}. \citet{how81wzsge} ;
   \refstepcounter{hvref}\label{count:confirmed:wzsge-39}
    \arabic{hvref}. \citet{lei81wzsge} ;
   \refstepcounter{hvref}\label{count:confirmed:wzsge-40}
    \arabic{hvref}. \citet{voi83wzsge} ;
   \refstepcounter{hvref}\label{count:confirmed:wzsge-41}
    \arabic{hvref}. \citet{gil86wzsge} ;
   \refstepcounter{hvref}\label{count:confirmed:wzsge-42}
    \arabic{hvref}. \citet{sim87wzsge} ;
   \refstepcounter{hvref}\label{count:confirmed:wzsge-43}
    \arabic{hvref}. \citet{nay89wzsgeIUEoutburst} ;
   \refstepcounter{hvref}\label{count:confirmed:wzsge-44}
    \arabic{hvref}. \citet{sio90wzsge} ;
   \refstepcounter{hvref}\label{count:confirmed:wzsge-45}
                         \label{count:confirmed:alcom-1}
    \arabic{hvref}. \citet{ric92wzsgedip} ;
   \refstepcounter{hvref}\label{count:confirmed:wzsge-46}
    \arabic{hvref}. \citet{sio95wzsgeHST} ;
   \refstepcounter{hvref}\label{count:confirmed:wzsge-47}
    \arabic{hvref}. \citet{ski97wzsge} ;
   \refstepcounter{hvref}\label{count:confirmed:wzsge-48}
    \arabic{hvref}. \citet{wel97wzsgeUVoscillation} ;
   \refstepcounter{hvref}\label{count:confirmed:wzsge-49}
    \arabic{hvref}. \pasjcitetsub{pat98wzsge}{b} ;
   \refstepcounter{hvref}\label{count:confirmed:wzsge-50}
    \arabic{hvref}. \citet{neu98wzsge} ;
   \refstepcounter{hvref}\label{count:confirmed:wzsge-51}
    \arabic{hvref}. \citet{spr98wzsge} ;
   \refstepcounter{hvref}\label{count:confirmed:wzsge-52}
    \arabic{hvref}. \citet{ski99wzsge} ;
   \refstepcounter{hvref}\label{count:confirmed:wzsge-53}
    \arabic{hvref}. \citet{sle99wzsge} ;
   \refstepcounter{hvref}\label{count:confirmed:wzsge-54}
    \arabic{hvref}. \citet{mas00wzsge} ;
   \refstepcounter{hvref}\label{count:confirmed:wzsge-55}
    \arabic{hvref}. \citet{ski00wzsge} ;
   \refstepcounter{hvref}\label{count:confirmed:wzsge-56}
    \arabic{hvref}. \citet{sle00wzsge} ;
   \refstepcounter{hvref}\label{count:confirmed:wzsge-57}
    \arabic{hvref}. \pasjcitetsub{kat01wzsgeiauc7672}{b} ;
   \refstepcounter{hvref}\label{count:confirmed:wzsge-58}
    \arabic{hvref}. \citet{ste01wzsgeiauc7675} ;
   \refstepcounter{hvref}\label{count:confirmed:wzsge-59}
    \arabic{hvref}. \citet{whe01wzsgeiauc7677} ;
   \refstepcounter{hvref}\label{count:confirmed:wzsge-60}
    \arabic{hvref}. \citet{bab01wzsgeiauc7678} ;
   \refstepcounter{hvref}\label{count:confirmed:wzsge-61}
    \arabic{hvref}. \pasjcitetsub{kat01wzsgeiauc7678}{f} ;
   \refstepcounter{hvref}\label{count:confirmed:wzsge-62}
    \arabic{hvref}. \pasjcitetsub{kat01wzsgeiauc7680}{e} ;
   \refstepcounter{hvref}\label{count:confirmed:wzsge-63}
    \arabic{hvref}. \pasjcitetsub{kat01wzsgealert6285}{c} ;
   \refstepcounter{hvref}\label{count:confirmed:wzsge-64}
    \arabic{hvref}. Ishioka et al. in preparation;
      (See also $\langle$
      http://www.kusastro.kyoto-u.ac.jp/vsnet/DNe/wzsge01.html
      $\rangle$) ;
   \refstepcounter{hvref}\label{count:confirmed:alcom-2}
    \arabic{hvref}. \citet{ros61alcomiauc} ;
   \refstepcounter{hvref}\label{count:confirmed:alcom-3}
    \arabic{hvref}. \citet{ber64alcom} ;
   \refstepcounter{hvref}\label{count:confirmed:alcom-4}
    \arabic{hvref}. \citet{zwi65alcom} ;
   \refstepcounter{hvref}\label{count:confirmed:alcom-5}
    \arabic{hvref}. \pasjcitetsub{wen65alcom1}{a} ;
   \refstepcounter{hvref}\label{count:confirmed:alcom-6}
    \arabic{hvref}. \pasjcitetsub{wen65alcom2}{b} ;
   \refstepcounter{hvref}\label{count:confirmed:alcom-7}
    \arabic{hvref}. \citet{moo65alcom} ;
   \refstepcounter{hvref}\label{count:confirmed:alcom-8}
    \arabic{hvref}. \citet{muk90faintCV} ;
   \refstepcounter{hvref}\label{count:confirmed:alcom-9}
                         \label{count:confirmed:rzleo-1}
    \arabic{hvref}. \citet{how88faintCV1} ;
   \refstepcounter{hvref}\label{count:confirmed:alcom-10}
    \arabic{hvref}. \citet{szk89faintCV2} ;
   \refstepcounter{hvref}\label{count:confirmed:alcom-11}
    \arabic{hvref}. \citet{how91alcom} ;
   \refstepcounter{hvref}\label{count:confirmed:alcom-12}
    \arabic{hvref}. \pasjcitetsub{ric91alcom}{a} ;
   \refstepcounter{hvref}\label{count:confirmed:alcom-13}
    \arabic{hvref}. \citet{abb92alcomcperi} ;
   \refstepcounter{hvref}\label{count:confirmed:alcom-14}
    \arabic{hvref}. \pasjcitetsub{mat95alcomiauc}{a} ; 
   \refstepcounter{hvref}\label{count:confirmed:alcom-15}
    \arabic{hvref}. \pasjcitetsub{pat95alcomiauc}{a} ; 
   \refstepcounter{hvref}\label{count:confirmed:alcom-16}
    \arabic{hvref}. \citet{aug95alcomiauc} ; 
   \refstepcounter{hvref}\label{count:confirmed:alcom-17}
    \arabic{hvref}. \pasjcitetsub{pyc95alcomiauc}{a} ; 
   \refstepcounter{hvref}\label{count:confirmed:alcom-18}
    \arabic{hvref}. \citet{nog95alcomiauc} ; 
   \refstepcounter{hvref}\label{count:confirmed:alcom-19}
    \arabic{hvref}. \pasjcitetsub{pyc95alcom}{b} ;
   \refstepcounter{hvref}\label{count:confirmed:alcom-20}
    \arabic{hvref}. \pasjcitetsub{kat96alcomproc}{a} ;
   \refstepcounter{hvref}\label{count:confirmed:alcom-21}
    \arabic{hvref}. \pasjcitetsub{kat96alcom}{b} ;
   \refstepcounter{hvref}\label{count:confirmed:alcom-22}
    \arabic{hvref}. \citet{pat96alcom} ;
   \refstepcounter{hvref}\label{count:confirmed:alcom-23}
    \arabic{hvref}. \pasjcitetsub{how96alcom}{a} ;
   \refstepcounter{hvref}\label{count:confirmed:alcom-24}
    \arabic{hvref}. \citet{szk96alcomIUE} ;
   \refstepcounter{hvref}\label{count:confirmed:alcom-25}
    \arabic{hvref}. \citet{poy96alcom} ;
   \refstepcounter{hvref}\label{count:confirmed:alcom-26}
    \arabic{hvref}. \pasjcitetsub{nog97alcom}{a} ;
   \refstepcounter{hvref}\label{count:confirmed:alcom-27}
    \arabic{hvref}. \citet{how98alcom} ;
   \refstepcounter{hvref}\label{count:confirmed:alcom-28}
    \arabic{hvref}. \citet{szk98alcom} ;
   \refstepcounter{hvref}\label{count:confirmed:alcom-29}
    \arabic{hvref}. \citet{spo98alcomv544herv660herv516cygdxand} ;
   \refstepcounter{hvref}\label{count:confirmed:alcom-30}
    \arabic{hvref}. \pasjcitetsub{mat01alcomiauc7669}{a} ;
   \refstepcounter{hvref}\label{count:confirmed:alcom-31}
    \arabic{hvref}. Ishioka et al. in preparation ;
      (See also $\langle$
      http://www.kusastro.kyoto-u.ac.jp/vsnet/DNe/alcom0105.html
      $\rangle$) ;
   \refstepcounter{hvref}\label{count:confirmed:egcnc-1}
    \arabic{hvref}. \citet{hur83egcnc} ;
   \refstepcounter{hvref}\label{count:confirmed:egcnc-2}
    \arabic{hvref}. \citet{mcn86egcnc} ;
   \refstepcounter{hvref}\label{count:confirmed:egcnc-3}
    \arabic{hvref}. \pasjcitetsub{kat97egcnc}{a} ;
   \refstepcounter{hvref}\label{count:confirmed:egcnc-4}
    \arabic{hvref}. \pasjcitetsub{mat98egcncqui}{a} ;
   \refstepcounter{hvref}\label{count:confirmed:egcnc-5}
    \arabic{hvref}. \pasjcitetsub{mat98egcnc}{b} ;
   \refstepcounter{hvref}\label{count:confirmed:egcnc-6}
    \arabic{hvref}. \pasjcitetsub{pat98egcnc}{a} ;
   \refstepcounter{hvref}\label{count:confirmed:egcnc-7}
    \arabic{hvref}. \citet{szk98multiwavelength} ;
   \refstepcounter{hvref}\label{count:confirmed:egcnc-8}
    \arabic{hvref}. \citet{liu98egcnc} ;
   \refstepcounter{hvref}\label{count:confirmed:egcnc-9}
                         \label{count:confirmed:v2176cyg-1}
                         \label{count:confirmed:hvvir-1}
    \arabic{hvref}. \pasjcitetsub{szk00TOADs}{a} ;
   \refstepcounter{hvref}\label{count:confirmed:v2176cyg-2}
    \arabic{hvref}. \citet{hu97v2176cygiauc} ;
   \refstepcounter{hvref}\label{count:confirmed:v2176cyg-3}
    \arabic{hvref}. \citet{van97v2176cygiauc} ;
   \refstepcounter{hvref}\label{count:confirmed:v2176cyg-4}
    \arabic{hvref}. \pasjcitetsub{kat97v2176cyg}{b} ;
   \refstepcounter{hvref}\label{count:confirmed:v2176cyg-5}
    \arabic{hvref}. \citet{nov01v2176cyg} ;
   \refstepcounter{hvref}\label{count:confirmed:hvvir-2}
    \arabic{hvref}. this paper and references therein ;
   \refstepcounter{hvref}\label{count:confirmed:hvvir-3}
    \arabic{hvref}. \citet{lei94hvvir} ;
   \refstepcounter{hvref}\label{count:confirmed:rzleo-2}
    \arabic{hvref}. \citet{wol19rzleo} ;
   \refstepcounter{hvref}\label{count:confirmed:rzleo-3}
    \arabic{hvref}. \citet{mun19rzleo} ;
   \refstepcounter{hvref}\label{count:confirmed:rzleo-4}
    \arabic{hvref}. \citet{her58VSchart} ;
   \refstepcounter{hvref}\label{count:confirmed:rzleo-5}
    \arabic{hvref}. \pasjcitetsub{mat85rzleoiauc}{a} ; 
   \refstepcounter{hvref}\label{count:confirmed:rzleo-6}
    \arabic{hvref}. \citet{cri85rzleoiauc} ; 
   \refstepcounter{hvref}\label{count:confirmed:rzleo-7}
    \arabic{hvref}. \citet{mcn85rzleoiauc} ; 
   \refstepcounter{hvref}\label{count:confirmed:rzleo-8}
    \arabic{hvref}. \citet{ric85rzleo} ;
   \refstepcounter{hvref}\label{count:confirmed:rzleo-9}
    \arabic{hvref}. \citet{szk87shortPCV} ;
   \refstepcounter{hvref}\label{count:confirmed:rzleo-10}
    \arabic{hvref}. \pasjcitetsub{hur87rzleoiauc}{b} ; 
   \refstepcounter{hvref}\label{count:confirmed:rzleo-11}
    \arabic{hvref}. \citet{mat87rzleoiauc} ; 
   \refstepcounter{hvref}\label{count:confirmed:rzleo-12}
    \arabic{hvref}. \citet{nar89rzleoiauc} ; 
   \refstepcounter{hvref}\label{count:confirmed:rzleo-13}
    \arabic{hvref}. \citet{how90faintCV3} ;
   \refstepcounter{hvref}\label{count:confirmed:rzleo-14}
    \arabic{hvref}. \citet{szk92CVspec} ;
   \refstepcounter{hvref}\label{count:confirmed:rzleo-15}
    \arabic{hvref}. \citet{men99rzleo} ;
   \refstepcounter{hvref}\label{count:confirmed:rzleo-16}
    \arabic{hvref}. \citet{ish00rzleoiauc} ;
   \refstepcounter{hvref}\label{count:confirmed:rzleo-17}
    \arabic{hvref}. \citet{men01rzleo} ;
   \refstepcounter{hvref}\label{count:confirmed:rzleo-18}
    \arabic{hvref}. \pasjcitetsub{ish01rzleo}{a}
}
\end{table*}

\begin{table*}
\caption{Properties of probable members of WZ Sge-type dwarf
        novae.}\label{tab:probablewzsge}
\setcounter{hvref}{0}
\setcounter{hvrem}{0}
\begin{center}
\begin{tabular}{lccccccl}
\hline\hline
Object    & Max  & Min & A\commenta & P$_{\rm SH}$\commentb & Remarks
          & References \\
\hline
UZ Boo    & 11.5 & 20.5 &  9.0 & 0.0619 & \ref{rem:prob:uzboo-1}
                                 & \ref{count:prob:uzboo-1},
                                   \ref{count:prob:uzboo-2},
                                   \ref{count:prob:uzboo-3},
                                   \ref{count:prob:uzboo-4} \\
V592 Her  & 12.3 & 21.5:& 9.2: & 0.06007 & \ref{rem:prob:v592her-1}
                                 & \ref{count:prob:v592her-1},
                                   \ref{count:prob:v592her-2},
                                   \ref{count:prob:v592her-3},
                                   \ref{count:prob:v592her-4},
                                   \ref{count:prob:v592her-5},
                                   \ref{count:prob:v592her-6} \\
UW Tri    &   15 & [21  &$>$6  & 0.0569 & \ref{rem:prob:uwtri-1}
                                 & \ref{count:prob:uwtri-1},
                                   \ref{count:prob:uwtri-2},
                                   \ref{count:prob:uwtri-3},
                                   \ref{count:prob:uwtri-4} \\
CG CMa    & 13.7 & [20? &$>$6  & 0.0636 & \ref{rem:prob:cgcma-1}
                                 & \ref{count:prob:cgcma-1},
                                   \ref{count:prob:cgcma-2},
                                   \ref{count:prob:cgcma-3} \\
\hline
 \multicolumn{6}{l}{\commenta Amplitude of outburst (mag).} \\
 \multicolumn{6}{l}{\commentb Reported superhump period (d).} \\
\end{tabular}
\end{center}
{\footnotesize
  {\bf Remarks:}
   \refstepcounter{hvrem}\label{rem:prob:uzboo-1}
    \arabic{hvrem}. -- Most likely superhump period is given (Kato et al.,
    in preparation).  Other one-day aliases are not completely excluded.
   \refstepcounter{hvrem}\label{rem:prob:v592her-1}
    \arabic{hvrem}. -- Superhump period from \citet{due98v592her}.
    \citet{due98v592her} also listed a possible period of 0.06391 d.
    \citet{kat98v592her} listed possible periods of 0.05705 and less likely
    0.06049 d, based on the observations by G. Garradd.
   \refstepcounter{hvrem}\label{rem:prob:uwtri-1}
    \arabic{hvrem}. -- Most likely superhump period
    \pasjcitepsub{kat01uwtri}{d}.
    Other one-day aliases are not completely excluded. 
   \refstepcounter{hvrem}\label{rem:prob:cgcma-1}
    \arabic{hvrem}. -- Possibly slow emergence of superhumps.  The superhump
    period is the most likely signal by \pasjcitetsub{kat99cgcma}{b}.

  {\bf References:}
   \refstepcounter{hvref}\label{count:prob:uzboo-1}
    \arabic{hvref}. \citet{bai79wzsge} ;
   \refstepcounter{hvref}\label{count:prob:uzboo-2}
                         \label{count:prob:v592her-1}
    \arabic{hvref}. \citet{ric92wzsgedip} ;
   \refstepcounter{hvref}\label{count:prob:uzboo-3}
    \arabic{hvref}. \citet{kuu96TOAD} ;
   \refstepcounter{hvref}\label{count:prob:uzboo-4}
    \arabic{hvref}. Kato et al. in preparation ;
   \refstepcounter{hvref}\label{count:prob:v592her-2}
    \arabic{hvref}. \citet{ric68v592her} ;
   \refstepcounter{hvref}\label{count:prob:v592her-3}
    \arabic{hvref}. \pasjcitetsub{ric91v592her}{b} ;
   \refstepcounter{hvref}\label{count:prob:v592her-4}
    \arabic{hvref}. \citet{due98v592her} ;
   \refstepcounter{hvref}\label{count:prob:v592her-5}
    \arabic{hvref}. \citet{vantee99v592her} ;
   \refstepcounter{hvref}\label{count:prob:v592her-6}
    \arabic{hvref}. Kato et al. in preparation ;
   \refstepcounter{hvref}\label{count:prob:uwtri-1}
    \arabic{hvref}. \citet{aks83uwtriiauc} ; 
   \refstepcounter{hvref}\label{count:prob:uwtri-2}
    \arabic{hvref}. \citet{arg83uwtriiauc} ; 
   \refstepcounter{hvref}\label{count:prob:uwtri-3}
    \arabic{hvref}. \citet{rob00oldnova} ;
   \refstepcounter{hvref}\label{count:prob:uwtri-4}
    \arabic{hvref}. \pasjcitetsub{kat01uwtri}{d} ;
   \refstepcounter{hvref}\label{count:prob:cgcma-1}
    \arabic{hvref}. \citet{due87novaatlas} ;
   \refstepcounter{hvref}\label{count:prob:cgcma-2}
    \arabic{hvref}. \pasjcitetsub{kat99cgcma}{b} ;
   \refstepcounter{hvref}\label{count:prob:cgcma-3}
    \arabic{hvref}. \citet{due99cgcma}
}
\end{table*}

\begin{table*}
\caption{Properties of candidates for WZ Sge-type dwarf novae
         and related systems.}\label{tab:candidate}
\setcounter{hvref}{0}
\setcounter{hvrem}{0}
\begin{center}
\begin{tabular}{lccccccl}
\hline\hline
Object    & Max  & Min & Amplitude & $P_{\rm orb}$ & $P_{\rm SH}$ & Remarks & References \\
\hline
LL And    & 13   & 20   & 7:   & - & 0.0567 &
                                 \ref{remcount:cand:lland}
                                 & \ref{count:cand:lland-1},
                                   \ref{count:cand:lland-2},
                                   \ref{count:cand:lland-3},
                                   \ref{count:cand:lland-4},
                                   \ref{count:cand:lland-5},
                                   \ref{count:cand:lland-6} \\
GW Lib    &  9.0 & 18.5 & 9.5  & 0.0551 & - &
                                 \ref{remcount:cand:oldnova},
                                 \ref{remcount:cand:wzsgelikespec},
                                 \ref{remcount:cand:gwlib}
                                 & \ref{count:cand:gwlib-1},
                                   \ref{count:cand:gwlib-2},
                                   \ref{count:cand:gwlib-3},
                                   \ref{count:cand:gwlib-4},
                                   \ref{count:cand:gwlib-5} \\
V358 Lyr  & 13.2 & [20  &$>$6.8& - & - &
                                 \ref{remcount:cand:oldnova},
                                 \ref{remcount:cand:candidate}
                                 & \ref{count:cand:v358lyr-1} \\
LS And    & 11.7 & 20.5 &  8.8 & - & - &
                                 \ref{remcount:cand:oldnova},
                                 \ref{remcount:cand:lsand}
                                 & \ref{count:cand:lsand-1},
                                   \ref{count:cand:lsand-2},
                                   \ref{count:cand:lsand-3},
                                   \ref{count:cand:lsand-4} \\
PQ And    & 10.0 & 18.8 &  8.8 & - & - &
                                 \ref{remcount:cand:oldnova},
                                 \ref{remcount:cand:dnspec},
                                 \ref{remcount:cand:mulout}
                                 & \ref{count:cand:pqand-1},
                                   \ref{count:cand:pqand-2},
                                   \ref{count:cand:pqand-3},
                                   \ref{count:cand:pqand-4},
                                   \ref{count:cand:pqand-5},
                                   \ref{count:cand:pqand-6},
                                   \ref{count:cand:pqand-7} \\
UZ Tri    & 14.2 & [21  &$>$7 & - & - &
                                \ref{remcount:cand:oldnova},
                                \ref{remcount:cand:lowfreq}
                                 & \ref{count:cand:uztri-1} \\
UW Per    & 15   & [20? &$>$5? & - & - &
                                \ref{remcount:cand:oldnova},
                                \ref{remcount:cand:mulout},
                                \ref{remcount:cand:lowfreq}
                                 & \ref{count:cand:uwper-1},
                                   \ref{count:cand:uwper-2},
                                   \ref{count:cand:uwper-3},
                                   \ref{count:cand:uwper-4},
                                   \ref{count:cand:uwper-5} \\
RY Dor    & 12.4 & [18  &$>$5  & - & - &
                                 \ref{remcount:cand:oldnova},
                                 \ref{remcount:cand:rydor}
                                 & \ref{count:cand:rydor-1},
                                   \ref{count:cand:rydor-2},
                                   \ref{count:cand:rydor-3} \\
VZ Tuc    & 11.4 & [18  &$>$6.6& - & - &
                                 \ref{remcount:cand:oldnova},
                                 \ref{remcount:cand:vztuc}
                                 & \ref{count:cand:vztuc-1},
                                   \ref{count:cand:vztuc-2},
                                   \ref{count:cand:vztuc-3},
                                   \ref{count:cand:vztuc-4} \\
KY Ara    & 15.1 & [21  &$>$6  & - & - &
                                 \ref{remcount:cand:oldnova},
                                 \ref{remcount:cand:kyara}
                                 & \ref{count:cand:kyara-1},
                                   \ref{count:cand:kyara-2} \\
V359 Cen  & 13.8 & 21.0 & 7.2  & - & 0.078: &
                                 \ref{remcount:cand:oldnova},
                                 \ref{remcount:cand:toofreq},
                                 \ref{remcount:cand:v359cen}
                                 & \ref{count:cand:v359cen-1},
                                   \ref{count:cand:v359cen-2},
                                   \ref{count:cand:v359cen-3} \\
KX Aql    & 11.5 & 18.4 & 6.9  & - & - &
                                 \ref{remcount:cand:lowfreq},
                                 \ref{remcount:cand:kxaql}
                                 & \ref{count:cand:kxaql-1},
                                   \ref{count:cand:kxaql-2} \\
EL UMa    & 14   & 19   & 5    & - & - &
                                 \ref{remcount:cand:lowfreq}
                                 & \ref{count:cand:eluma-1},
                                   \ref{count:cand:eluma-2},
                                   \ref{count:cand:eluma-3} \\
V336 Per  & 14.3 & 19.8 & 5.5  & - & - &
                                 \ref{remcount:cand:lowfreq}
                                 & \ref{count:cand:v336per-1},
                                   \ref{count:cand:v336per-2} \\
IO Del    & 15.5 & [20 &$>$4.5 & - & - &
                                 \ref{remcount:cand:lowfreq}
                                 & \ref{count:cand:iodel-1},
                                   \ref{count:cand:iodel-2} \\
SV Ari    &   12 &  22  & 10   & - & - &
                                 \ref{remcount:cand:oldnova},
                                 \ref{remcount:cand:svari}
                                 & \ref{count:cand:svari-1},
                                   \ref{count:cand:svari-2} \\
BC Cas    & 10.7 & 17.4 &  6.7 & - & - &
                                 \ref{remcount:cand:oldnova},
                                 \ref{remcount:cand:posnova},
                                 \ref{remcount:cand:bccas}
                                 & \ref{count:cand:bccas-1},
                                   \ref{count:cand:bccas-2},
                                   \ref{count:cand:bccas-3},
                                   \ref{count:cand:bccas-4},
                                   \ref{count:cand:bccas-5} \\
AP Cru    & 10.7 & 21.7 & 11.0 & - & - &
                                 \ref{remcount:cand:oldnova},
                                 \ref{remcount:cand:posnova},
                                 \ref{remcount:cand:apcru}
                                 & \ref{count:cand:apcru-1},
                                   \ref{count:cand:apcru-2} \\
EY Aql    & 10.5 & 20:  & 9.5: & - & - &
                                 \ref{remcount:cand:oldnova},
                                 \ref{remcount:cand:posnova},
                                 \ref{remcount:cand:eyaql}
                                 & \ref{count:cand:eyaql-1},
                                   \ref{count:cand:eyaql-2} \\
DV Dra    & 15.0 & [21  & $>$6 & - & - &
                                 \ref{remcount:cand:lowfreq},
                                 \ref{remcount:cand:dvdra}
                                 & \ref{count:cand:dvdra-1},
                                   \ref{count:cand:dvdra-2},
                                   \ref{count:cand:dvdra-3},
                                   \ref{count:cand:dvdra-4},
                                   \ref{count:cand:dvdra-5},
                                   \ref{count:cand:dvdra-6} \\
V632 Her  & 15.4 &  21  & 5.6: & - & - &
                                 \ref{remcount:cand:oldnova},
                                 \ref{remcount:cand:mulout}
                                 & \ref{count:cand:v632her-1},
                                   \ref{count:cand:v632her-2} \\
V369 Lyr  & 15.2 & [20.0&$>$4.8& - & - &
                                 \ref{remcount:cand:oldnova}
                                 & \ref{count:cand:v369lyr-1} \\
GR Ori    & 11.5 & 22.8 & 11.3 & - & - &
                                 \ref{remcount:cand:oldnova}
                                 & \ref{count:cand:grori-1},
                                   \ref{count:cand:grori-2},
                                   \ref{count:cand:grori-3} \\
V522 Sgr  & 12.9 & 17?  &$>$4.1& - & - &
                                 \ref{remcount:cand:oldnova},
                                 \ref{remcount:cand:v522sgr}
                                 & \ref{count:cand:v522sgr-1},
                                   \ref{count:cand:v522sgr-2} \\
SS LMi    &  15  & 21.6 & 6.6  & - & - &
                                 \ref{remcount:cand:oldnova},
                                 \ref{remcount:cand:sslmi}
                                 & \ref{count:cand:sslmi-1},
                                   \ref{count:cand:sslmi-2},
                                   \ref{count:cand:sslmi-3},
                                   \ref{count:cand:sslmi-4} \\
CI Gem    & 14.7 & 21.7: & 7.0:& - & - &
                                 \ref{remcount:cand:oldnova},
                                 \ref{remcount:cand:cigem}
                                 & \ref{count:cand:cigem-1},
                                   \ref{count:cand:cigem-2},
                                   \ref{count:cand:cigem-3},
                                   \ref{count:cand:cigem-4},
                                   \ref{count:cand:cigem-5},
                                   \ref{count:cand:cigem-6},
                                   \ref{count:cand:cigem-7} \\
PT And    & 16.3 & [20  &$>$3.7& - & - &
                                 \ref{remcount:cand:oldnova},
                                 \ref{remcount:cand:ptand}
                                 & \ref{count:cand:ptand-1},
                                   \ref{count:cand:ptand-2},
                                   \ref{count:cand:ptand-3} \\
V4338 Sgr &  8.0 & [21 & $>$13 & - & - &
                                 \ref{remcount:cand:oldnova},
                                 \ref{remcount:cand:dnspec},
                                 \ref{remcount:cand:v4338sgr}
                                 & \ref{count:cand:v4338sgr-1},
                                   \ref{count:cand:v4338sgr-2},
                                   \ref{count:cand:v4338sgr-3},
                                   \ref{count:cand:v4338sgr-4},
                                   \ref{count:cand:v4338sgr-5},
                                   \ref{count:cand:v4338sgr-6} \\
AS Psc    & 16.6 & 21.5: & 4.9: & - & - &
                                 \ref{remcount:cand:lowfreq}
                                 & \ref{count:cand:aspsc-1},
                                   \ref{count:cand:aspsc-2},
                                   \ref{count:cand:aspsc-3},
                                   \ref{count:cand:aspsc-4},
                                   \ref{count:cand:aspsc-5},
                                   \ref{count:cand:aspsc-6},
                                   \ref{count:cand:aspsc-7} \\
NSV 15820 & 15.3 & [21  &$>$6  & - & - &
                                 \ref{remcount:cand:dnspec},
                                 \ref{remcount:cand:flarestar}
                                 & \ref{count:cand:nsv15820-1} \\
VX For    & 12.7 & [19  &$>$6  & - & - &
                                 \ref{remcount:cand:dnspec}
                                 & \ref{count:cand:vxfor-1} \\
XY Psc    & 13.0 & 21.1 & 8.1  & - & - &
                                 \ref{remcount:cand:snsuspect},
                                 \ref{remcount:cand:xypsc}
                                 & \ref{count:cand:xypsc-1},
                                   \ref{count:cand:xypsc-2},
                                   \ref{count:cand:xypsc-3} \\
CE UMa    & 15.5 & [21  &$>6$  & - & - &
                                 \ref{remcount:cand:snsuspect}
                                 & \ref{count:cand:ceuma-1},
                                   \ref{count:cand:ceuma-2} \\
NSV 00895 & 11.7 & [20  &$>$8  & - & - &
                                 \ref{remcount:cand:snsuspect},
                                 \ref{remcount:cand:nsv00895}
                                 & \ref{count:cand:nsv00895-1} \\
SN 1964O  & 15   & [20  &$>$5  & - & - &
                                 \ref{remcount:cand:snsuspect},
                                 \ref{remcount:cand:sn1964o}
                                 & \ref{count:cand:sn1964o-1} \\
FBS 1719+834 & 13.5 & [21 &$>$7.5 & - & - &
                                 \ref{remcount:cand:novahighgal}
                                 & \ref{count:cand:fbs1719-1},
                                   \ref{count:cand:fbs1719-2} \\
FBS 1735+825 & 14 & [21 &$>$7  & - & - &
                                 -
                                 & \ref{count:cand:fbs1735-1},
                                   \ref{count:cand:fbs1735-2} \\
AO Oct    & 13.5 &  21  & 7.5: & - & - &
                                 \ref{remcount:cand:candidate},
                                 \ref{remcount:cand:toofreq}
                                 & \ref{count:cand:aooct-1} \\
V551 Sgr  & 13.5 &  20  & 6.5: & - & - &
                                 \ref{remcount:cand:candidate}
                                 & \ref{count:cand:v551sgr-1} \\
FV Ara    & 12   &  20: &  8:  & - & - &
                                 -
                                 & \ref{count:cand:fvara-1} \\
GO Com    & 13.1 &  20: &  7:  & - & - &
                                 \ref{remcount:cand:candidate},
                                 \ref{remcount:cand:toofreq}
                                 & \ref{count:cand:gocom-1},
                                   \ref{count:cand:gocom-2},
                                   \ref{count:cand:gocom-3},
                                   \ref{count:cand:gocom-4} \\
YY Tel    & 14.4 & 19.3 &  4.9 & - & - &
                                 \ref{remcount:cand:candidate},
                                 \ref{remcount:cand:yytel}
                                 & \ref{count:cand:yytel-1},
                                   \ref{count:cand:yytel-2},
                                   \ref{count:cand:yytel-3} \\
NSV 15272 & 10?  & 18?  &  8?  & - & - &
                                 \ref{remcount:cand:nsv15272}
                                 & \ref{count:cand:nsv15272-1} \\
RX J0643.9$-$2052 & 11.4 &[20   &$>$8.6 & - & - &
                                 \ref{remcount:cand:j0643}
                                 & \ref{count:cand:j0643-1} \\
V1289 Aql & 13   &[15.5 &$>$2.5 & - & - &
                                 \ref{remcount:cand:v1289aql}
                                 & \ref{count:cand:v1289aql-1},
                                   \ref{count:cand:v1289aql-2},
                                   \ref{count:cand:v1289aql-3},
                                   \ref{count:cand:v1289aql-4} \\
GD 552    &  -   & 17.4 &  -   & 0.0713 & - &
                                 \ref{remcount:cand:wzsgelikespec},
                                 \ref{remcount:cand:noburst}
                                 & \ref{count:cand:gd552-1},
                                   \ref{count:cand:gd552-2},
                                   \ref{count:cand:gd552-3} \\
BW Scl    &  -   & 16.5 &  -   & 0.0543 & - &
                                 \ref{remcount:cand:wzsgelikespec},
                                 \ref{remcount:cand:noburst},
                                 \ref{remcount:cand:bwscl}
                                 & \ref{count:cand:bwscl-1},
                                   \ref{count:cand:bwscl-2},
                                   \ref{count:cand:bwscl-3} \\
RE J1255+266 & - & 18   &  -   & - & - &
                                 \ref{remcount:cand:wzsgelikespec},
                                 \ref{remcount:cand:noburst},
                                 \ref{remcount:cand:euvtransient}
                                 & \ref{count:cand:rej1255-1},
                                   \ref{count:cand:rej1255-2},
                                   \ref{count:cand:rej1255-3},
                                   \ref{count:cand:rej1255-4},
                                   \ref{count:cand:rej1255-5}, \\
          &      &      &  & & & & \ref{count:cand:rej1255-6},
                                   \ref{count:cand:rej1255-7} \\
1RXS J105010 & - & 17.0 & -    & 0.0615 & - &
                                 \ref{remcount:cand:wzsgelikespec},
                                 \ref{remcount:cand:noburst},
                                 \ref{remcount:cand:j1050}
                                 & \ref{count:cand:j1050-1},
                                   \ref{count:cand:j1050-2} \\
GP Com    &  -   & 16.0 &  -   & 0.0323 & - &
                                 \ref{remcount:cand:noburst},
                                 \ref{remcount:cand:IBWD}
                                 & \ref{count:cand:gpcom-1},
                                   \ref{count:cand:gpcom-2},
                                   \ref{count:cand:gpcom-3},
                                   \ref{count:cand:gpcom-4},
                                   \ref{count:cand:gpcom-5}, \\
          &      &      &  & & & & \ref{count:cand:gpcom-6},
                                   \ref{count:cand:gpcom-7},
                                   \ref{count:cand:gpcom-8},
                                   \ref{count:cand:gpcom-9},
                                   \ref{count:cand:gpcom-10}, \\
          &      &      &  & & & & \ref{count:cand:gpcom-11} \\
CE-315    &  -   & 17.6 &  -   & 0.0452 & - &
                                 \ref{remcount:cand:noburst},
                                 \ref{remcount:cand:IBWD}
                                 & \ref{count:cand:ce315-1},
                                   \ref{count:cand:ce315-2} \\
\hline
\end{tabular}
\end{center}
\end{table*}

\begin{table*}
{\footnotesize
  {\bf Remarks to table \ref{tab:candidate}:}
   \refstepcounter{hvrem}\label{remcount:cand:lland}
    \arabic{hvrem}. -- Large-amplitude SU UMa-type dwarf nova
    (superhumps detected).  Unusual system with a short superhump
    period; possibly related to a WZ Sge-type star.  The superhump
    period is taken from Kato et al. in preparation, whose preliminary
    result is cited in \citet{how96lland}.
   \refstepcounter{hvrem}\label{remcount:cand:oldnova}
    \arabic{hvrem}. -- Originally discovered as a possible nova.
   \refstepcounter{hvrem}\label{remcount:cand:wzsgelikespec}
    \arabic{hvrem}. -- Optical spectrum resembles that of WZ Sge
    in quiescence.
   \refstepcounter{hvrem}\label{remcount:cand:gwlib}
    \arabic{hvrem}. -- Orbital period 79.4 min \pasjcitepsub{szk00gwlib}{b}.
    Very likely a WZ Sge-type star.
   \refstepcounter{hvrem}\label{remcount:cand:candidate}
    \arabic{hvrem}. -- Proposed as a possible WZ Sge-type star in the
    past literature.
   \refstepcounter{hvrem}\label{remcount:cand:lsand}
    \arabic{hvrem}. -- Initially proposed as an intergalactic fast nova.
    The published light curve strongly resembles that of the 1946 outburst
    of WZ Sge.
   \refstepcounter{hvrem}\label{remcount:cand:dnspec}
    \arabic{hvrem}. -- Spectroscopy in outburst suggests a dwarf nova.
   \refstepcounter{hvrem}\label{remcount:cand:mulout}
    \arabic{hvrem}. -- Although originally classified as a possible nova,
    multiple outbursts are known.
   \refstepcounter{hvrem}\label{remcount:cand:lowfreq}
    \arabic{hvrem}. -- Low outburst frequency.
   \refstepcounter{hvrem}\label{remcount:cand:rydor}
    \arabic{hvrem}. -- Possible nova recorded in 1926 in the LMC region
    without spectroscopic information.  Some authors suggest the dwarf nova
    classification.  The observed $t_{\rm 3}\sim100$ d
    (e.g. \cite{devac78distancescale}) might be too long for a dwarf nova.
   \refstepcounter{hvrem}\label{remcount:cand:vztuc}
    \arabic{hvrem}. -- Possible nova recorded in 1927 in the SMC region
    without spectroscopic information.  Some authors suggest the dwarf nova
    classification, and has been monitored for further outbursts
    (see e.g. \cite{how91CVamateur}).  The fastest decline
    ($t_{\rm 3}\sim20$ d) among historical novae and suspected novae in
    the SMC \citep{devac78distancescale} makes the classification as
    a large-amplitude dwarf nova not unlikely.
   \refstepcounter{hvrem}\label{remcount:cand:kyara}
    \arabic{hvrem}. -- Misidentified in \citet{due87novaatlas}.  The new
    identification by \citet{sch94kyara} suggests a dwarf nova with a large
    outburst amplitude.
   \refstepcounter{hvrem}\label{remcount:cand:toofreq}
    \arabic{hvrem}. -- Rather frequent outburst (from VSNET data).
    Not very likely a WZ Sge-type star.
   \refstepcounter{hvrem}\label{remcount:cand:v359cen}
    \arabic{hvrem}. -- Dwarf nova-like outbursts.  Possible superhumps
    were detected.  112-min photometric periodicity in post-superoutburst
    quiescence has been reported, which \citet{wou01v359cenxzeriyytel}
    interpreted as persisting superhumps.
   \refstepcounter{hvrem}\label{remcount:cand:kxaql}
    \arabic{hvrem}. -- The maximum magnitude taken from the VSOLJ database,
    which recorded a bright, long-lasting outburst in 1980 November.
    The last confirmed outburst was a short outburst reaching 15 mag
    reported in 1994 December (VSNET).  Very good candidate for a
    WZ Sge-type dwarf nova.
   \refstepcounter{hvrem}\label{remcount:cand:svari}
    \arabic{hvrem}. -- The outburst light variation \citep{wol05svari}
    resembles that of a WZ Sge-type dwarf nova.
   \refstepcounter{hvrem}\label{remcount:cand:posnova}
    \arabic{hvrem}. -- No spectroscopic confirmation during outburst;
    observations during quiescence more suggests a classical nova
    rather than a WZ Sge-type star.
   \refstepcounter{hvrem}\label{remcount:cand:bccas}
    \arabic{hvrem}. -- \citet{zwi94CVspec1} detected absorption lines, while
    \citet{rin96oldnovaspec} detected weak Balmer emission lines.  Both
    features are not inconsistent with a high-$\dot{M}$ postnova spectrum.
   \refstepcounter{hvrem}\label{remcount:cand:apcru}
    \arabic{hvrem}. -- Poorly observed object during the outburst in 1935.
    Some authors suggest the dwarf nova classification.  \citet{mun98CVspec5}
    reported the detection of weak H$\alpha$ emission in quiescence,
    not inconsistent with a high-$\dot{M}$ postnova spectrum.
   \refstepcounter{hvrem}\label{remcount:cand:dvdra}
    \arabic{hvrem}. -- The 1991 outburst lasted $\sim$20 d.  A short
    brightness dip in the early stage of the outburst, and a post-outburst
    brightening were observed (\pasjcitesub{wen91dvdra}{a};
    \cite{ric92wzsgedip}).
    Both of these features are very characteristic of an SU UMa-type
    dwarf nova with very rare outbursts.
   \refstepcounter{hvrem}\label{remcount:cand:eyaql}
    \arabic{hvrem}. -- Poorly observed object during the outburst in 1926.
    Some authors suggest the dwarf nova classification.  No spectroscopic
    information.  The light curve \pasjcitepsub{due84eyaqlbccasmtcenv745sco}{a}
    may not be unlike that of a WZ Sge-type star.
   \refstepcounter{hvrem}\label{remcount:cand:v522sgr}
    \arabic{hvrem}. -- Originally discovered as a possible nova in 1931.
    The light curve \citep{fer35v522sgr} may be interpreted as the plateau
    portion of an SU UMa-type superoutburst.  Short-term variations also
    seem to have been present in the data of \citet{fer35v522sgr}, which
    may suggest the presence of superhumps.  \citet{rin96oldnovaspec}
    classified the proposed quiescent counterpart as a G2--K5 star,
    which seems to be inconsistent with the eruptive nature of the object.
    The true counterpart may be an undetected, fainter companion to this
    star, as in the case of CG CMa (\pasjcitesub{kat99cgcma}{b};
    \cite{due99cgcma}).
    The star has been monitored by the members of the VSOLJ and VSNET,
    without any further outburst detection.
    \\
}
\end{table*}
\begin{table*}
{\footnotesize
  {\bf Remarks to table \ref{tab:candidate} (continued):}
   \refstepcounter{hvrem}\label{remcount:cand:sslmi}
    \arabic{hvrem}. -- Originally discovered as a possible nova in 1980.
    The light curve \citep{alk80sslmi} resembles the plateau portion of
    an SU UMa-type superoutburst.  \citet{har91sslmiiauc} reported
    an activity in 1991, when the object was observed red.
    \citet{how91sslmiiauc}, however, suggested that the object may belong
    to a large-amplitude dwarf nova.  Unpublished $I$-band photometry
    by the authors also indicates a red object.  The classification is
    still uncertain.
   \refstepcounter{hvrem}\label{remcount:cand:cigem}
    \arabic{hvrem}. -- Originally discovered as a dwarf nova or a nova
    in 1940.  \citet{wen90cigem} suggested it to be a short-period
    SU UMa-type dwarf nova based on archival plate search.  Low outburst
    frequency (VSNET).  The only known recent outburst (short outburst)
    occurred in 1999 February (\cite{sch99cigem}; \cite{kat99cigem}).
    \citet{sch99cigem} identified a relatively red quiescent counterpart
    with $V$=21.66.  Likely a large-amplitude dwarf nova, possibly
    related to WZ Sge-type stars.
   \refstepcounter{hvrem}\label{remcount:cand:ptand}
    \arabic{hvrem}. -- Originally discovered as a nova in M31
    \citep{gru59ptand}.  \citet{sha89ptand} considered the object to be
    an SU UMa-type dwarf nova based on the second outburst detection.
    Three well-documented (1957, 1986, 1998) and two poorly recorded
    (1983, 1988) are known.  Although \citet{alk00ptand} suggested the
    possibility of the object being a recurrent nova in M31, the shortest
    interval of the outbursts is too short [for recent theoretical
    calculations, see \authorcite{hac99tcrb}
    (\yearcite{hac99tcrb},\yearcite{hac00rsoph}a,b,\yearcite{hac01ciaql}a),
    \citet{hac00uscoburst};
    see also a review: \pasjcitetsub{hac01RN}{b}].
    The outburst light curve resembles that of a WZ Sge-type dwarf nova.
   \refstepcounter{hvrem}\label{remcount:cand:v4338sgr}
    \arabic{hvrem}. -- \citet{wag90v4338sgriauc} presented a spectrum
    taken during the 1990 outburst, which showed Balmer, HeI, He II and
    C III/N III emission lines, and no forbidden lines.  The spectrum
    was unlike those of novae.  \citet{wag90v4338sgriauc} suggested the
    object to be a large-amplitude dwarf nova like WZ Sge.  The minimum
    magnitude is taken from \citet{sum97v4338sgr} (see also $\langle$
    http://www.kusastro.kyoto-u.ac.jp/vsnet/DNe/nv1990sgr.html
    $\rangle$).  If the dwarf nova nature and the quiescent identification
    are confirmed, this object makes the largest record of the outburst
    amplitude of dwarf novae (see also the discussion in
    \citet{vantee99v592her} for the implication from the existence of
    such a large-amplitude dwarf nova.
    The field was recently observed by the 2MASS survey and ISOGAL variable
    star survey \citet{sch00ISOGAL}, yielding no promising quiescent
    counterpart.
   \refstepcounter{hvrem}\label{remcount:cand:flarestar}
    \arabic{hvrem}. -- Originally discovered as a large-amplitude flare
    star.  The long duration of the observed flare suggests a dwarf nova-type
    outburst.
   \refstepcounter{hvrem}\label{remcount:cand:snsuspect}
    \arabic{hvrem}. -- Originally discovered as a possible supernova.
   \refstepcounter{hvrem}\label{remcount:cand:xypsc}
    \arabic{hvrem}. -- The minimum magnitude is taken from \citet{hen01xypsc}.
    No outburst has been reported since the discovery in 1972.  Very good
    candidate for a WZ Sge-type dwarf nova.
   \refstepcounter{hvrem}\label{remcount:cand:nsv00895}
    \arabic{hvrem}. -- The existence of the object is well established
    on multiple plates \citep{wen90nsv00895}.  The light curve resembles
    that of a long-lasting WZ Sge-type outburst.
   \refstepcounter{hvrem}\label{remcount:cand:sn1964o}
    \arabic{hvrem}. -- Also called as SN s1964a.  Other (poorly studied)
    supernova candidates which can be large-amplitude dwarf novae
    include SN 1985J and NSV 05285.
   \refstepcounter{hvrem}\label{remcount:cand:novahighgal}
    \arabic{hvrem}. -- Originally discovered as a possible nova.
    Faint, high galactic lattitude object.
   \refstepcounter{hvrem}\label{remcount:cand:yytel}
    \arabic{hvrem}. -- Outburst in 1998 October (VSNET).  Outbursts are
    not frequent.
   \refstepcounter{hvrem}\label{remcount:cand:nsv15272}
    \arabic{hvrem}. -- Found at magnitude 10 on one photograph.
    \citet{iid90nsv15272} suspected a rare outburst of a cataclysmic
    variable.  Wenzel (private communication) could not find further
    evidence for the outburst on Sonneberg plates.
   \refstepcounter{hvrem}\label{remcount:cand:j0643}
    \arabic{hvrem}. -- Recorded in AC 2000 catalogue, from a photograph
    taken in 1917.  Possibly identifed with RX J0643.9$-$2052.
    Large-amplitude dwarf nova or nova \citep{gre00j0643id210}.
    No further outburst has been found in VSNET reports.
   \refstepcounter{hvrem}\label{remcount:cand:v1289aql}
    \arabic{hvrem}. -- Discovered by \citet{wol04v1289aql}.
    \citet{mei70v1289aql} suggested it to be a ``nova-like" object.
    Although the star has been monitored by the VSOLJ and VSNET members,
    no further outburst has been detected.  Quiescent identification is
    still uncertain \citep{dow95CVspec}.
   \refstepcounter{hvrem}\label{remcount:cand:noburst}
    \arabic{hvrem}. -- No outbursts were observed.
   \refstepcounter{hvrem}\label{remcount:cand:bwscl}
    \arabic{hvrem}. -- Short-period (P$_{orb }$=78 min) system resembling
    quiescent WZ Sge \citep{aug97bwscl}.
   \refstepcounter{hvrem}\label{remcount:cand:euvtransient}
    \arabic{hvrem}. -- EUV transient.
   \refstepcounter{hvrem}\label{remcount:cand:j1050}
    \arabic{hvrem}. -- Full name: 1RXS J105010.3$-$140431.  Reported orbital
    period of 88.6 minutes and the spectrum closely resemble those of WZ Sge
    \citep{men01j1050}.
    The object is identified with a NLTT high proper-motion object
    NLTT 731-60, which implies a very small distance and a low intrinsic
    luminosity \pasjcitepsub{kat01j1050chat4601}{f}.
   \refstepcounter{hvrem}\label{remcount:cand:IBWD}
    \arabic{hvrem}. -- Belongs to interacting binary white dwarfs (IBWD)
    or AM CVn stars (helium CVs).  The objects listed in the table
    are considered to correspond to WZ Sge-type stars in hydrogen-rich
    CVs (cf. \cite{sma83HeCV}; \pasjcitesub{war95amcvn}{a};
    \cite{tsu97amcvn}).
    \\
}
\end{table*}
\begin{table*}
{\footnotesize
  {\bf References to table \ref{tab:candidate}:}
   \refstepcounter{hvref}\label{count:cand:lland-1}
    \arabic{hvref}. \citet{wil79lland} ;
   \refstepcounter{hvref}\label{count:cand:lland-2}
    \arabic{hvref}. \citet{how94lland} ;
   \refstepcounter{hvref}\label{count:cand:lland-3}
    \arabic{hvref}. \citet{how96lland} ;
   \refstepcounter{hvref}\label{count:cand:lland-4}
    \arabic{hvref}. \pasjcitetsub{szk00TOADs}{a} ;
   \refstepcounter{hvref}\label{count:cand:lland-5}
    \arabic{hvref}. \citet{how01llandeferi} ;
   \refstepcounter{hvref}\label{count:cand:lland-6}
    \arabic{hvref}. Kato et al. in preparation ;
   \refstepcounter{hvref}\label{count:cand:gwlib-1}
    \arabic{hvref}. \citet{maz83gwlibiauc} ;
   \refstepcounter{hvref}\label{count:cand:gwlib-2}
    \arabic{hvref}. \citet{due87gwlib} ;
   \refstepcounter{hvref}\label{count:cand:gwlib-3}
    \arabic{hvref}. \pasjcitetsub{szk00gwlib}{b} ;
   \refstepcounter{hvref}\label{count:cand:gwlib-4}
    \arabic{hvref}. \citet{vanzyl00gwlib} ;
   \refstepcounter{hvref}\label{count:cand:gwlib-5}
    \arabic{hvref}. \citet{vanzyl01gwlib} ;
   \refstepcounter{hvref}\label{count:cand:v358lyr-1}
    \arabic{hvref}. \pasjcitetsub{ric86v358lyr}{b} ;
   \refstepcounter{hvref}\label{count:cand:lsand-1}
    \arabic{hvref}. \citet{vandenber73lsand} ;
   \refstepcounter{hvref}\label{count:cand:lsand-2}
    \arabic{hvref}. \citet{rom77lsand} ;
   \refstepcounter{hvref}\label{count:cand:lsand-3}
    \arabic{hvref}. \citet{mei77lsand} ;
   \refstepcounter{hvref}\label{count:cand:lsand-4}
    \arabic{hvref}. \citet{sha78lsand} ;
   \refstepcounter{hvref}\label{count:cand:pqand-1}
    \arabic{hvref}. \pasjcitetsub{hur88pqandiauc1}{a} ;
   \refstepcounter{hvref}\label{count:cand:pqand-2}
    \arabic{hvref}. \pasjcitetsub{hur88pqandiauc2}{b} ;
   \refstepcounter{hvref}\label{count:cand:pqand-3}
    \arabic{hvref}. \pasjcitetsub{hur88pqandiauc3}{c} ;
   \refstepcounter{hvref}\label{count:cand:pqand-4}
    \arabic{hvref}. \citet{mca88pqandiauc} ;
   \refstepcounter{hvref}\label{count:cand:pqand-5}
    \arabic{hvref}. \citet{roy88pqandiauc} ;
   \refstepcounter{hvref}\label{count:cand:pqand-6}
    \arabic{hvref}. \citet{wad88pqandiauc} ;
   \refstepcounter{hvref}\label{count:cand:pqand-7}
    \arabic{hvref}. \pasjcitetsub{ric90pqand}{a} ;
   \refstepcounter{hvref}\label{count:cand:uztri-1}
    \arabic{hvref}. \citet{mei86uztri} ;
   \refstepcounter{hvref}\label{count:cand:uwper-1}
    \arabic{hvref}. \citet{dest12uwper} ;
   \refstepcounter{hvref}\label{count:cand:uwper-2}
    \arabic{hvref}. \citet{ric87uwper} ;
   \refstepcounter{hvref}\label{count:cand:uwper-3}
                         \label{count:cand:v359cen-1}
                         \label{count:cand:kyara-1}
                         \label{count:cand:v632her-1}
    \arabic{hvref}. \citet{due87novaatlas} ;
   \refstepcounter{hvref}\label{count:cand:uwper-4}
    \arabic{hvref}. \citet{mob97uwper} ;
   \refstepcounter{hvref}\label{count:cand:uwper-5}
                         \label{count:cand:svari-1}
                         \label{count:cand:grori-1}
                         \label{count:cand:sslmi-1}
    \arabic{hvref}. \citet{rob00oldnova} ;
   \refstepcounter{hvref}\label{count:cand:rydor-1}
                         \label{count:cand:vztuc-1}
                         \label{count:cand:ceuma-1}
                         \label{count:cand:cpdra-1}
    \arabic{hvref}. \citet{how90highgalCV} ;
   \refstepcounter{hvref}\label{count:cand:rydor-2}
                         \label{count:cand:vztuc-2}
    \arabic{hvref}. \citet{gra71LMCnova} ;
   \refstepcounter{hvref}\label{count:cand:rydor-3}
                         \label{count:cand:vztuc-3}
    \arabic{hvref}. \citet{devac78distancescale} ;
   \refstepcounter{hvref}\label{count:cand:vztuc-4}
    \arabic{hvref}. \citet{how91CVamateur} ;
    \refstepcounter{hvref}\label{count:cand:kyara-2}
     \arabic{hvref}. \citet{sch94kyara} ;
    \refstepcounter{hvref}\label{count:cand:v359cen-2}
     \arabic{hvref}. \citet{gil98v359cen} ;
    \refstepcounter{hvref}\label{count:cand:v359cen-3}
     \arabic{hvref}. \citet{wou01v359cenxzeriyytel} ;
   \refstepcounter{hvref}\label{count:cand:kxaql-1}
    \arabic{hvref}. \citet{GCVS} ;
   \refstepcounter{hvref}\label{count:cand:kxaql-2}
    \arabic{hvref}. \citet{tap01kxaql} ;
   \refstepcounter{hvref}\label{count:cand:eluma-1}
    \arabic{hvref}. \citet{pes87eluma} ;
   \refstepcounter{hvref}\label{count:cand:eluma-2}
    \arabic{hvref}. \citet{pes88elumaCasesurvey} ;
   \refstepcounter{hvref}\label{count:cand:eluma-3}
    \arabic{hvref}. \citet{wag88CaseSurvey} ;
   \refstepcounter{hvref}\label{count:cand:v336per-1}
    \arabic{hvref}. \citet{bus79VS17} ;
   \refstepcounter{hvref}\label{count:cand:v336per-2}
                         \label{count:cand:bccas-1}
    \arabic{hvref}. \citet{liu00CVspec3} ;
   \refstepcounter{hvref}\label{count:cand:iodel-1}
    \arabic{hvref}. \citet{ric70newvar} ;
   \refstepcounter{hvref}\label{count:cand:iodel-2}
    \arabic{hvref}. \citet{ric94iodel} ;
   \refstepcounter{hvref}\label{count:cand:svari-2}
    \arabic{hvref}. \citet{wol05svari} ;
   \refstepcounter{hvref}\label{count:cand:bccas-2}
                         \label{count:cand:eyaql-1}
    \arabic{hvref}. \pasjcitetsub{due84eyaqlbccasmtcenv745sco}{a} ;
   \refstepcounter{hvref}\label{count:cand:bccas-3}
    \arabic{hvref}. \citet{sha90bccas} ;
   \refstepcounter{hvref}\label{count:cand:bccas-4}
    \arabic{hvref}. \citet{zwi94CVspec1} ;
   \refstepcounter{hvref}\label{count:cand:bccas-5}
    \arabic{hvref}. \citet{rin96oldnovaspec} ;
   \refstepcounter{hvref}\label{count:cand:apcru-1}
    \arabic{hvref}. \citet{oco36apcru} ;
   \refstepcounter{hvref}\label{count:cand:apcru-2}
    \arabic{hvref}. \citet{mun98CVspec5} ;
   \refstepcounter{hvref}\label{count:cand:eyaql-2}
    \arabic{hvref}. \citet{alb29eyaql} ;
   \refstepcounter{hvref}\label{count:cand:dvdra-1}
    \arabic{hvref}. \citet{pav85dvdra} ;
   \refstepcounter{hvref}\label{count:cand:dvdra-2}
    \arabic{hvref}. \pasjcitetsub{wen91dvdra}{a} ;
   \refstepcounter{hvref}\label{count:cand:dvdra-3}
    \arabic{hvref}. \citet{ric92wzsgedip} ;
   \refstepcounter{hvref}\label{count:cand:dvdra-4}
    \arabic{hvref}. \citet{iid95dvdra} ;
   \refstepcounter{hvref}\label{count:cand:dvdra-5}
    \arabic{hvref}. \citet{liu99CVspec2} ;
   \refstepcounter{hvref}\label{count:cand:dvdra-6}
    \arabic{hvref}. \citet{sam00dvdra} ;
   \refstepcounter{hvref}\label{count:cand:v632her-2}
    \arabic{hvref}. \citet{dor67v632her} ;
   \refstepcounter{hvref}\label{count:cand:v632her-3}
    \arabic{hvref}. \citet{dor68v632her} ;
   \refstepcounter{hvref}\label{count:cand:v369lyr-1}
    \arabic{hvref}. \citet{kur68newvar} ;
   \refstepcounter{hvref}\label{count:cand:grori-2}
    \arabic{hvref}. \citet{thi16grori} ;
   \refstepcounter{hvref}\label{count:cand:grori-3}
    \arabic{hvref}. \citet{her58VSchart} ;
   \refstepcounter{hvref}\label{count:cand:v522sgr-1}
    \arabic{hvref}. \citet{fer35v522sgr} ;
   \refstepcounter{hvref}\label{count:cand:v522sgr-2}
    \arabic{hvref}. \citet{bia92CVspec} ;
   \refstepcounter{hvref}\label{count:cand:v522sgr-3}
    \arabic{hvref}. \citet{rin96oldnovaspec} ;
   \refstepcounter{hvref}\label{count:cand:sslmi-2}
    \arabic{hvref}. \citet{alk80sslmi} ;
   \refstepcounter{hvref}\label{count:cand:sslmi-3}
    \arabic{hvref}. \citet{har91sslmiiauc} ;
   \refstepcounter{hvref}\label{count:cand:sslmi-4}
    \arabic{hvref}. \citet{how91sslmiiauc} ;
   \refstepcounter{hvref}\label{count:cand:cigem-1}
    \arabic{hvref}. \pasjcitetsub{hof43cigem}{a} ;
   \refstepcounter{hvref}\label{count:cand:cigem-2}
    \arabic{hvref}. \citet{hof47cigem} ;
   \refstepcounter{hvref}\label{count:cand:cigem-3}
    \arabic{hvref}. \citet{wen90cigem} ;
   \refstepcounter{hvref}\label{count:cand:cigem-4}
    \arabic{hvref}. \citet{zwi94CVspec1} ;
   \refstepcounter{hvref}\label{count:cand:cigem-5}
    \arabic{hvref}. \citet{mun98CVspec5} ;
   \refstepcounter{hvref}\label{count:cand:cigem-6}
    \arabic{hvref}. \citet{kat99cigem} ;
   \refstepcounter{hvref}\label{count:cand:cigem-7}
    \arabic{hvref}. \citet{sch99cigem} ;
   \refstepcounter{hvref}\label{count:cand:ptand-1}
    \arabic{hvref}. \citet{gru59ptand} ;
   \refstepcounter{hvref}\label{count:cand:ptand-2}
    \arabic{hvref}. \citet{sha89ptand} ;
   \refstepcounter{hvref}\label{count:cand:ptand-3}
    \arabic{hvref}. \citet{alk00ptand} ;
   \refstepcounter{hvref}\label{count:cand:v4338sgr-1}
    \arabic{hvref}. \pasjcitetsub{lil90v4338sgriauc1}{a} ;
   \refstepcounter{hvref}\label{count:cand:v4338sgr-2}
    \arabic{hvref}. \pasjcitetsub{lil90v4338sgriauc2}{b} ;
   \refstepcounter{hvref}\label{count:cand:v4338sgr-3}
    \arabic{hvref}. \citet{mcn90v4338sgriauc} ;
   \refstepcounter{hvref}\label{count:cand:v4338sgr-4}
    \arabic{hvref}. \citet{har90v4338sgriauc} ;
   \refstepcounter{hvref}\label{count:cand:v4338sgr-5}
    \arabic{hvref}. \citet{wag90v4338sgriauc} ;
   \refstepcounter{hvref}\label{count:cand:v4338sgr-6}
    \arabic{hvref}. \citet{sum97v4338sgr} ;
   \refstepcounter{hvref}\label{count:cand:aspsc-1}
    \arabic{hvref}. \citet{ric79m33blueobjectaspsc} ;
   \refstepcounter{hvref}\label{count:cand:aspsc-2}
    \arabic{hvref}. \citet{ric81aspsc} ;
   \refstepcounter{hvref}\label{count:cand:aspsc-3}
    \arabic{hvref}. \pasjcitetsub{ric83aspsc}{b} ;
   \refstepcounter{hvref}\label{count:cand:aspsc-4}
    \arabic{hvref}. \pasjcitetsub{ric86CVamplitudecyclelength}{a} ;
   \refstepcounter{hvref}\label{count:cand:aspsc-5}
    \arabic{hvref}. \citet{sha87aspsc} ;
   \refstepcounter{hvref}\label{count:cand:aspsc-6}
    \arabic{hvref}. \citet{ric89aspsc} ;
   \refstepcounter{hvref}\label{count:cand:aspsc-7}
    \arabic{hvref}. \citet{kat01uvgemfsandaspsc} ;
   \refstepcounter{hvref}\label{count:cand:nsv15820-1}
    \arabic{hvref}. \citet{par83nsv15820} ;
   \refstepcounter{hvref}\label{count:cand:vxfor-1}
    \arabic{hvref}. \pasjcitetsub{lil90vxforiauc}{xx} ;
   \refstepcounter{hvref}\label{count:cand:xypsc-1}
    \arabic{hvref}. \pasjcitetsub{ros72xypsciauc1}{a} ;
   \refstepcounter{hvref}\label{count:cand:xypsc-2}
    \arabic{hvref}. \pasjcitetsub{ros72xypsciauc2}{b} ;
   \refstepcounter{hvref}\label{count:cand:xypsc-3}
    \arabic{hvref}. \citet{hen01xypsc} ;
   \refstepcounter{hvref}\label{count:cand:ceuma-2}
    \arabic{hvref}. \citet{and67ceumaiauc} ;
   \refstepcounter{hvref}\label{count:cand:nsv00895-1}
    \arabic{hvref}. \citet{wen90nsv00895} ;
   \refstepcounter{hvref}\label{count:cand:sn1964o-1}
    \arabic{hvref}. \citet{DownesCVatlas2} ;
   \refstepcounter{hvref}\label{count:cand:fbs1719-1}
                         \label{count:cand:fbs1735-1}
    \arabic{hvref}. \citet{abr96FBS11} ;
   \refstepcounter{hvref}\label{count:cand:fbs1719-2}
                         \label{count:cand:fbs1735-2}
    \arabic{hvref}. \pasjcitetsub{kat97FBS}{c} ;
   \refstepcounter{hvref}\label{count:cand:aooct-1}
                         \label{count:cand:v551sgr-1}
                         \label{count:cand:gocom-1}
                         \label{count:cand:yytel-1}
                         \label{count:cand:fvara-1}
    \arabic{hvref}. \citet{vog82atlas} ;
   \refstepcounter{hvref}\label{count:cand:gocom-2}
    \arabic{hvref}. \citet{ush81gocom} ;
   \refstepcounter{hvref}\label{count:cand:gocom-3}
    \arabic{hvref}. \citet{kat90gocom} ;
   \refstepcounter{hvref}\label{count:cand:gocom-4}
    \arabic{hvref}. \citet{kat95gocom} ;
   \refstepcounter{hvref}\label{count:cand:yytel-2}
    \arabic{hvref}. \citet{odo91wzsge} ;
   \refstepcounter{hvref}\label{count:cand:yytel-3}
    \arabic{hvref}. \citet{wou01v359cenxzeriyytel} ;
   \refstepcounter{hvref}\label{count:cand:nsv15272-1}
    \arabic{hvref}. \citet{iid90nsv15272} ;
   \refstepcounter{hvref}\label{count:cand:j0643-1}
    \arabic{hvref}. \citet{gre00j0643id210} ;
   \refstepcounter{hvref}\label{count:cand:v1289aql-1}
    \arabic{hvref}. \citet{wol04v1289aql} ;
   \refstepcounter{hvref}\label{count:cand:v1289aql-2}
    \arabic{hvref}. \citet{mei70v1289aql} ;
   \refstepcounter{hvref}\label{count:cand:v1289aql-3}
    \arabic{hvref}. \citet{har92CVIRAS3} ;
   \refstepcounter{hvref}\label{count:cand:v1289aql-4}
    \arabic{hvref}. \citet{dow95CVspec} ;
   \refstepcounter{hvref}\label{count:cand:gd552-1}
    \arabic{hvref}. \citet{gre78gd552} ;
   \refstepcounter{hvref}\label{count:cand:gd552-2}
    \arabic{hvref}. \citet{hes90gd552} ;
   \refstepcounter{hvref}\label{count:cand:gd552-3}
    \arabic{hvref}. \pasjcitetsub{ric90gd552}{b} ;
   \refstepcounter{hvref}\label{count:cand:bwscl-1}
    \arabic{hvref}. \citet{abb97bwscl} ;
   \refstepcounter{hvref}\label{count:cand:bwscl-2}
    \arabic{hvref}. \citet{aug97bwscl} ;
   \refstepcounter{hvref}\label{count:cand:bwscl-3}
    \arabic{hvref}. \citet{aug98bwscl} ;
   \refstepcounter{hvref}\label{count:cand:rej1255-1}
    \arabic{hvref}. \citet{dah94j1255iauc} ;
   \refstepcounter{hvref}\label{count:cand:rej1255-2}
    \arabic{hvref}. \citet{wen94rej1255iauc} ;
   \refstepcounter{hvref}\label{count:cand:rej1255-3}
    \arabic{hvref}. \citet{dah95j1255} ;
   \refstepcounter{hvref}\label{count:cand:rej1255-4}
    \arabic{hvref}. \citet{wat95j1255iauc} ;
   \refstepcounter{hvref}\label{count:cand:rej1255-5}
    \arabic{hvref}. \citet{wat96j1255} ;
   \refstepcounter{hvref}\label{count:cand:rej1255-6}
    \arabic{hvref}. \citet{dra98j1255EUVE} ;
   \refstepcounter{hvref}\label{count:cand:rej1255-7}
    \arabic{hvref}. \citet{whe00j1255} ;
   \refstepcounter{hvref}\label{count:cand:j1050-1}
    \arabic{hvref}. \citet{men01j1050} ;
   \refstepcounter{hvref}\label{count:cand:j1050-2}
    \arabic{hvref}. \pasjcitetsub{kat01j1050chat4601}{f} ;
   \refstepcounter{hvref}\label{count:cand:gpcom-1}
    \arabic{hvref}. \citet{bur71gpcom} ;
   \refstepcounter{hvref}\label{count:cand:gpcom-2}
    \arabic{hvref}. \citet{war72gpcom} ;
   \refstepcounter{hvref}\label{count:cand:gpcom-3}
    \arabic{hvref}. \citet{ric73gpcom} ;
   \refstepcounter{hvref}\label{count:cand:gpcom-4}
    \arabic{hvref}. \citet{sma75gpcom} ;
   \refstepcounter{hvref}\label{count:cand:gpcom-5}
    \arabic{hvref}. \citet{nat81gpcom} ;
   \refstepcounter{hvref}\label{count:cand:gpcom-6}
    \arabic{hvref}. \citet{lam81gpcom} ;
   \refstepcounter{hvref}\label{count:cand:gpcom-7}
    \arabic{hvref}. \citet{sto83gpcom} ;
   \refstepcounter{hvref}\label{count:cand:gpcom-8}
    \arabic{hvref}. \citet{ull90amcvncrboov803cengpcom} ;
   \refstepcounter{hvref}\label{count:cand:gpcom-9}
    \arabic{hvref}. \citet{mar91gpcom} ;
   \refstepcounter{hvref}\label{count:cand:gpcom-10}
    \arabic{hvref}. \citet{mar95gpcom} ;
   \refstepcounter{hvref}\label{count:cand:gpcom-11}
    \arabic{hvref}. \citet{mar99gpcom} ;
   \refstepcounter{hvref}\label{count:cand:ce315-1}
    \arabic{hvref}. \citet{rui01ce315} ;
   \refstepcounter{hvref}\label{count:cand:ce315-2}
    \arabic{hvref}. \citet{wou01v359cenxzeriyytel} 
}
\end{table*}

\begin{table*}
\caption{Properties of large-amplitude SU UMa-type dwarf novae and related
         systems.}\label{tab:largeampsuuma}
\setcounter{hvref}{0}
\setcounter{hvrem}{0}
\begin{center}
\begin{tabular}{lccccccl}
\hline\hline
Object    & Max  & Min & Amplitude & $P_{\rm orb}$ & $P_{\rm SH}$ &
            Remarks & References \\
\hline
WX Cet    &  9.3 & 18.0 &  8.7 & 0.05827 & 0.05949 &
                                 \ref{remcount:suuma:noearlysuperhump},
                                 \ref{remcount:suuma:wxcet}
                                 & \ref{count:suuma:wxcet-1},
                                   \ref{count:suuma:wxcet-2},
                                   \ref{count:suuma:wxcet-3},
                                   \ref{count:suuma:wxcet-4},
                                   \ref{count:suuma:wxcet-5},
                                   \ref{count:suuma:wxcet-6}, \\
          &      &      &  & & & & \ref{count:suuma:wxcet-7},
                                   \ref{count:suuma:wxcet-8},
                                   \ref{count:suuma:wxcet-9},
                                   \ref{count:suuma:wxcet-10},
                                   \ref{count:suuma:wxcet-11},
                                   \ref{count:suuma:wxcet-12}, \\
          &      &      &  & & & & \ref{count:suuma:wxcet-13},
                                   \ref{count:suuma:wxcet-14},
                                   \ref{count:suuma:wxcet-15},
                                   \ref{count:suuma:wxcet-16},
                                   \ref{count:suuma:wxcet-17}, \\
          &      &      &  & & & & \ref{count:suuma:wxcet-18},
                                   \ref{count:suuma:wxcet-19},
                                   \ref{count:suuma:wxcet-20},
                                   \ref{count:suuma:wxcet-21} \\
VY Aqr    &  9.0 & 17.5 &  8.5 & 0.06309 & 0.06436 &
                                 \ref{remcount:suuma:vyaqr}
                                 & \ref{count:suuma:vyaqr-1},
                                   \ref{count:suuma:vyaqr-2},
                                   \ref{count:suuma:vyaqr-3},
                                   \ref{count:suuma:vyaqr-4},
                                   \ref{count:suuma:vyaqr-5},
                                   \ref{count:suuma:vyaqr-6}, \\
          &      &      &  & & & & \ref{count:suuma:vyaqr-7},
                                   \ref{count:suuma:vyaqr-8},
                                   \ref{count:suuma:vyaqr-9},
                                   \ref{count:suuma:vyaqr-10},
                                   \ref{count:suuma:vyaqr-11},
                                   \ref{count:suuma:vyaqr-12}, \\
          &      &      &  & & & & \ref{count:suuma:vyaqr-13},
                                   \ref{count:suuma:vyaqr-14},
                                   \ref{count:suuma:vyaqr-15},
                                   \ref{count:suuma:vyaqr-16},
                                   \ref{count:suuma:vyaqr-17}, \\
          &      &      &  & & & & \ref{count:suuma:vyaqr-18},
                                   \ref{count:suuma:vyaqr-19},
                                   \ref{count:suuma:vyaqr-20} \\
BC UMa    & 10.9 & 18.3 &  7.4 & 0.06261 & 0.06452 &
                                 \ref{remcount:suuma:bcuma}
                                 & \ref{count:suuma:bcuma-1},
                                   \ref{count:suuma:bcuma-2},
                                   \ref{count:suuma:bcuma-3},
                                   \ref{count:suuma:bcuma-4},
                                   \ref{count:suuma:bcuma-5},
                                   \ref{count:suuma:bcuma-6} \\
V1251 Cyg & 12.5 & 19:  & 6.5: & - & 0.07604 &
                                 -
                                 & \ref{count:suuma:v1251cyg-1},
                                   \ref{count:suuma:v1251cyg-2},
                                   \ref{count:suuma:v1251cyg-3},
                                   \ref{count:suuma:v1251cyg-4},
                                   \ref{count:suuma:v1251cyg-5},
                                   \ref{count:suuma:v1251cyg-6} \\
V1028 Cyg & 12.8 & 18.0 &  5.2 & - & 0.06154 &
                                 -
                                 & \ref{count:suuma:v1028cyg-1},
                                   \ref{count:suuma:v1028cyg-2},
                                   \ref{count:suuma:v1028cyg-3},
                                   \ref{count:suuma:v1028cyg-4},
                                   \ref{count:suuma:v1028cyg-5},
                                   \ref{count:suuma:v1028cyg-6}, \\
          &      &      &  & & & & \ref{count:suuma:v1028cyg-7},
                                   \ref{count:suuma:v1028cyg-8} \\
UV Per    & 11.7 & 17.9 &  6.2 & 0.06489 & 0.06641 &
                                 \ref{remcount:suuma:noearlysuperhump},
                                 \ref{remcount:suuma:uvper}
                                 & \ref{count:suuma:uvper-1},
                                   \ref{count:suuma:uvper-2},
                                   \ref{count:suuma:uvper-3},
                                   \ref{count:suuma:uvper-4},
                                   \ref{count:suuma:uvper-5} \\
CP Dra    & 15.1 & 20   &  5   & - & 0.08474 &
                                 \ref{remcount:suuma:noearlysuperhump},
                                 \ref{remcount:suuma:cpdra}
                                 & \ref{count:suuma:cpdra-1},
                                   \ref{count:suuma:cpdra-2},
                                   \ref{count:suuma:cpdra-3},
                                   \ref{count:suuma:cpdra-4},
                                   \ref{count:suuma:cpdra-5},
                                   \ref{count:suuma:cpdra-6}, \\
          &      &      &  & & & & \ref{count:suuma:cpdra-7},
                                   \ref{count:suuma:cpdra-8} \\
DV UMa    & 13.9 & 19.1  & 5.2 & 0.08585 & 0.08869 &
                                 \ref{remcount:suuma:noearlysuperhump},
                                 \ref{remcount:suuma:longp},
                                 \ref{remcount:suuma:dvuma}
                                 & \ref{count:suuma:dvuma-1},
                                   \ref{count:suuma:dvuma-2},
                                   \ref{count:suuma:dvuma-3},
                                   \ref{count:suuma:dvuma-4},
                                   \ref{count:suuma:dvuma-5},
                                   \ref{count:suuma:dvuma-6}, \\
          &      &      &  & & & & \ref{count:suuma:dvuma-7},
                                   \ref{count:suuma:dvuma-8},
                                   \ref{count:suuma:dvuma-9},
                                   \ref{count:suuma:dvuma-10} \\
EF Peg    & 10.7 & 18.5  & 7.8 & - & 0.08705 &
                                 \ref{remcount:suuma:noearlysuperhump},
                                 \ref{remcount:suuma:longp},
                                 \ref{remcount:suuma:efpeg}
                                 & \ref{count:suuma:efpeg-1},
                                   \ref{count:suuma:efpeg-2},
                                   \ref{count:suuma:efpeg-3},
                                   \ref{count:suuma:efpeg-4},
                                   \ref{count:suuma:efpeg-5},
                                   \ref{count:suuma:efpeg-6}, \\
          &      &      &  & & & & \ref{count:suuma:efpeg-7},
                                   \ref{count:suuma:efpeg-8},
                                   \ref{count:suuma:efpeg-9},
                                   \ref{count:suuma:efpeg-10} \\
V725 Aql  & 13.2 & 19.3 & 6.1 & - & 0.09909 &
                                 \ref{remcount:suuma:longp},
                                 \ref{remcount:suuma:v725aql}
                                 & \ref{count:suuma:v725aql-1},
                                   \ref{count:suuma:v725aql-2},
                                   \ref{count:suuma:v725aql-3} \\
TV Crv    & 12.5 & 19   & 6.5: & - & 0.0650 &
                                 -
                                 & \ref{count:suuma:tvcrv-1},
                                   \ref{count:suuma:tvcrv-2},
                                   \ref{count:suuma:tvcrv-3},
                                   \ref{count:suuma:tvcrv-4},
                                   \ref{count:suuma:tvcrv-5},
                                   \ref{count:suuma:tvcrv-6} \\
PU Per    & 15.2 & [20  &$>$4.8& - & 0.0733 &
                                 \ref{remcount:suuma:puper}
                                 & \ref{count:suuma:puper-1},
                                   \ref{count:suuma:puper-2},
                                   \ref{count:suuma:puper-3},
                                   \ref{count:suuma:puper-4},
                                   \ref{count:suuma:puper-5},
                                   \ref{count:suuma:puper-6} \\
QY Per    & 14.2 & [20  &$>$5.8& - & 0.07681 &
                                 \ref{remcount:suuma:noearlysuperhump}
                                 & \ref{count:suuma:qyper-1},
                                   \ref{count:suuma:qyper-2},
                                   \ref{count:suuma:qyper-3},
                                   \ref{count:suuma:qyper-4} \\
MM Hya    & 14.2 & 18.9 & 4.7  & 0.05759 & 0.05865 &
                                 \ref{remcount:suuma:mmhya}
                                 & \ref{count:suuma:mmhya-1},
                                   \ref{count:suuma:mmhya-2},
                                   \ref{count:suuma:mmhya-3},
                                   \ref{count:suuma:mmhya-4} \\
KV And    & 13.6 & 20.5:& 6.9: & - & 0.07434 &
                                 \ref{remcount:suuma:kvand}
                                 & \ref{count:suuma:kvand-1},
                                   \ref{count:suuma:kvand-2},
                                   \ref{count:suuma:kvand-3} \\
SW UMa    &  9.0 & 17.0 & 8.0  & 0.05681 & 0.05818 &
                                 \ref{remcount:suuma:noearlysuperhump}
                                 & \ref{count:suuma:swuma-1},
                                   \ref{count:suuma:swuma-2},
                                   \ref{count:suuma:swuma-3},
                                   \ref{count:suuma:swuma-4},
                                   \ref{count:suuma:swuma-5},
                                   \ref{count:suuma:swuma-6}, \\
          &      &      &  & & & & \ref{count:suuma:swuma-7},
                                   \ref{count:suuma:swuma-8},
                                   \ref{count:suuma:swuma-9},
                                   \ref{count:suuma:swuma-10},
                                   \ref{count:suuma:swuma-11},
                                   \ref{count:suuma:swuma-12}, \\
          &      &      &  & & & & \ref{count:suuma:swuma-13},
                                   \ref{count:suuma:swuma-14} \\
CT Hya    & 14.0 & 19.9 & 5.9  & - & 0.06643 &
                                 \ref{remcount:suuma:cthya}
                                 & \ref{count:suuma:cthya-1},
                                   \ref{count:suuma:cthya-2},
                                   \ref{count:suuma:cthya-3} \\
T Leo     &  9.9 & 15.9 & 6.0  & 0.05882 & 0.0602 &
                                 \ref{remcount:suuma:tleo}
                                 & \ref{count:suuma:tleo-1},
                                   \ref{count:suuma:tleo-2},
                                   \ref{count:suuma:tleo-3},
                                   \ref{count:suuma:tleo-4},
                                   \ref{count:suuma:tleo-5},
                                   \ref{count:suuma:tleo-6}, \\
          &      &      &  & & & & \ref{count:suuma:tleo-7},
                                   \ref{count:suuma:tleo-8},
                                   \ref{count:suuma:tleo-9},
                                   \ref{count:suuma:tleo-10},
                                   \ref{count:suuma:tleo-11},
                                   \ref{count:suuma:tleo-12}, \\
          &      &      &  & & & & \ref{count:suuma:tleo-13},
                                   \ref{count:suuma:tleo-14},
                                   \ref{count:suuma:tleo-15},
                                   \ref{count:suuma:tleo-16},
                                   \ref{count:suuma:tleo-17},
                                   \ref{count:suuma:tleo-18}, \\
          &      &      &  & & & & \ref{count:suuma:tleo-19},
                                   \ref{count:suuma:tleo-20},
                                   \ref{count:suuma:tleo-21},
                                   \ref{count:suuma:tleo-22} \\
EK TrA    & 10.4 & 16.3 & 5.9  & 0.06455 & 0.0649 &
                                 \ref{remcount:suuma:ektra}
                                 & \ref{count:suuma:ektra-1},
                                   \ref{count:suuma:ektra-2},
                                   \ref{count:suuma:ektra-3},
                                   \ref{count:suuma:ektra-4},
                                   \ref{count:suuma:ektra-5},
                                   \ref{count:suuma:ektra-6} \\
V844 Her  & 12.5 & 17.5 & 5.0  & 0.0547 & 0.05592 &
                                 \ref{remcount:suuma:v844her}
                                 & \ref{count:suuma:v844her-1},
                                   \ref{count:suuma:v844her-2} \\
KV Dra    & 11.8 & 17.1 & 5.3  & 0.05898 & 0.0601 &
                                 \ref{remcount:suuma:kvdra}
                                 & \ref{count:suuma:kvdra-1},
                                   \ref{count:suuma:kvdra-2} \\
HT Cam    & 12.9 & 18.2 & 5.3  & 0.0598 & - &
                                 \ref{remcount:suuma:htcam}
                                 & \ref{count:suuma:htcam-1},
                                   \ref{count:suuma:htcam-2},
                                   \ref{count:suuma:htcam-3},
                                   \ref{count:suuma:htcam-4} \\
HT Cas    & 10.8 & 16.4 & 5.6  & 0.07367 & 0.07608 &
                                 \ref{remcount:suuma:htcas}
                                 & \ref{count:suuma:htcas-1},
                                   \ref{count:suuma:htcas-2},
                                   \ref{count:suuma:htcas-3},
                                   \ref{count:suuma:htcas-4},
                                   \ref{count:suuma:htcas-5},
                                   \ref{count:suuma:htcas-6}, \\
          &      &      &  & & & & \ref{count:suuma:htcas-7},
                                   \ref{count:suuma:htcas-8},
                                   \ref{count:suuma:htcas-9},
                                   \ref{count:suuma:htcas-10},
                                   \ref{count:suuma:htcas-11},
                                   \ref{count:suuma:htcas-12}, \\
          &      &      &  & & & & \ref{count:suuma:htcas-13},
                                   \ref{count:suuma:htcas-14},
                                   \ref{count:suuma:htcas-15},
                                   \ref{count:suuma:htcas-16},
                                   \ref{count:suuma:htcas-17},
                                   \ref{count:suuma:htcas-18}, \\
          &      &      &  & & & & \ref{count:suuma:htcas-19},
                                   \ref{count:suuma:htcas-20},
                                   \ref{count:suuma:htcas-21},
                                   \ref{count:suuma:htcas-22},
                                   \ref{count:suuma:htcas-23},
                                   \ref{count:suuma:htcas-24}, \\
          &      &      &  & & & & \ref{count:suuma:htcas-25} \\
IR Com    & 13.5 & 17.0 & 3.5  & 0.08704 & - &
                                 \ref{remcount:suuma:ircom}
                                 & \ref{count:suuma:ircom-1},
                                   \ref{count:suuma:ircom-2},
                                   \ref{count:suuma:ircom-3},
                                   \ref{count:suuma:ircom-4},
                                   \ref{count:suuma:ircom-5} \\
\hline
\end{tabular}
\end{center}
\end{table*}

\begin{table*}
{\footnotesize
  {\bf Remarks to table \ref{tab:largeampsuuma}:}
   \refstepcounter{hvrem}\label{remcount:suuma:noearlysuperhump}
    \arabic{hvrem}. -- Negative detection of early superhumps has been
    reported at least during one superoutburst.
   \refstepcounter{hvrem}\label{remcount:suuma:wxcet}
    \arabic{hvrem}. -- Large-amplitude SU UMa-type dwarf nova
    with a typical cycle length of 1--2 years.
    \citet{rog01wxcet} pointed out some similarities to WZ Sge in the
    behavior of quiescent humps.  Extensive observations by
    \pasjcitetsub{kat01wxcet}{c} during the 1998 superoutburst suggested
    more resemblance to ordinary SU UMa-type dwarf nova.
   \refstepcounter{hvrem}\label{remcount:suuma:vyaqr}
    \arabic{hvrem}. -- Large-amplitude SU UMa-type dwarf nova
    with a typical cycle length of 1--2 years until the superoutburst
    in 1990.  No outburst has been detected up to 2001 (VSNET).
    The system needs to be closely examined with respect to the outburst
    cycle length and outburst amplitude.
   \refstepcounter{hvrem}\label{remcount:suuma:bcuma}
    \arabic{hvrem}. -- Large-amplitude SU UMa-type dwarf nova
    with a typical cycle length of 1 to a few years.
    \citet{rom64bcuma} reported three types of outbursts: very bright and
    long, intermediately bright and moderately long, and short ones.
    The brightest outbursts occur very infrequently (once in $\sim$10 years).
    There may have been a positive detection of early superhumps
    \pasjcitepsub{kat00bcumaalert4562}{b} or possibly negative superhumps at
    $P$=0.0621 d (for a discussion, see \pasjcitesub{kat00bcumaalert4586}{d}).
    Possibly an intermediate object between a WZ Sge-type and SU UMa-type
    dwarf nova.  The superhump period is taken from
    \citet{uem00bcumaalert4601}.  The orbital period is taken from
    the table of \citet{pat01SH}, while the well-documented period in the
    literature is 0.063 d \citep{how90faintCV3}.
   \refstepcounter{hvrem}\label{remcount:suuma:uvper}
    \arabic{hvrem}. -- One of prototypical large-amplitude SU UMa-type dwarf
    novae with a low outburst frequency.  A well-observed post-superoutburst
    rebrightening occurred in 1989, which was also cited in
    \citet{kuu01j1118} showing the similarity with outbursts of black-hole
    X-ray transients.
   \refstepcounter{hvrem}\label{remcount:suuma:cpdra}
    \arabic{hvrem}. -- The superhump period is taken from
    \citet{uem01cpdracampaigndn556}.
   \refstepcounter{hvrem}\label{remcount:suuma:longp}
    \arabic{hvrem}. -- Aside from the long $P_{\rm orb}$, the outburst
    parameters inferred from recent observations are intermediate between
    usual SU UMa-type dwarf novae and WZ Sge-type dwarf novae.
   \refstepcounter{hvrem}\label{remcount:suuma:dvuma}
    \arabic{hvrem}. -- Eclipsing large-amplitude SU UMa-type dwarf nova.
    The maximum magnitude is taken from VSNET; the quoted value in
    \citet{RitterCV} is largely in error.  The minimum magnitude in the
    table is our unpublished measurement of an averaged quiescent magnitude
    outside eclipses.  The superhump period is taken from Uemura et al.
    in preparation.  Outbursts are relatively infrequent (once in 1--2 yr),
    with a mean supercycle length of 770 d \citet{nog01dvuma}.
    The record of archival plates \citep{ush82dvuma} may suggest that
    outbursts were more frequent in the past.  Early development of
    superhumps was precisely recorded during the 1999 December superoutburst
    (Uemura et al. in preparation).
   \refstepcounter{hvrem}\label{remcount:suuma:efpeg}
    \arabic{hvrem}. -- Large-amplitude SU UMa-type dwarf nova.
    Outbursts are relatively infrequent (once in $>$3 yr).
    The maximum and minimum magnitudes are taken from \citet{how93efpeg};
    the superhump period is from \pasjcitetsub{kat01efpeg}{a}.
    Superhumps had already appeared at the beginning of the past observations
    (\pasjcitesub{kat01efpeg}{a}; Matsumoto et al. in prepartion),
    suggestive of the lack of early superhumps.  A post-superoutburst
    rebrightening has been suggested \pasjcitepsub{kat97efpegalert1344}{b},
    although the confirmation was very difficult
    because of the existence of a bright close companion.
    \citet{pol98TOAD} suggested this object to be a post-period minimum
    object.
   \refstepcounter{hvrem}\label{remcount:suuma:v725aql}
    \arabic{hvrem}. -- Large-amplitude SU UMa-type dwarf nova.
    Outbursts are relatively infrequent (once in $\sim$1 yr).  Two confirmed
    superoutbursts were separated by 4 yr.  System parameters are taken
    from \citet{uem01v725aql}.  A confirmed post-superoutburst rebrightening
    was observed \citep{uem01v725aql}.  A system similar to EF Peg.
   \refstepcounter{hvrem}\label{remcount:suuma:puper}
    \arabic{hvrem}. -- The superhump period is taken from \citet{kat99puper}.
    Normal outbursts are relatively frequently seen (\cite{rom76DN},
    \cite{kat95puper}).
   \refstepcounter{hvrem}\label{remcount:suuma:mmhya}
    \arabic{hvrem}. -- Originally suggested to be a WZ Sge-type system
    based on the short orbital period \citep{mis95PGCV}.  The superhump
    period is taken from the table in \citet{pat98evolution}; more recent
    period determination ($P_{\rm SH}$=0.05785 d) during the 2001
    superoutburst \citep{war01mmhyaalert5884} is also known.  Possible early
    superhumps? (Ishioka et al. in preparation).
   \refstepcounter{hvrem}\label{remcount:suuma:kvand}
    \arabic{hvrem}. -- Large-amplitude SU UMa-type dwarf nova.  Also suggested
    to be a TOAD \citep{how95TOADproc}.  However, the minimum magnitude
    (22.5p) given in \citet{kur77kvandkwand} seems to be too
    faint as compared with the DSS 2 image and the unpublished quiescent
    CCD photometry by the authors, which is listed in the table.
    The maximum magnitude is from observations reported to VSNET.
    Outbursts are relatively frequent (a fairly normal SU UMa-type object
    in terms of its outburst activity and a superhump period), in spite of
    the large outburst amplitude.
   \refstepcounter{hvrem}\label{remcount:suuma:cthya}
    \arabic{hvrem}. -- The superhump period is taken from
    \pasjcitetsub{kat99cthya}{a}, which corresponds to the longer alias
    listed in \citet{nog96cthya}.
    Although \citet{nog96cthya} suggested the possible intermediate nature
    between usual SU UMa-type stars and WZ Sge-type stars, outbursts are
    relatively frequent: the shortest interval between normal outbursts
    is $\sim$30 d, and the likely supercycle is 250--300 d (VSNET).
   \refstepcounter{hvrem}\label{remcount:suuma:tleo}
    \arabic{hvrem}. -- System with a short orbital period.  Based on the
    research of past outbursts, \pasjcitetsub{wen83tleo}{a} already
    pointed out the similarity with WZ Sge.  Pickering and Wendell recorded
    an outburst at magnitude 7.4 in 1882, which was questioned by
    \citet{wen84tleo} because of its extreme brightness.  Although recent
    superoutbursts of this object usually reach $V$=10, there may have
    been in the past high-amplitude outburst(s) like those of WZ Sge.
    Superhumps in the early stage of superoutbursts have been relatively
    well observed:
    \pasjcitetsub{kat97tleo}{a} recorded a smooth increase of the superhump
    period starting with the nearly orbital period.  This phenomenon may be
    related to early superhumps in WZ Sge-type stars
    \pasjcitepsub{kat97tleo}{a}.
   \refstepcounter{hvrem}\label{remcount:suuma:ektra}
    \arabic{hvrem}. -- System with a short orbital period.  Resembles T Leo
    in its outburst behavior.  Photometric studies of superhumps in the
    past were not comprehensive enough to draw a conclusion regarding
    early superhumps.  The minimum magnitude is taken from \citet{gan01ektra},
    while \citet{RitterCV} gave an upper limit of $V$=17.  The maximum
    magnitude is an extreme value listed in \citet{GCVS}.  Recent observations
    reported to VSNET indicate that the usual magnitudes of superoutbursts
    reach $V$=11.4.
   \\
}
\end{table*}
\begin{table*}
{\footnotesize
  {\bf Remarks to table \ref{tab:largeampsuuma} (continued):}
   \refstepcounter{hvrem}\label{remcount:suuma:v844her}
    \arabic{hvrem}. -- System with a very short superhump period.
    Only superoutbursts were known, but with a short recurrence time
    (220--290 d: \cite{kat00v844her}).  The orbital period in the table
    is from quiescent photometry (Kato in preparation).
   \refstepcounter{hvrem}\label{remcount:suuma:kvdra}
    \arabic{hvrem}. -- Other names: RX J1450.5+6403, HS 1449+6415.
    System with a short orbital period ($P_{\rm orb}$=0.05898 d,
    $P_{\rm SH}$=0.0601 d).  \citet{nog00kvdra} suggested the object to be
    an intermediate object between WZ Sge-type and ER UMa-type stars.
    One well-established superoutburst in 2000 May and subsequent normal
    outbursts are known.  The lack of further outbursts (VSNET) may suggest
    a lower typical frequency of outbursts than proposed by
    \citet{nog00kvdra}.  \pasjcitetsub{kat00kvdraalert4858}{a}
    reported double-wave modulations in quiescence, which are reminiscent
    of those of WZ Sge.  More observations are needed to establish the
    outburst behavior and the evolution of superhumps.
   \refstepcounter{hvrem}\label{remcount:suuma:htcam}
    \arabic{hvrem}. -- Peculiar system with a short orbital period of
    81$\pm$5 min \citep{tov98htcam}.  Although the object was suggested to
    be an intermediate polar \citep{tov98htcam}, its classification has
    not yet been established.  The object shows very brief
    (less than $\sim$1 d) dwarf nova-like outbursts (\cite{wat98htcam};
    see also \cite{sch00htcam} and \pasjcitesub{kat01htcam}{b} for recent
    observations).
    Possible relation to WZ Sge has been suggested \citep{las99wzsgeIP}.
   \refstepcounter{hvrem}\label{remcount:suuma:htcas}
    \arabic{hvrem}. -- Eclipsing system (the eclipsing nature was discovered
    by H. E. Bond in 1978, and published by \citet{why80CVevolution} and
    \citet{pat81DNOhtcas}).
    The minimum magnitude listed in the table is a typical magnitude outside
    eclipses.  Extreme magnitude of $V$=18.4 was reported.  Low/high-state
    transitions have been observed in quiescence \citep{rob96htcas}.
    Superoutbursts are very rare (the only secure superoutburst in recent
    years was the 1985 one (\pasjcitesub{mat85htcasiauc}{b};
    \cite{waa85htcasiauc}; \cite{wen85htcas};
    \cite{zha86htcas})), while normal outbursts occur with a typical
    recurrence time of 1--2 yr (VSNET data; see also \citet{wen87htcas}
    for the discussion of historical outbursts).  The quoted outburst
    period in \citet{GCVS} came from \citet{GlasbyDNbook}, which is
    most likely erroneous.
    The only superhump detection was done during the 1985 superoutburst
    \citep{zha86htcas}.  No systematic observation of the early stage
    of this superoutburst was available.  There was evidence of
    large-amplitude variations outside eclipses from photographic
    observations (Wakuda, private communication).
    A similarity of the outburst mechanism with WZ Sge-type systems and
    X-ray transients has been proposed \citep{las95wzsge}.
   \refstepcounter{hvrem}\label{remcount:suuma:ircom}
    \arabic{hvrem}. -- Eclipsing system.  The minimum magnitude listed in
    the table is a typical magnitude outside eclipses.  Very similar to
    HT Cas, despite that no superoutburst has been yet observed
    \pasjcitepsub{kat01ircom}{a}.
   \\
}
\end{table*}
\begin{table*}
{\footnotesize
  {\bf References to table \ref{tab:largeampsuuma}:}
   \refstepcounter{hvref}\label{count:suuma:wxcet-1}
    \arabic{hvref}. \citet{str64wxcet} ;
   \refstepcounter{hvref}\label{count:suuma:wxcet-2}
    \arabic{hvref}. \citet{mei76wxcet} ;
   \refstepcounter{hvref}\label{count:suuma:wxcet-3}
    \arabic{hvref}. \citet{gap76wxcet} ;
   \refstepcounter{hvref}\label{count:suuma:wxcet-4}
    \arabic{hvref}. \citet{wyc78oldnovaID} ;
   \refstepcounter{hvref}\label{count:suuma:wxcet-5}
    \arabic{hvref}. \citet{bai79wzsge} ;
   \refstepcounter{hvref}\label{count:suuma:wxcet-6}
    \arabic{hvref}. \citet{vanpar89wxcet} ;
   \refstepcounter{hvref}\label{count:suuma:wxcet-7}
    \arabic{hvref}. \citet{mcn89wxcetiauc} ; 
   \refstepcounter{hvref}\label{count:suuma:wxcet-8}
    \arabic{hvref}. \citet{ove89wxcetiauc} ; 
   \refstepcounter{hvref}\label{count:suuma:wxcet-9}
    \arabic{hvref}. \citet{pea89wxcetiauc} ; 
   \refstepcounter{hvref}\label{count:suuma:wxcet-10}
    \arabic{hvref}. \citet{odo90wxcetiauc} ; 
   \refstepcounter{hvref}\label{count:suuma:wxcet-11}
    \arabic{hvref}. \citet{dow90wxcet} ;
   \refstepcounter{hvref}\label{count:suuma:wxcet-12}
    \arabic{hvref}. \citet{bat91wxcetiauc} ; 
   \refstepcounter{hvref}\label{count:suuma:wxcet-13}
    \arabic{hvref}. \citet{gil91wxcetiauc} ; 
   \refstepcounter{hvref}\label{count:suuma:wxcet-14}
    \arabic{hvref}. \citet{odo91wzsge} ;
   \refstepcounter{hvref}\label{count:suuma:wxcet-15}
                         \label{count:suuma:vyaqr-1}
    \arabic{hvref}. \citet{ric92wzsgedip} ;
   \refstepcounter{hvref}\label{count:suuma:wxcet-16}
    \arabic{hvref}. \citet{men93wxcetaqericuvel} ;
   \refstepcounter{hvref}\label{count:suuma:wxcet-17}
    \arabic{hvref}. \citet{men94wxcet} ;
   \refstepcounter{hvref}\label{count:suuma:wxcet-18}
    \arabic{hvref}. \pasjcitetsub{kat95wxcet}{c} ;
   \refstepcounter{hvref}\label{count:suuma:wxcet-19}
                         \label{count:suuma:swuma-1}
    \arabic{hvref}. \pasjcitetsub{szk00TOADs}{a} ;
   \refstepcounter{hvref}\label{count:suuma:wxcet-20}
    \arabic{hvref}. \citet{rog01wxcet} ;
   \refstepcounter{hvref}\label{count:suuma:wxcet-21}
    \arabic{hvref}. \pasjcitetsub{kat01wxcet}{c} ;
   \refstepcounter{hvref}\label{count:suuma:vyaqr-2}
    \arabic{hvref}. \citet{ros25vyaqr} ;
   \refstepcounter{hvref}\label{count:suuma:vyaqr-3}
    \arabic{hvref}. \citet{wyc78oldnovaID} ;
   \refstepcounter{hvref}\label{count:suuma:vyaqr-4}
    \arabic{hvref}. \citet{mcn82vyaqr} ;
   \refstepcounter{hvref}\label{count:suuma:vyaqr-5}
    \arabic{hvref}. \citet{mcn83vyaqriauc} ; 
   \refstepcounter{hvref}\label{count:suuma:vyaqr-6}
    \arabic{hvref}. \citet{mca83vyaqriauc} ; 
   \refstepcounter{hvref}\label{count:suuma:vyaqr-7}
    \arabic{hvref}. \pasjcitetsub{wen83vyaqr}{b} ; 
   \refstepcounter{hvref}\label{count:suuma:vyaqr-8}
    \arabic{hvref}. \pasjcitetsub{ric83vyaqr}{c} ; 
   \refstepcounter{hvref}\label{count:suuma:vyaqr-9}
    \arabic{hvref}. \citet{lil83vyaqr} ; 
   \refstepcounter{hvref}\label{count:suuma:vyaqr-10}
    \arabic{hvref}. \pasjcitetsub{ric83vyaqraditional}{a} ; 
   \refstepcounter{hvref}\label{count:suuma:vyaqr-11}
    \arabic{hvref}. \citet{hen83vyaqr} ; 
   \refstepcounter{hvref}\label{count:suuma:vyaqr-12}
    \arabic{hvref}. \citet{hur84vyaqr} ;
   \refstepcounter{hvref}\label{count:suuma:vyaqr-13}
    \arabic{hvref}. \citet{lub86vyaqriauc} ; 
   \refstepcounter{hvref}\label{count:suuma:vyaqr-14}
    \arabic{hvref}. \pasjcitetsub{hur87vyaqriauc}{a} ; 
   \refstepcounter{hvref}\label{count:suuma:vyaqr-15}
    \arabic{hvref}. \citet{lei87vyaqriauc} ; 
   \refstepcounter{hvref}\label{count:suuma:vyaqr-16}
    \arabic{hvref}. \citet{mun90vyaqriauc} ; 
   \refstepcounter{hvref}\label{count:suuma:vyaqr-17}
    \arabic{hvref}. \citet{dellaval90vyaqr} ;
   \refstepcounter{hvref}\label{count:suuma:vyaqr-18}
    \arabic{hvref}. \citet{pat93vyaqr} ;
   \refstepcounter{hvref}\label{count:suuma:vyaqr-19}
    \arabic{hvref}. \citet{aug94vyaqr} ;
   \refstepcounter{hvref}\label{count:suuma:vyaqr-20}
    \arabic{hvref}. \citet{tho97uvpervyaqrv1504cyg} ;
   \refstepcounter{hvref}\label{count:suuma:bcuma-1}
    \arabic{hvref}. \citet{rom64bcuma} ;
   \refstepcounter{hvref}\label{count:suuma:bcuma-2}
                         \label{count:suuma:cpdra-1}
                         \label{count:suuma:dvuma-1}
    \arabic{hvref}. \citet{muk90faintCV} ;
   \refstepcounter{hvref}\label{count:suuma:bcuma-3}
    \arabic{hvref}. \citet{how90faintCV3} ;
   \refstepcounter{hvref}\label{count:suuma:bcuma-4}
                         \label{count:suuma:tvcrv-1}
                         \label{count:suuma:swuma-2}
    \arabic{hvref}. \pasjcitetsub{how95swumabcumatvcrv}{b} ;
   \refstepcounter{hvref}\label{count:suuma:bcuma-5}
    \arabic{hvref}. \citet{waa00bcumaiauc} ;
   \refstepcounter{hvref}\label{count:suuma:bcuma-6}
    \arabic{hvref}. Uemura et al. in preparation; see also
      $\langle$
      http://www.kusastro.kyoto-u.ac.jp/vsnet/DNe/bcuma.html
      $\rangle$ for the past superhump detection by M. Iida ;
   \refstepcounter{hvref}\label{count:suuma:v1251cyg-1}
    \arabic{hvref}. \citet{web66v1251cyg} ;
   \refstepcounter{hvref}\label{count:suuma:v1251cyg-2}
    \arabic{hvref}. \citet{BruchCVatlas} ;
   \refstepcounter{hvref}\label{count:suuma:v1251cyg-3}
    \arabic{hvref}. \citet{mor91v1251cygiauc} ;
   \refstepcounter{hvref}\label{count:suuma:v1251cyg-4}
    \arabic{hvref}. \pasjcitetsub{kat91v1251cygiauc}{b} ;
   \refstepcounter{hvref}\label{count:suuma:v1251cyg-5}
    \arabic{hvref}. \pasjcitetsub{wen91v1251cyg}{b} ;
   \refstepcounter{hvref}\label{count:suuma:v1251cyg-6}
    \arabic{hvref}. \pasjcitetsub{kat95v1251cyg}{a} ;
   \refstepcounter{hvref}\label{count:suuma:v1028cyg-1}
    \arabic{hvref}. \pasjcitetsub{hof63v1028cyg}{a} ;
   \refstepcounter{hvref}\label{count:suuma:v1028cyg-2}
    \arabic{hvref}. \pasjcitetsub{hof63an287169}{b} ;
   \refstepcounter{hvref}\label{count:suuma:v1028cyg-3}
    \arabic{hvref}. \citet{tch63v1028cyg} ;
   \refstepcounter{hvref}\label{count:suuma:v1028cyg-4}
    \arabic{hvref}. \citet{may68v1028cyg} ;
   \refstepcounter{hvref}\label{count:suuma:v1028cyg-5}
    \arabic{hvref}. \citet{may70v1028cyg} ;
   \refstepcounter{hvref}\label{count:suuma:v1028cyg-6}
    \arabic{hvref}. \citet{bru92CVspec2} ;
   \refstepcounter{hvref}\label{count:suuma:v1028cyg-7}
    \arabic{hvref}. \citet{mat95v1028cyg} ;
   \refstepcounter{hvref}\label{count:suuma:v1028cyg-8}
    \arabic{hvref}. \citet{bab00v1028cyg} ;
   \refstepcounter{hvref}\label{count:suuma:uvper-1}
    \arabic{hvref}. \citet{szk85CVmultiwavelength} ;
   \refstepcounter{hvref}\label{count:suuma:uvper-2}
    \arabic{hvref}. \citet{szk87shortPCV} ;
   \refstepcounter{hvref}\label{count:suuma:uvper-3}
    \arabic{hvref}. \citet{kat90uvper} ;
   \refstepcounter{hvref}\label{count:suuma:uvper-4}
    \arabic{hvref}. \citet{mob97uwper} ;
   \refstepcounter{hvref}\label{count:suuma:uvper-5}
    \arabic{hvref}. \citet{tho97uvpervyaqrv1504cyg} ;
   \refstepcounter{hvref}\label{count:suuma:cpdra-2}
                         \label{count:suuma:tleo-1}
    \arabic{hvref}. \citet{how90highgalCV} ;
   \refstepcounter{hvref}\label{count:suuma:cpdra-3}
    \arabic{hvref}. \citet{rob14cpdra} ;
   \refstepcounter{hvref}\label{count:suuma:cpdra-4}
    \arabic{hvref}. \citet{baa38cpdra} ;
   \refstepcounter{hvref}\label{count:suuma:cpdra-5}
    \arabic{hvref}. \citet{alt72cpdraiauc} ;
   \refstepcounter{hvref}\label{count:suuma:cpdra-6}
    \arabic{hvref}. \citet{kho72cpdra} ;
   \refstepcounter{hvref}\label{count:suuma:cpdra-7}
    \arabic{hvref}. \citet{kol79cpdraciuma} ;
   \refstepcounter{hvref}\label{count:suuma:cpdra-8}
    \arabic{hvref}. Ishioka et al. in preparation ;
   \refstepcounter{hvref}\label{count:suuma:dvuma-2}
    \arabic{hvref}. \citet{ush82dvuma} ;
   \refstepcounter{hvref}\label{count:suuma:dvuma-3}
    \arabic{hvref}. \citet{ush83dvuma} ;
   \refstepcounter{hvref}\label{count:suuma:dvuma-4}
    \arabic{hvref}. \citet{how87dvuma} ;
   \refstepcounter{hvref}\label{count:suuma:dvuma-5}
    \arabic{hvref}. \citet{how88dvuma} ;
   \refstepcounter{hvref}\label{count:suuma:dvuma-6}
    \arabic{hvref}. \citet{how93arcncaypscdvuma} ;
   \refstepcounter{hvref}\label{count:suuma:dvuma-7}
    \arabic{hvref}. \citet{szk93dvumaaypscv503cyg} ;
   \refstepcounter{hvref}\label{count:suuma:dvuma-8}
    \arabic{hvref}. \citet{pat00dvuma} ;
   \refstepcounter{hvref}\label{count:suuma:dvuma-9}
    \arabic{hvref}. \citet{nog01dvuma} ;
   \refstepcounter{hvref}\label{count:suuma:dvuma-10}
    \arabic{hvref}. Uemura et al. in preparation ;
     See also $\langle$
     http://www.kusastro.kyoto-u.ac.jp/vsnet/DNe/dvuma9912.html
     $\rangle$ ;
   \refstepcounter{hvref}\label{count:suuma:efpeg-1}
    \arabic{hvref}. \citet{esc37efpeg} ;
   \refstepcounter{hvref}\label{count:suuma:efpeg-2}
    \arabic{hvref}. \citet{san50efpeg} ;
   \refstepcounter{hvref}\label{count:suuma:efpeg-3}
    \arabic{hvref}. \citet{tse79efpeg} ;
   \refstepcounter{hvref}\label{count:suuma:efpeg-4}
    \arabic{hvref}. \citet{ges88efpeg} ;
   \refstepcounter{hvref}\label{count:suuma:efpeg-5}
    \arabic{hvref}. \citet{kat91efpegiauc} ; 
   \refstepcounter{hvref}\label{count:suuma:efpeg-6}
    \arabic{hvref}. \citet{how91efpegiauc} ; 
   \refstepcounter{hvref}\label{count:suuma:efpeg-7}
    \arabic{hvref}. \citet{dey91efpegiauc} ; 
   \refstepcounter{hvref}\label{count:suuma:efpeg-8}
    \arabic{hvref}. \citet{how93efpeg} ;
   \refstepcounter{hvref}\label{count:suuma:efpeg-9}
    \arabic{hvref}. \pasjcitetsub{kat01efpeg}{a} ;
   \refstepcounter{hvref}\label{count:suuma:efpeg-10}
    \arabic{hvref}. Matsumoto et al. in preparation ;
   \refstepcounter{hvref}\label{count:suuma:v725aql-1}
    \arabic{hvref}. \citet{nog95v725aql} ;
   \refstepcounter{hvref}\label{count:suuma:v725aql-2}
    \arabic{hvref}. \citet{haz96v725aql} ;
   \refstepcounter{hvref}\label{count:suuma:v725aql-3}
    \arabic{hvref}. \citet{uem01v725aql} ;
   \refstepcounter{hvref}\label{count:suuma:tvcrv-2}
    \arabic{hvref}. \pasjcitetsub{lev90tvcrv}{a} ;
   \refstepcounter{hvref}\label{count:suuma:tvcrv-3}
    \arabic{hvref}. \pasjcitetsub{lev90tvcrviauc}{b} ;
   \refstepcounter{hvref}\label{count:suuma:tvcrv-4}
    \arabic{hvref}. \citet{hud92tvcrv} ;
   \refstepcounter{hvref}\label{count:suuma:tvcrv-5}
    \arabic{hvref}. \pasjcitetsub{how96tvcrv}{b} ;
   \refstepcounter{hvref}\label{count:suuma:tvcrv-6}
    \arabic{hvref}. \citet{lev00tvcrv} ;
   \refstepcounter{hvref}\label{count:suuma:puper-1}
    \arabic{hvref}. \citet{hof67an29043} ;
   \refstepcounter{hvref}\label{count:suuma:puper-2}
    \arabic{hvref}. \citet{rom76DN} ;
   \refstepcounter{hvref}\label{count:suuma:puper-3}
    \arabic{hvref}. \citet{bus79VS17} ;
   \refstepcounter{hvref}\label{count:suuma:puper-4}
    \arabic{hvref}. \citet{BruchCVatlas} ;
   \refstepcounter{hvref}\label{count:suuma:puper-5}
    \arabic{hvref}. \citet{kat95puper} ;
   \refstepcounter{hvref}\label{count:suuma:puper-6}
    \arabic{hvref}. \citet{kat99puper} ;
   \refstepcounter{hvref}\label{count:suuma:qyper-1}
    \arabic{hvref}. \citet{bus79VS17} ;
   \refstepcounter{hvref}\label{count:suuma:qyper-2}
    \arabic{hvref}. \citet{ros89qyperiauc} ;
   \refstepcounter{hvref}\label{count:suuma:qyper-3}
    \arabic{hvref}. \citet{mat99qyperiauc} ;
   \refstepcounter{hvref}\label{count:suuma:qyper-4}
    \arabic{hvref}. \pasjcitetsub{kat00qyperiauc}{c};
      See also $\langle$
      http://www.kusastro.kyoto-u.ac.jp/vsnet/DNe/qyper.html
      $\rangle$ ;
   \refstepcounter{hvref}\label{count:suuma:mmhya-1}
    \arabic{hvref}. \pasjcitetsub{kat91mmhya}{a} ;
   \refstepcounter{hvref}\label{count:suuma:mmhya-2}
    \arabic{hvref}. \citet{mis95PGCV} ;
   \refstepcounter{hvref}\label{count:suuma:mmhya-3}
    \arabic{hvref}. \citet{pat98evolution} ;
   \refstepcounter{hvref}\label{count:suuma:mmhya-4}
    \arabic{hvref}. Ishioka et al. in preparation ;
   \refstepcounter{hvref}\label{count:suuma:kvand-1}
    \arabic{hvref}. \citet{kur77kvandkwand} ;
   \refstepcounter{hvref}\label{count:suuma:kvand-2}
    \arabic{hvref}. \citet{kat94kvand} ;
   \refstepcounter{hvref}\label{count:suuma:kvand-3}
    \arabic{hvref}. \pasjcitetsub{kat95kvand}{b} ;
   \refstepcounter{hvref}\label{count:suuma:swuma-3}
    \arabic{hvref}. \citet{her60swuma} ;
   \refstepcounter{hvref}\label{count:suuma:swuma-4}
    \arabic{hvref}. \citet{sha83swuma} ;
   \refstepcounter{hvref}\label{count:suuma:swuma-5}
    \arabic{hvref}. \citet{sha86swumaXray} ;
   \refstepcounter{hvref}\label{count:suuma:swuma-6}
    \arabic{hvref}. \citet{sha87swuma} ;
   \refstepcounter{hvref}\label{count:suuma:swuma-7}
    \arabic{hvref}. \citet{rob87swumaQPO} ;
   \refstepcounter{hvref}\label{count:suuma:swuma-8}
    \arabic{hvref}. \citet{szk88swumaEXOSATIUE} ;
   \refstepcounter{hvref}\label{count:suuma:swuma-9}
    \arabic{hvref}. \citet{kat92swumasuperQPO} ;
   \refstepcounter{hvref}\label{count:suuma:swuma-10}
    \arabic{hvref}. \citet{ros94v426ophswumav348pup} ;
   \refstepcounter{hvref}\label{count:suuma:swuma-11}
    \arabic{hvref}. \pasjcitetsub{sem97swuma}{b} ;
   \refstepcounter{hvref}\label{count:suuma:swuma-12}
    \arabic{hvref}. \citet{nog98swuma} ;
   \refstepcounter{hvref}\label{count:suuma:swuma-13}
    \arabic{hvref}. \citet{pav00swuma} ;
   \refstepcounter{hvref}\label{count:suuma:swuma-14}
    \arabic{hvref}. Uemura et al. in preparation ;
     See also $\langle$
     http://www.kusastro.kyoto-u.ac.jp/vsnet/DNe/swuma00.html
     $\rangle$ ;
   \refstepcounter{hvref}\label{count:suuma:cthya-1}
    \arabic{hvref}. \citet{hof36an25937} ;
   \refstepcounter{hvref}\label{count:suuma:cthya-2}
    \arabic{hvref}. \citet{nog96cthya} ;
   \refstepcounter{hvref}\label{count:suuma:cthya-3}
    \arabic{hvref}. \pasjcitetsub{kat99cthya}{a}
   \refstepcounter{hvref}\label{count:suuma:tleo-2}
    \arabic{hvref}. \citet{pet1865tleo} ;
   \refstepcounter{hvref}\label{count:suuma:tleo-3}
    \arabic{hvref}. \citet{kra62ugem} ;
   \refstepcounter{hvref}\label{count:suuma:tleo-4}
    \arabic{hvref}. \citet{rob69tleo} ;
   \refstepcounter{hvref}\label{count:suuma:tleo-5}
    \arabic{hvref}. \citet{oke82CVspec} ;
   \refstepcounter{hvref}\label{count:suuma:tleo-6}
    \arabic{hvref}. \citet{wil83CVspec1} ;
   \refstepcounter{hvref}\label{count:suuma:tleo-7}
    \arabic{hvref}. \pasjcitetsub{wen83tleo}{a} ;
   \refstepcounter{hvref}\label{count:suuma:tleo-8}
    \arabic{hvref}. \citet{sha84tleo} ;
   \refstepcounter{hvref}\label{count:suuma:tleo-9}
    \arabic{hvref}. \citet{wen84tleo} ;
   \refstepcounter{hvref}\label{count:suuma:tleo-10}
    \arabic{hvref}. \citet{kat87tleo} ;
   \refstepcounter{hvref}\label{count:suuma:tleo-11}
    \arabic{hvref}. \citet{slo87tleoiauc} ;
   \refstepcounter{hvref}\label{count:suuma:tleo-12}
    \arabic{hvref}. \citet{szk87shortPCV} ;
   \refstepcounter{hvref}\label{count:suuma:tleo-13}
    \arabic{hvref}. \citet{szk91interoutburst} ;
   \refstepcounter{hvref}\label{count:suuma:tleo-14}
    \arabic{hvref}. \citet{lem93tleo} ;
   \refstepcounter{hvref}\label{count:suuma:tleo-15}
    \arabic{hvref}. \citet{kun94tleo} ;
   \refstepcounter{hvref}\label{count:suuma:tleo-16}
    \arabic{hvref}. \citet{hum96ugemtleo} ;
   \refstepcounter{hvref}\label{count:suuma:tleo-17}
    \arabic{hvref}. \citet{tho96Porb} ;
   \refstepcounter{hvref}\label{count:suuma:tleo-18}
    \arabic{hvref}. \pasjcitetsub{kat97tleo}{a} ;
   \refstepcounter{hvref}\label{count:suuma:tleo-19}
    \arabic{hvref}. \citet{bel98tleoIUE} ;
   \refstepcounter{hvref}\label{count:suuma:tleo-20}
    \arabic{hvref}. \citet{how99tleo} ;
   \refstepcounter{hvref}\label{count:suuma:tleo-21}
    \arabic{hvref}. \citet{dhi00CVIRspec} ;
   \refstepcounter{hvref}\label{count:suuma:tleo-22}
    \arabic{hvref}. \citet{szk01lspegtleoXray} ;
   \refstepcounter{hvref}\label{count:suuma:ektra-1}
    \arabic{hvref}. \citet{vog80ektra} ;
   \refstepcounter{hvref}\label{count:suuma:ektra-2}
    \arabic{hvref}. \citet{has85ektra} ;
   \refstepcounter{hvref}\label{count:suuma:ektra-3}
    \arabic{hvref}. \citet{men98ektra} ;
   \refstepcounter{hvref}\label{count:suuma:ektra-4}
    \arabic{hvref}. \citet{gan97ektra} ;
   \refstepcounter{hvref}\label{count:suuma:ektra-5}
    \arabic{hvref}. \citet{mun98CVspec5} ;
   \refstepcounter{hvref}\label{count:suuma:ektra-6}
    \arabic{hvref}. \citet{gan01ektra} ;
   \refstepcounter{hvref}\label{count:suuma:v844her-1}
    \arabic{hvref}. \citet{ant96newvar} ;
   \refstepcounter{hvref}\label{count:suuma:v844her-2}
    \arabic{hvref}. \citet{kat00v844her} ;
   \refstepcounter{hvref}\label{count:suuma:kvdra-1}
    \arabic{hvref}. \citet{nog00kvdra} ;
   \refstepcounter{hvref}\label{count:suuma:kvdra-2}
    \arabic{hvref}. \pasjcitetsub{kat00kvdraalert4858}{a} ;
   \refstepcounter{hvref}\label{count:suuma:htcam-1}
    \arabic{hvref}. \citet{tov98htcam} ;
   \refstepcounter{hvref}\label{count:suuma:htcam-2}
    \arabic{hvref}. \citet{wat98htcam} ;
   \refstepcounter{hvref}\label{count:suuma:htcam-3}
    \arabic{hvref}. \citet{sch00htcam} ;
   \refstepcounter{hvref}\label{count:suuma:htcam-4}
    \arabic{hvref}. \pasjcitetsub{kat01htcam}{b} ;
   \refstepcounter{hvref}\label{count:suuma:htcas-1}
    \arabic{hvref}. \pasjcitetsub{hof43htcas}{b} ;
   \refstepcounter{hvref}\label{count:suuma:htcas-2}
    \arabic{hvref}. \pasjcitetsub{pat79htcas}{a} ;
   \refstepcounter{hvref}\label{count:suuma:htcas-3}
    \arabic{hvref}. \citet{pat81DNOhtcas} ;
   \refstepcounter{hvref}\label{count:suuma:htcas-4}
    \arabic{hvref}. \citet{you81htcas} ;
   \refstepcounter{hvref}\label{count:suuma:htcas-5}
    \arabic{hvref}. \pasjcitetsub{mat85htcasiauc}{b} ; 
   \refstepcounter{hvref}\label{count:suuma:htcas-6}
    \arabic{hvref}. \citet{waa85htcasiauc} ; 
   \refstepcounter{hvref}\label{count:suuma:htcas-7}
    \arabic{hvref}. \citet{rob85htcas} ;
   \refstepcounter{hvref}\label{count:suuma:htcas-8}
    \arabic{hvref}. \citet{wen85htcas} ;
   \refstepcounter{hvref}\label{count:suuma:htcas-9}
    \arabic{hvref}. \citet{zha86htcas} ;
   \refstepcounter{hvref}\label{count:suuma:htcas-10}
    \arabic{hvref}. \citet{wlo86htcas} ;
   \refstepcounter{hvref}\label{count:suuma:htcas-11}
    \arabic{hvref}. \citet{wen87htcas} ;
   \refstepcounter{hvref}\label{count:suuma:htcas-12}
    \arabic{hvref}. \citet{ber87htcasIR} ;
   \refstepcounter{hvref}\label{count:suuma:htcas-13}
    \arabic{hvref}. \citet{mar90htcas} ;
   \refstepcounter{hvref}\label{count:suuma:htcas-14}
    \arabic{hvref}. \citet{hor91htcas} ;
   \refstepcounter{hvref}\label{count:suuma:htcas-15}
    \arabic{hvref}. \citet{woo92htcas} ;
   \refstepcounter{hvref}\label{count:suuma:htcas-16}
    \arabic{hvref}. \pasjcitetsub{mat95htcasiauc}{b} ;
   \refstepcounter{hvref}\label{count:suuma:htcas-17}
    \arabic{hvref}. \citet{wel95htcas} ;
   \refstepcounter{hvref}\label{count:suuma:htcas-18}
    \arabic{hvref}. \citet{woo95htcasXray} ;
   \refstepcounter{hvref}\label{count:suuma:htcas-19}
    \arabic{hvref}. \citet{rob96htcas} ;
   \refstepcounter{hvref}\label{count:suuma:htcas-20}
    \arabic{hvref}. \citet{wel96flickeringhtcas} ;
   \refstepcounter{hvref}\label{count:suuma:htcas-21}
    \arabic{hvref}. \citet{muk97htcas} ;
   \refstepcounter{hvref}\label{count:suuma:htcas-22}
    \arabic{hvref}. \citet{bab99htcas} ;
   \refstepcounter{hvref}\label{count:suuma:htcas-23}
    \arabic{hvref}. \citet{ioa99htcas} ;
   \refstepcounter{hvref}\label{count:suuma:htcas-24}
    \arabic{hvref}. \citet{cat99htcasqui} ;
   \refstepcounter{hvref}\label{count:suuma:htcas-25}
    \arabic{hvref}. \citet{bru00htcasv2051ophippeguxumaflickering} ;
   \refstepcounter{hvref}\label{count:suuma:ircom-1}
    \arabic{hvref}. \citet{ric95ircom} ;
   \refstepcounter{hvref}\label{count:suuma:ircom-2}
    \arabic{hvref}. \citet{wen95ircom} ;
   \refstepcounter{hvref}\label{count:suuma:ircom-3}
    \arabic{hvref}. \citet{ric96ircomiauc} ;
   \refstepcounter{hvref}\label{count:suuma:ircom-4}
    \arabic{hvref}. \citet{ric97ircom} ;
   \refstepcounter{hvref}\label{count:suuma:ircom-5}
    \arabic{hvref}. \pasjcitetsub{kat01ircom}{a} 
}
\end{table*}

\setcounter{hvref}{0}
\setcounter{hvrem}{0}


\newpage

\begin{thebibliography}{}

\bibitem[Abbott et~al.\labelspace(1997)]{abb97bwscl}
  Abbott, T. M.~C., Fleming, T.~A., \& Pasquini, L.\ 1997, \aap, 318, 134

\bibitem[Abbott et~al.\labelspace(1992)]{abb92alcomcperi}
  Abbott, T. M.~C., Robinson, E.~L., Hill, G.~J., \& Haswell, C.~A.\ 1992,
  \apj, 399, 680

\bibitem[Abrahamian, Mickaelian\labelspace(1996)]{abr96FBS11}
  Abrahamian, H.~V., \& Mickaelian, A.~M.\ 1996, \Ap, 39, 315

\bibitem[Aksenov, Kurochkin\labelspace(1983)]{aks83uwtriiauc}
  Aksenov, E.~P., \& Kurochkin, N.~E.\ 1983, \iaucirc, 3869

\bibitem[Albitzky\labelspace(1929)]{alb29eyaql}
  Albitzky, B.\ 1929, \an, 235, 317

\bibitem[Alksnis, Zacs\labelspace(1981)]{alk80sslmi}
  Alksnis, A., \& Zacs, L.\ 1981, \ibvs, 1972

\bibitem[Alksnis, Zharova\labelspace(2000)]{alk00ptand}
  Alksnis, A., \& Zharova, A.~V.\ 2000, \ibvs, 4909

\bibitem[Altizer\labelspace(1972)]{alt72cpdraiauc}
  Altizer, R.\ 1972, \iaucirc, 2381

\bibitem[Anderson et~al.\labelspace(1967)]{and67ceumaiauc}
  Anderson, J.~H., Luyten, W.~J., \& Sandage, A.~R.\ 1967, \iaucirc, 2011

\bibitem[Antipin\labelspace(1996)]{ant96newvar}
  Antipin, S.~V.\ 1996, \ibvs, 4360

\bibitem[Argyle, Green\labelspace(1983)]{arg83uwtriiauc}
  Argyle, R.~W., \& Green, S.\ 1983, \iaucirc, 3878

\bibitem[Augusteijn\labelspace(1994)]{aug94vyaqr}
  Augusteijn, T.\ 1994, \aap, 292, 481

\bibitem[Augusteijn et~al.\labelspace(1995)]{aug95alcomiauc}
  Augusteijn, T., Schmeer, P., \& Dillon, W.~G.\ 1995, \iaucirc, 6160

\bibitem[Augusteijn, Wisotzki\labelspace(1997)]{aug97bwscl}
  Augusteijn, T., \& Wisotzki, L.\ 1997, \aap, 324, L57

\bibitem[Augusteijn, Wisotzki\labelspace(1998)]{aug98bwscl}
  Augusteijn, T., \& Wisotzki, L.\ 1998, in \ASPConf{137}{Wild Stars in the Old
  West}, ed. S. Howell, E. Kuulkers, \& C. Woodward (San Francisco: ASP),
  p.~434

\bibitem[Baade\labelspace(1938)]{baa38cpdra}
  Baade, W.\ 1938, \apj, 88, 285

\newpage

\bibitem[Baba et~al.\labelspace(1999)]{bab99htcas}
  Baba, H., Kato, T., Nogami, D., \& Hirata, R.\ 1999, in Disk Instabilities in
  Close Binary Systems, ed. S. Mineshige, \& J.~C. Wheeler (Universal Academy
  Press, Tokyo), p.~123

\bibitem[Baba et~al.\labelspace(2000)]{bab00v1028cyg}
  Baba, H., Kato, T., Nogami, D., Hirata, R., Matsumoto, K., \& Sadakane, K.\
  2000, \pasj, 52, 429

\bibitem[Baba et~al.\labelspace(2001)]{bab01wzsgeiauc7678}
  Baba, H., Sadakane, K., Norimoto, Y., Nogami, D., Matsumoto, K., Makita, M.,
  \& T., K.\ 2001, \iaucirc, 7678

\bibitem[Bailey\labelspace(1979)]{bai79wzsge}
  Bailey, J.\ 1979, \mnras, 189, 41P

\bibitem[Barwig et~al.\labelspace(1992)]{bar92hvvir}
  Barwig, H., Mantel, K.~H., \& Ritter, H.\ 1992, \aap, 266, L5

\bibitem[Bateson, Jones\labelspace(1991)]{bat91wxcetiauc}
  Bateson, F.~M., \& Jones, A. F. A.~L.\ 1991, \iaucirc, 5302

\bibitem[Belle et~al.\labelspace(1998)]{bel98tleoIUE}
  Belle, K., Nguyen, Q., Fabian, D., Sion, E.~M., \& Huang, M.\ 1998, \pasp,
  110, 47

\bibitem[Berriman et~al.\labelspace(1987)]{ber87htcasIR}
  Berriman, G., Kenyon, S., \& Boyle, C.\ 1987, \aj, 94, 1291

\bibitem[Bertola\labelspace(1964)]{ber64alcom}
  Bertola, F.\ 1964, Ann. d'Astrop., 27, 298

\bibitem[Beyer\labelspace(1951)]{bey51wzsge}
  Beyer, M.\ 1951, \an, 280, 274

\bibitem[Bianchini et~al.\labelspace(1992)]{bia92CVspec}
  Bianchini, A., Dellavalle, M., Duerbeck, H.~W., \& Orio, M.\ 1992, \Msngr,
  69, 42

\bibitem[Bohusz, Udalski\labelspace(1979)]{boh79wzsge}
  Bohusz, E., \& Udalski, A.\ 1979, \ibvs, 1583

\bibitem[Bronnikova\labelspace(1979)]{bro79wzsgeatsir}
  Bronnikova, N.~M.\ 1979, \ATsir, 1034, 8

\bibitem[Brosch\labelspace(1979)]{bro79wzsge}
  Brosch, N.\ 1979, \ibvs, 1693

\bibitem[Brosch, Leibowitz\labelspace(1978)]{bro78wzsgeiauc3313}
  Brosch, N., \& Leibowitz, E.~M.\ 1978, \iaucirc, 3313

\bibitem[Brosch et~al.\labelspace(1980)]{bro80wzsgespec}
  Brosch, N., Leibowitz, E.~M., \& Mazeh, T.\ 1980, \apjl, 236, L29

\bibitem[Bruch\labelspace(1992)]{bru92hvvir}
  Bruch, A.\ 1992, \ibvs, 3745

\bibitem[Bruch\labelspace(2000)]{bru00htcasv2051ophippeguxumaflickering}
  Bruch, A.\ 2000, \aap, 359, 998

\bibitem[Bruch et~al.\labelspace(1987)]{BruchCVatlas}
  Bruch, A., Fischer, F., \& Wilmsen, U.\ 1987, \aaps, 70, 481

\bibitem[Bruch, Schimpke\labelspace(1992)]{bru92CVspec2}
  Bruch, A., \& Schimpke, T.\ 1992, \aaps, 93, 419

\bibitem[Burbidge, Strittmatter\labelspace(1971)]{bur71gpcom}
  Burbidge, E.~M., \& Strittmatter, P.~A.\ 1971, \apjl, 170, L39

\bibitem[Busch et~al.\labelspace(1979)]{bus79VS17}
  Busch, H., H\"{a}ussler, K., \& Splittgerber, E.\ 1979, \VeSon, 9, 125

\bibitem[Campbell\labelspace(1947)]{cam47wzsge}
  Campbell, L.\ 1947, Pop. Astron., 55, 220

\bibitem[Cannizzo\labelspace(1998)]{can98DNabsmag}
  Cannizzo, J.~K.\ 1998, \apj, 493, 426

\bibitem[Capaccioli et~al.\labelspace(1990a)]{cap90comanovaabsmag}
  Capaccioli, M., Cappellaro, E., della Valle, M., D'Onofrio, M., Rosino, L.,
  \& Turatto, M.\ 1990a, \apj, 350, 110

\bibitem[Capaccioli et~al.\labelspace(1990b)]{cap90LMCnovaabsmag}
  Capaccioli, M., della Valle, M., D'Onofrio, M., \& Rosino, L.\ 1990b, \apj,
  360, 63

\bibitem[Catal\'{a}n, Cannon~Smith\labelspace(1999)]{cat99htcasqui}
  Catal\'{a}n, M.~S., \& Cannon~Smith, R.\ 1999, in Disk Instabilities in Close
  Binary Systems, ed. S. Mineshige, \& J.~C. Wheeler (Universal Academy Press,
  Tokyo), p.~143

\bibitem[Cohen\labelspace(1985)]{coh85novaabsmag}
  Cohen, J.~G.\ 1985, apj, 292, 90

\bibitem[Crampton et~al.\labelspace(1979)]{cra79wzsgespec}
  Crampton, D., Hutchings, J.~B., \& Cowley, A.~P.\ 1979, \apj, 234, 182

\bibitem[Cristiani et~al.\labelspace(1985)]{cri85rzleoiauc}
  Cristiani, H., Duerbeck, H.~W., \& Seitter, W.~C.\ 1985, \iaucirc, 4027

\bibitem[Dahlem, Kreysing\labelspace(1994)]{dah94j1255iauc}
  Dahlem, M., \& Kreysing, H.\ 1994, \iaucirc, 6085

\bibitem[Dahlem et~al.\labelspace(1995)]{dah95j1255}
  Dahlem, M., Kreysing, H., White, S.~M., Engels, D., Condon, J.~J., Harmon,
  B.~A., Zhang, S.~N., Kouveliotou, C., Paciesas, W.~S., \& Voges, W.\ 1995,
  \aap, 295, L13

\bibitem[de~Vaucouleurs\labelspace(1978)]{devac78distancescale}
  de Vaucouleurs, G.\ 1978, \apj, 223, 730

\bibitem[della Valle, Augusteijn\labelspace(1990)]{dellaval90vyaqr}
  della Valle, M., \& Augusteijn, T.\ 1990, \Msngr, 61, 41

\bibitem[della Valle, Livio\labelspace(1995)]{dellaval95novaabsmag}
  della Valle, M., \& Livio, M.\ 1995, \apj, 452, 704

\bibitem[d'Esterre\labelspace(1912)]{dest12uwper}
  d'Esterre, C.~R.\ 1912, \an, 191, 63

\bibitem[Deyoung et~al.\labelspace(1991)]{dey91efpegiauc}
  Deyoung, J., Schmidt, R., \& Schmeer, P.\ 1991, \iaucirc, 5378

\bibitem[Dhillon et~al.\labelspace(2000)]{dhi00CVIRspec}
  Dhillon, V.~S., Littlefair, S.~P., Howell, S.~B., Ciardi, D.~R.,
  Harrop-Allin, M.~K., \& Marsh, T.~R.\ 2000, \mnras, 314, 826

\bibitem[Dorschner, Friedemann\labelspace(1968)]{dor68v632her}
  Dorschner, J., \& Friedemann, C.\ 1968, \an, 291, 7

\bibitem[Dorschner et~al.\labelspace(1967)]{dor67v632her}
  Dorschner, J., Friedemann, C., \& Pfau, W.\ 1967, \ATsir, 430, 1

\bibitem[Downes et~al.\labelspace(1995)]{dow95CVspec}
  Downes, R., Hoard, D.~W., Szkody, P., \& Wachter, S.\ 1995, \aj, 110, 1824

\bibitem[Downes et~al.\labelspace(1997)]{DownesCVatlas2}
  Downes, R., Webbink, R.~F., \& Shara, M.~M.\ 1997, \pasp, 109, 345

\bibitem[Downes\labelspace(1990)]{dow90wxcet}
  Downes, R.~A.\ 1990, \aj, 99, 339

\bibitem[Downes, Margon\labelspace(1981)]{dow81wzsge}
  Downes, R.~A., \& Margon, B.\ 1981, \mnras, 197, 35P

\bibitem[Drake et~al.\labelspace(1998)]{dra98j1255EUVE}
  Drake, J.~J., Fruscione, A., Hoare, M.~G., \& Callanan, P.\ 1998, \apj, 493,
  926

\bibitem[Duerbeck\labelspace(1984a)]{due84eyaqlbccasmtcenv745sco}
  Duerbeck, H.~W.\ 1984a, \ibvs, 2490

\bibitem[Duerbeck\labelspace(1984b)]{due84hvvir}
  Duerbeck, H.~W.\ 1984b, \ibvs, 2502

\bibitem[Duerbeck\labelspace(1987)]{due87novaatlas}
  Duerbeck, H.~W.\ 1987, \ssr, 45, 1

\bibitem[Duerbeck, Mennickent\labelspace(1998)]{due98v592her}
  Duerbeck, H.~W., \& Mennickent, R.~E.\ 1998, \ibvs, 4637

\bibitem[Duerbeck et~al.\labelspace(1999)]{due99cgcma}
  Duerbeck, H.~W., Schmeer, P., Knapen, J.~H., \& Pollacco, D.\ 1999, \ibvs,
  4759

\bibitem[Duerbeck, Seitter\labelspace(1987)]{due87gwlib}
  Duerbeck, H.~W., \& Seitter, W.~C.\ 1987, \apss, 131, 467

\bibitem[Esch\labelspace(1937)]{esc37efpeg}
  Esch, M.\ 1937, \an, 264, 305

\bibitem[Eskioglu\labelspace(1963)]{esk63wzsge}
  Eskioglu, A.~N.\ 1963, \aap, 26, 331

\bibitem[Fabian et~al.\labelspace(1978)]{fab78wzsgeparameter}
  Fabian, A.~C., Lin, D. N.~C., Papaloizou, J., Pringle, J.~E., \& Whelan, J.
  A.~J.\ 1978, \mnras, 184, 835

\bibitem[Fabian et~al.\labelspace(1980)]{fab80wzsgeUV}
  Fabian, A.~C., Pringle, J.~E., Whelan, J. A.~J., \& Stickland, D.~J.\ 1980,
  \mnras, 191, 457

\bibitem[Ferwerda\labelspace(1935)]{fer35v522sgr}
  Ferwerda, J.~G.\ 1935, \bain, 7, 273

\bibitem[Friedjung\labelspace(1981)]{fri81wzsge}
  Friedjung, M.\ 1981, \aap, 99, 226

\bibitem[Gammie, Menou\labelspace(1998)]{gam98}
  Gammie, C.~F., \& Menou, K.\ 1998, \apjl, 492, L75

\bibitem[Gaposhkin\labelspace(1976)]{gap76wxcet}
  Gaposhkin, S.~I.\ 1976, \ibvs, 1204

\bibitem[Gessner\labelspace(1988)]{ges88efpeg}
  Gessner, H.\ 1988, \ibvs, 3209

\bibitem[Gill, O'Brien\labelspace(1998)]{gil98v359cen}
  Gill, C.~D., \& O'Brien, T.~J.\ 1998, \mnras, 300, 221

\bibitem[Gilliland, Kemper\labelspace(1980)]{gil80wzsgeSH}
  Gilliland, R.~L., \& Kemper, E.\ 1980, \apj, 236, 854

\bibitem[Gilliland et~al.\labelspace(1986)]{gil86wzsge}
  Gilliland, R.~L., Kemper, E., \& Suntzeff, N.\ 1986, \apj, 301, 252

\bibitem[Gilmore\labelspace(1991)]{gil91wxcetiauc}
  Gilmore, A.~C.\ 1991, \iaucirc, 5308

\bibitem[Glasby\labelspace(1970)]{GlasbyDNbook}
  Glasby, J.~S.\ 1970, The Dwarf Novae (Constable, London)

\bibitem[Graham, Araya\labelspace(1971)]{gra71LMCnova}
  Graham, J.~A., \& Araya, G.\ 1971, \aj, 76, 768

\bibitem[Greaves\labelspace(2000)]{gre00j0643id210}
  Greaves, J.\ 2000, \vsnetid{210}, \\
  http://www.kusastro.kyoto-u.ac.jp/vsnet/Mail/\\
  vsnet-id/msg00210.html

\bibitem[Greenstein\labelspace(1957)]{gre57wzsge}
  Greenstein, J.~L.\ 1957, \apj, 126, 23

\bibitem[Greenstein, Giclas\labelspace(1978)]{gre78gd552}
  Greenstein, J.~L., \& Giclas, H.\ 1978, \pasp, 90, 460

\bibitem[Grubissich, Rosino\labelspace(1959)]{gru59ptand}
  Grubissich, C., \& Rosino, L.\ 1959, Asiago Contr., 93

\bibitem[Guinan, McCook\labelspace(1979)]{gui79wzsgeiauc3319}
  Guinan, E.~F., \& McCook, G.~P.\ 1979, \iaucirc, 3319

\bibitem[G\"{a}nsicke et~al.\labelspace(1997)]{gan97ektra}
  G\"{a}nsicke, B.~T., Beuermann, K., \& Thomas, H.\ 1997, \mnras, 289,
  388

\bibitem[G\"{a}nsicke et~al.\labelspace(2001)]{gan01ektra}
  G\"{a}nsicke, B.~T., Szkody, P., Sion, E.~M., Hoard, D.~W., Howell, S.,
  Cheng, F.~H., \& Hubeny, I.\ 2001, \aap, 374, 656

\bibitem[Hachisu, Kato\labelspace(1999)]{hac99tcrb}
  Hachisu, I., \& Kato, M.\ 1999, \apjl, 517, L47

\bibitem[Hachisu, Kato\labelspace(2000a)]{hac00rsoph}
  Hachisu, I., \& Kato, M.\ 2000a, \apjl, 536, L93

\bibitem[Hachisu, Kato\labelspace(2000b)]{hac00v394cra}
  Hachisu, I., \& Kato, M.\ 2000b, \apj, 540, 447

\bibitem[Hachisu, Kato\labelspace(2001a)]{hac01ciaql}
  Hachisu, I., \& Kato, M.\ 2001a, \apjl, 553, L161

\bibitem[Hachisu, Kato\labelspace(2001b)]{hac01RN}
  Hachisu, I., \& Kato, M.\ 2001b, \apj, in press (astro-ph/0104040)

\bibitem[Hachisu et~al.\labelspace(2000)]{hac00uscoburst}
  Hachisu, I., Kato, M., Kato, T., \& Matsumoto, K.\ 2000, \apjl, 528, L97

\bibitem[Haefner et~al.\labelspace(1979)]{hae79lateSH}
  Haefner, R., Schoembs, R., \& Vogt, N.\ 1979, \aap, 77, 7

\bibitem[Harrison, McNaught\labelspace(1990)]{har90v4338sgriauc}
  Harrison, T., \& McNaught, R.~H.\ 1990, \iaucirc, 5006

\bibitem[Harrison\labelspace(1991)]{har91sslmiiauc}
  Harrison, T.~E.\ 1991, \iaucirc, 5233

\bibitem[Harrison, Gehrz\labelspace(1992)]{har92CVIRAS3}
  Harrison, T.~E., \& Gehrz, R.~D.\ 1992, \aj, 103, 243

\bibitem[Harvey, Patterson\labelspace(1995)]{har95cyuma}
  Harvey, D.~A., \& Patterson, J.\ 1995, \pasp, 107, 1055

\bibitem[Hassall\labelspace(1985)]{has85ektra}
  Hassall, B. J.~M.\ 1985, \mnras, 216, 335

\bibitem[Hazen\labelspace(1996)]{haz96v725aql}
  Hazen, M.~L.\ 1996, \JAVSO, 24, 14

\bibitem[Heiser, Henry\labelspace(1979)]{hei79wzsge}
  Heiser, A.~M., \& Henry, G.~W.\ 1979, \ibvs, 1559

\bibitem[Henden et~al.\labelspace(2001)]{hen01xypsc}
  Henden, A.~A., Munari, U., \& Sumner, B.\ 2001, \ibvs, 5140

\bibitem[Hendry\labelspace(1983)]{hen83vyaqr}
  Hendry, E.~M.\ 1983, \ibvs, 2381

\bibitem[Herbig\labelspace(1958)]{her58VSchart}
  Herbig, G.~H.\ 1958, \pasp, 70, 605

\bibitem[Herbig\labelspace(1960)]{her60swuma}
  Herbig, G.~H.\ 1960, \apj, 131, 632

\bibitem[Hessman, Hopp\labelspace(1990)]{hes90gd552}
  Hessman, F.~V., \& Hopp, U.\ 1990, \aap, 228, 387

\bibitem[Hessman et~al.\labelspace(1992)]{hes92lateSH}
  Hessman, F.~V., Mantel, K.-H., Barwig, H., \& Schoembs, R.\ 1992, \aap, 263,
  147

\bibitem[Himpel\labelspace(1946)]{him46wzsge}
  Himpel, K.\ 1946, \iaucirc, 1054

\bibitem[Hirose, Osaki\labelspace(1990)]{hir90SHexcess}
  Hirose, M., \& Osaki, Y.\ 1990, \pasj, 42, 135

\bibitem[Hoffmeister\labelspace(1936)]{hof36an25937}
  Hoffmeister, C.\ 1936, \an, 259, 37

\bibitem[Hoffmeister\labelspace(1943a)]{hof43cigem}
  Hoffmeister, C.\ 1943a, \an, 274, 37

\bibitem[Hoffmeister\labelspace(1943b)]{hof43htcas}
  Hoffmeister, C.\ 1943b, Publ. Berlin-Babelsburg Univ. Obs., 28

\bibitem[Hoffmeister\labelspace(1947)]{hof47cigem}
  Hoffmeister, C.\ 1947, \VeSon, 1, 107

\bibitem[Hoffmeister\labelspace(1963a)]{hof63v1028cyg}
  Hoffmeister, C.\ 1963a, \ibvs, 24

\bibitem[Hoffmeister\labelspace(1963b)]{hof63an287169}
  Hoffmeister, C.\ 1963b, \an, 287, 169

\bibitem[Hoffmeister\labelspace(1967)]{hof67an29043}
  Hoffmeister, C.\ 1967, \an, 290, 43

\bibitem[Horne et~al.\labelspace(1991)]{hor91htcas}
  Horne, K., Wood, J.~H., \& Stiening, R.~F.\ 1991, \apj, 378, 271

\bibitem[Howarth\labelspace(1981)]{how81wzsge}
  Howarth, I.~D.\ 1981, \ibvs, 1925

\bibitem[Howell, Szkody\labelspace(1988)]{how88faintCV1}
  Howell, S., \& Szkody, P.\ 1988, \pasp, 100, 224

\bibitem[Howell et~al.\labelspace(1992)]{how92hvviriauc}
  Howell, S., Wagner, R.~M., Bertram, R., \& Szkody, P.\ 1992, \iaucirc, 5505

\bibitem[Howell\labelspace(1991)]{how91CVamateur}
  Howell, S.~B.\ 1991, \JAVSO, 20, 159

\bibitem[Howell, Blanton\labelspace(1993)]{how93arcncaypscdvuma}
  Howell, S.~B., \& Blanton, S.~A.\ 1993, \aj, 106, 311

\bibitem[Howell, Ciardi\labelspace(2001)]{how01llandeferi}
  Howell, S.~B., \& Ciardi, D.~R.\ 2001, \apjl, 550, 57

\bibitem[Howell et~al.\labelspace(1999)]{how99tleo}
  Howell, S.~B., Ciardi, D.~R., Szkody, P., van Paradijs, J., Kuulkers, E.,
  Cash, J., Sirk, M., \& Long, K.~S.\ 1999, \pasp, 111, 342

\bibitem[Howell et~al.\labelspace(1996a)]{how96alcom}
  Howell, S.~B., DeYoung, J.~A., Mattei, J.~A., Foster, G., Szkody, P.,
  Cannizzo, J.~K., Walker, G., \& Fierce, E.\ 1996a, \aj, 111, 2367

\bibitem[Howell et~al.\labelspace(1991)]{how91efpegiauc}
  Howell, S.~B., Fried, R., \& Schmeer, P.\ 1991, \iaucirc, 5372

\bibitem[Howell et~al.\labelspace(1998)]{how98alcom}
  Howell, S.~B., Hauschildt, P., \& Dhillon, V.~S.\ 1998, \apjl, 494, L223

\bibitem[Howell, Hurst\labelspace(1994)]{how94lland}
  Howell, S.~B., \& Hurst, G.~M.\ 1994, \ibvs, 4043

\bibitem[Howell, Hurst\labelspace(1996)]{how96lland}
  Howell, S.~B., \& Hurst, G.~M.\ 1996, \JBAA, 106, 29

\bibitem[Howell, Kreidl\labelspace(1991)]{how91sslmiiauc}
  Howell, S.~B., \& Kreidl, T.~J.\ 1991, \iaucirc, 5235

\bibitem[Howell et~al.\labelspace(1987)]{how87dvuma}
  Howell, S.~B., Mitchell, K.~J., \& Warnock, A.~I.\ 1987, \pasp, 99, 126

\bibitem[Howell et~al.\labelspace(1996b)]{how96tvcrv}
  Howell, S.~B., Reyes, A.~L., Ashley, R., Harrop-Allin, M.~K., \& Warner, B.\
  1996b, \mnras, 282, 623

\bibitem[Howell et~al.\labelspace(1993)]{how93efpeg}
  Howell, S.~B., Schmidt, R., Deyoung, J.~A., Fried, R., Schmeer, P., \& Gritz,
  L.\ 1993, \pasp, 105, 579

\bibitem[Howell, Szkody\labelspace(1990)]{how90highgalCV}
  Howell, S.~B., \& Szkody, P.\ 1990, \apj, 356, 623

\bibitem[Howell, Szkody\labelspace(1991)]{how91alcom}
  Howell, S.~B., \& Szkody, P.\ 1991, \ibvs, 3653

\bibitem[Howell, Szkody\labelspace(1995)]{how95TOADproc}
  Howell, S.~B., \& Szkody, P.\ 1995, in Cataclysmic Variables, ed. A.
  Bianchini, M. della Valle, \& M. Orio (Kluwer Academic Publishers,
  Dordrecht), p.~335

\bibitem[Howell et~al.\labelspace(1995a)]{how95TOAD}
  Howell, S.~B., Szkody, P., \& Cannizzo, J.~K.\ 1995a, \apj, 439, 337

\bibitem[Howell et~al.\labelspace(1990)]{how90faintCV3}
  Howell, S.~B., Szkody, P., Kreidl, T.~J., Mason, K.~O., \& Puchnarewicz,
  E.~M.\ 1990, \pasp, 102, 758

\bibitem[Howell et~al.\labelspace(1995b)]{how95swumabcumatvcrv}
  Howell, S.~B., Szkody, P., Sonneborn, G., Fried, R., Mattei, J., Oliversen,
  R.~J., Ingram, D., \& Hurst, G.~M.\ 1995b, \apj, 453, 454

\bibitem[Howell et~al.\labelspace(1988)]{how88dvuma}
  Howell, S.~B., Warnock, A., Mason, K.~O., Reichert, G.~A., \& Kreidl, T.~J.\
  1988, \mnras, 233, 79

\bibitem[Hu et~al.\labelspace(1997)]{hu97v2176cygiauc}
  Hu, J.-Y., Qiu, Y.-L., Li, W.-D., Wei, J.-Y., \& Esamdin, A.\ 1997, \iaucirc,
  6731

\bibitem[Hudec\labelspace(1992)]{hud92tvcrv}
  Hudec, R.\ 1992, \ibvs, 3706

\bibitem[Humason\labelspace(1938)]{hum38}
  Humason, M.\ 1938, \apj, 88, 228

\bibitem[Hummel et~al.\labelspace(1996)]{hum96ugemtleo}
  Hummel, W., Horne, K., Marsh, T.~R., \& Wood, J.~H.\ 1996, in
  \IAUColloq{158}{Cataclysmic Variables and Related Objects}, ed. A. Evans, \&
  J.~H. Wood (Kluwer Academic Publishers, Dordrecht), p.~87

\bibitem[Hurst et~al.\labelspace(1987a)]{hur87vyaqriauc}
  Hurst, G.~M., Isles, J., \& McNaught, R.~H.\ 1987a, \iaucirc, 4413

\bibitem[Hurst et~al.\labelspace(1987b)]{hur87rzleoiauc}
  Hurst, G.~M., Lubbock, S., \& McNaught, R.~H.\ 1987b, \iaucirc, 4504

\bibitem[Hurst et~al.\labelspace(1988a)]{hur88pqandiauc1}
  Hurst, G.~M., McAdam, D., Mobberley, M., \& James, N.\ 1988a, \iaucirc, 4570

\bibitem[Hurst, Young\labelspace(1988)]{hur88pqandiauc3}
  Hurst, G.~M., \& Young, A.\ 1988, \iaucirc, 4579

\bibitem[Hurst et~al.\labelspace(1988b)]{hur88pqandiauc2}
  Hurst, G.~M., Young, A., Manning, B., Mobberley, M., Oates, M., Boattini, A.,
  \& Scovil, C.\ 1988b, \iaucirc, 4577

\bibitem[Huruhata\labelspace(1983)]{hur83egcnc}
  Huruhata, M.\ 1983, \ibvs, 2401

\bibitem[Huruhata\labelspace(1984)]{hur84vyaqr}
  Huruhata, M.\ 1984, \ibvs, 2470

\bibitem[Iida\labelspace(1990)]{iid90nsv15272}
  Iida, M.\ 1990, \VSOLJBul, 11, 42

\bibitem[Iida et~al.\labelspace(1995)]{iid95dvdra}
  Iida, M., Kato, T., Nogami, D., \& Baba, H.\ 1995, \ibvs, 4279

\bibitem[Ingram, Szkody\labelspace(1992)]{ing92hvvir}
  Ingram, D., \& Szkody, P.\ 1992, \ibvs, 3810

\bibitem[Ioannou et~al.\labelspace(1999)]{ioa99htcas}
  Ioannou, Z., Naylor, T., Welsh, W.~F., Catal\'{a}n, M.~S., Worraker, W.~J.,
  \& James, N.~D.\ 1999, \mnras, 310, 398

\bibitem[Ishioka et~al.\labelspace(2001a)]{ish01rzleo}
  Ishioka, R., Kato, T., Uemura, M., Iwamatsu, H., Matsumoto, K., Stubbings,
  R., Mennickent, R., Billings, G.~W., Kiyota, S., Masi, G., Pietz, J.,
  Nov\'{a}k, R., Martin, B., Oksanen, A., Moilanen, M., Torii, K., Kinugasa,
  K., \& Kawakita, H.\ 2001a, \pasj, in press

\bibitem[Ishioka et~al.\labelspace(2000)]{ish00rzleoiauc}
  Ishioka, R., Uemura, M., Kato, T., Iwamatsu, H., Matsumoto, K., Billings, G.,
  Masi, G., \& Kiyota, S.\ 2000, \iaucirc, 7552

\bibitem[Ishioka et~al.\labelspace(2001b)]{ish01wzsgeiauc7669}
  Ishioka, R., Uemura, M., Matsumoto, K., Kato, T., Ayani, K., Yamaoka, H.,
  Ohshima, T., Maehara, H., \& Watanabe, T.\ 2001b, \iaucirc, 7669

\bibitem[Kato, Fujino\labelspace(1987)]{kat87tleo}
  Kato, M., \& Fujino, S.\ 1987, \VSOLJBul, 3, 10

\bibitem[Kato\labelspace(1990)]{kat90uvper}
  Kato, T.\ 1990, \ibvs, 3522

\bibitem[Kato\labelspace(1991a)]{kat91mmhya}
  Kato, T.\ 1991a, \VSOLJBul, 13, 51

\bibitem[Kato\labelspace(1991b)]{kat91v1251cygiauc}
  Kato, T.\ 1991b, \iaucirc, 5379

\bibitem[Kato\labelspace(1993)]{kat93v344lyr}
  Kato, T.\ 1993, \pasj, 45, L67

\bibitem[Kato\labelspace(1995a)]{kat95v1251cyg}
  Kato, T.\ 1995a, \ibvs, 4152

\bibitem[Kato\labelspace(1995b)]{kat95kvand}
  Kato, T.\ 1995b, \ibvs, 4239

\bibitem[Kato\labelspace(1995c)]{kat95wxcet}
  Kato, T.\ 1995c, \ibvs, 4256

\bibitem[Kato\labelspace(1996)]{kat96rzsge}
  Kato, T.\ 1996, \ibvs, 4369

\bibitem[Kato\labelspace(1997a)]{kat97tleo}
  Kato, T.\ 1997a, \pasj, 49, 583

\bibitem[Kato\labelspace(1997b)]{kat97v2176cyg}
  Kato, T.\ 1997b, \vsnetalert{1195}, \\
  http://www.kusastro.kyoto-u.ac.jp/vsnet/Mail/\\
  alert1000/msg00195.html

\bibitem[Kato\labelspace(1997c)]{kat97FBS}
  Kato, T.\ 1997c, \vsnetchat{633}, \\
  http://www.kusastro.kyoto-u.ac.jp/vsnet/Mail/\\
  vsnet-chat/msg00633.html

\bibitem[Kato\labelspace(1998)]{kat98v592her}
  Kato, T.\ 1998, \vsnetalert{2137}, \\
  http://www.kusastro.kyoto-u.ac.jp/vsnet/Mail/\\
  alert2000/msg00137.html

\bibitem[Kato\labelspace(2001a)]{kat01efpeg}
  Kato, T.\ 2001a, \pasj, submitted

\bibitem[Kato\labelspace(2001b)]{kat01htcam}
  Kato, T.\ 2001b, \vsnetalert{6066}, \\
  http://www.kusastro.kyoto-u.ac.jp/vsnet/Mail/\\
  alert6000/msg00066.html

\bibitem[Kato\labelspace(2001c)]{kat01wzsgealert6285}
  Kato, T.\ 2001c, \vsnetalert{6285}, \\
  http://www.kusastro.kyoto-u.ac.jp/vsnet/Mail/\\
  alert6000/msg00285.html

\bibitem[Kato\labelspace(2001d)]{kat01wzsgealert6345}
  Kato, T.\ 2001d, \vsnetalert{6345}, \\
  http://www.kusastro.kyoto-u.ac.jp/vsnet/Mail/\\
  alert6000/msg00345.html

\bibitem[Kato\labelspace(2001e)]{kat01wzsgealert6432}
  Kato, T.\ 2001e, \vsnetalert{6432}, \\
  http://www.kusastro.kyoto-u.ac.jp/vsnet/Mail/\\
  alert6000/msg00432.html

\bibitem[Kato\labelspace(2001f)]{kat01j1050chat4601}
  Kato, T.\ 2001f, \vsnetchat{4601}, \\
  http://www.kusastro.kyoto-u.ac.jp/vsnet/Mail/\\
  chat6000/msg00601.html

\bibitem[Kato et~al.\labelspace(2001a)]{kat01ircom}
  Kato, T., Baba, H., \& Nogami, D.\ 2001a, \pasj, submitted

\bibitem[Kato, Hirata\labelspace(1990)]{kat90gocom}
  Kato, T., \& Hirata, R.\ 1990, \ibvs, 3489

\bibitem[Kato et~al.\labelspace(1992)]{kat92swumasuperQPO}
  Kato, T., Hirata, R., \& Mineshige, S.\ 1992, \pasj, 44, L215

\bibitem[Kato et~al.\labelspace(2001b)]{kat01wzsgeiauc7672}
  Kato, T., Ishioka, R., Uemura, M., Matsumoto, K., Ohashi, H., Masi, G., Good,
  G.~A., Pietz, J., \& Moilanen, M.\ 2001b, \iaucirc, 7672

\bibitem[Kato et~al.\labelspace(1999a)]{kat99cthya}
  Kato, T., Kiyota, S., Nov\'{a}k, R., \& Matsumoto, K.\ 1999a, \ibvs, 4794

\bibitem[Kato et~al.\labelspace(1994)]{kat94kvand}
  Kato, T., Kunjaya, C., Okyudo, M., \& Takahashi, A.\ 1994, \pasj, 46, L199

\bibitem[Kato, Matsumoto\labelspace(1999)]{kat99puper}
  Kato, T., \& Matsumoto, K.\ 1999, \ibvs, 4765

\bibitem[Kato et~al.\labelspace(2001c)]{kat01wxcet}
  Kato, T., Matsumoto, K., Nogami, D., Morikawa, K., \& Kiyota, S.\ 2001c,
  \pasj, in press

\bibitem[Kato et~al.\labelspace(1999b)]{kat99cgcma}
  Kato, T., Matsumoto, K., \& Stubbings, R.\ 1999b, \ibvs, 4760

\bibitem[Kato, Nogami\labelspace(1995)]{kat95puper}
  Kato, T., \& Nogami, D.\ 1995, \ibvs, 4260

\bibitem[Kato et~al.\labelspace(1995)]{kat95gocom}
  Kato, T., Nogami, D., \& Baba, H.\ 1995, \ibvs, 4228

\bibitem[Kato et~al.\labelspace(1998a)]{kat98super}
  Kato, T., Nogami, D., Baba, H., \& Matsumoto, K.\ 1998a, in
  \ASPConf{137}{Wild Stars in the Old West}, ed. S. Howell, E. Kuulkers, \& C.
  Woodward (San Francisco: ASP), p.~9

\bibitem[Kato et~al.\labelspace(1996a)]{kat96alcomproc}
  Kato, T., Nogami, D., Baba, H., Matsumoto, K., Arimoto, J., \& Tanabe, K.\
  1996a, in \IAUColloq{158}{Cataclysmic Variables and Related Objects}, ed. A.
  Evans, \& J.~H. Wood (Kluwer Academic Publishers, Dordrecht), p.~77

\bibitem[Kato et~al.\labelspace(1996b)]{kat96alcom}
  Kato, T., Nogami, D., Baba, H., Matsumoto, K., Arimoto, J., Tanabe, K., \&
  Ishikawa, K.\ 1996b, \pasj, 48, L21

\bibitem[Kato et~al.\labelspace(2001d)]{kat01uwtri}
  Kato, T., Nogami, D., Lockley, J.~J., \& Somers, M.\ 2001d, \ibvs, 5116

\bibitem[Kato et~al.\labelspace(1998b)]{kat98hsvir}
  Kato, T., Nogami, D., Masuda, S., \& Baba, H.\ 1998b, \pasp, 110, 1400

\bibitem[Kato et~al.\labelspace(1997a)]{kat97egcnc}
  Kato, T., Nogami, D., Matsumoto, K., \& Baba, H.\ 1997a, \\
  ftp://ftp.kusastro.kyoto-u.ac.jp/pub/vsnet/preprints/\\ EG\_Cnc/

\bibitem[Kato et~al.\labelspace(2001e)]{kat01wzsgeiauc7680}
  Kato, T., Ohashi, H., Ishioka, R., Uemura, M., Matsumoto, K., Masi, G.,
  Mallia, F., Starkey, D., Pietz, J., Martin, B., Good, A.~A., Richmond, M.,
  Davis, S., \& Davis, T.\ 2001e, \iaucirc, 7680

\bibitem[Kato et~al.\labelspace(2001f)]{kat01wzsgeiauc7678}
  Kato, T., Ohashi, H., Ishioka, R., Uemura, M., Matsumoto, K., Masi, G.,
  Starkey, D., Pietz, J., \& Martin, B.\ 2001f, \iaucirc, 7678

\bibitem[Kato et~al.\labelspace(2000a)]{kat00kvdraalert4858}
  Kato, T., Oksanen, A., Moilanen, M., \& Kinnunen, T.\ 2000a,
  \vsnetalert{4858}, \\
  http://www.kusastro.kyoto-u.ac.jp/vsnet/Mail/\\
  alert4000/msg00858.html

\bibitem[Kato, Schmeer\labelspace(1999)]{kat99cigem}
  Kato, T., \& Schmeer, P.\ 1999, \ibvs, 4757

\bibitem[Kato et~al.\labelspace(1997b)]{kat97efpegalert1344}
  Kato, T., Schmeer, P., \& Poyner, G.\ 1997b, \vsnetalert{1344}, \\
  http://www.kusastro.kyoto-u.ac.jp/vsnet/Mail/\\
  alert1000/msg00344.html

\bibitem[Kato, Takata\labelspace(1991)]{kat91efpegiauc}
  Kato, T., \& Takata, T.\ 1991, \iaucirc, 5371

\bibitem[Kato, Uemura\labelspace(2000)]{kat00v844her}
  Kato, T., \& Uemura, M.\ 2000, \ibvs, 4902

\bibitem[Kato, Uemura\labelspace(2001)]{kat01uvgemfsandaspsc}
  Kato, T., \& Uemura, M.\ 2001, \ibvs, 5158

\bibitem[Kato et~al.\labelspace(2000b)]{kat00bcumaalert4562}
  Kato, T., Uemura, M., \& Masi, G.\ 2000b, \vsnetalert{4562}, \\
  http://www.kusastro.kyoto-u.ac.jp/vsnet/Mail/\\
  alert4000/msg00562.html

\bibitem[Kato et~al.\labelspace(2000c)]{kat00qyperiauc}
  Kato, T., Uemura, M., Matsumoto, K., Masi, G., Cassetti, A., Cook, L.,
  Jensen, L.~T., Martin, B., Vanmunster, T., \& Buczynski, D.\ 2000c, \iaucirc,
  7343

\bibitem[Kato et~al.\labelspace(2000d)]{kat00bcumaalert4586}
  Kato, T., Uemura, M., Shugarov, S., Pavlenko, E., Schmeer, P., Nov\'{a}k, R.,
  \& Masi, G.\ 2000d, \vsnetalert{4586}, \\
  http://www.kusastro.kyoto-u.ac.jp/vsnet/Mail/\\
  alert4000/msg00586.html

\bibitem[Kholopov\labelspace(1972)]{kho72cpdra}
  Kholopov, P.~N.\ 1972, \ATsir, 700

\bibitem[Kholopov et~al.\labelspace(1985)]{GCVS}
  Kholopov, P.~N., Samus', N.~N., Frolov, M.~S., Goranskij, V.~P., Gorynya,
  N.~A., Kireeva, N.~N., Kukarkina, N.~P., Kurochkin, N.~E., I., M.~G., Perova,
  N.~B., \& Yu., S.~S.\ 1985, General Catalogue of Variable Stars, fourth
  edition (Nauka Publishing House, Moscow)

\bibitem[Kikuchi\labelspace(1988)]{kik88Dodaira}
  Kikuchi, S.\ 1988, Tokyo Astron. Bull., Second Ser., 281, 3267

\bibitem[Kilmartin et~al.\labelspace(1992)]{kil92hvviriauc}
  Kilmartin, P.~M., Gilmore, A.~C., della Valle, M., Duerbeck, H.~W., Motch,
  C., \& Poretti, E.\ 1992, \iaucirc, 5503

\bibitem[Kolotovkina\labelspace(1979)]{kol79cpdraciuma}
  Kolotovkina, S.~A.\ 1979, \PZP, 3, 665

\bibitem[Kraft\labelspace(1962)]{kra62ugem}
  Kraft, R.~P.\ 1962, \apj, 135, 408

\bibitem[Kraft et~al.\labelspace(1962)]{kra62wzsge}
  Kraft, R.~P., Mathews, J., \& Greenstein, J.~L.\ 1962, \apj, 136, 312

\bibitem[Kruszewski et~al.\labelspace(1978)]{kru78wzsgeiauc3312}
  Kruszewski, A., Krzeminski, W., Bohusz, E., \& Udalski, A.\ 1978, \iaucirc,
  3312

\bibitem[Kruszewski et~al.\labelspace(1979)]{kru79wzsgeiauc3318}
  Kruszewski, A., Krzeminski, W., Bohusz, E., \& Udalski, A.\ 1979, \iaucirc,
  3318

\bibitem[Krzeminski\labelspace(1962)]{krz62wzsge}
  Krzeminski, W.\ 1962, \pasp, 74, 66

\bibitem[Krzeminski, Kraft\labelspace(1964)]{krz64wzsge}
  Krzeminski, W., \& Kraft, R.~P.\ 1964, \apj, 140, 921

\bibitem[Kunjaya et~al.\labelspace(1994)]{kun94tleo}
  Kunjaya, C., Kato, T., \& Hirata, R.\ 1994, \ibvs, 4023

\bibitem[Kurochkin\labelspace(1968)]{kur68newvar}
  Kurochkin, N.~E.\ 1968, \PZ, 16, 460

\bibitem[Kurochkin\labelspace(1977)]{kur77kvandkwand}
  Kurochkin, N.~E.\ 1977, \ATsir, 974, 4

\bibitem[Kuulkers\labelspace(2000)]{kuu00wzsgeSXT}
  Kuulkers, E.\ 2000, \NewAR, 44, 27

\bibitem[Kuulkers\labelspace(2001)]{kuu01j1118}
  Kuulkers, E.\ 2001, \an, 322, 9

\bibitem[Kuulkers et~al.\labelspace(1996)]{kuu96TOAD}
  Kuulkers, E., Howell, S.~B., \& van Paradijs, J.\ 1996, \apjl, 462, L87

\bibitem[Lambert, Slovak\labelspace(1981)]{lam81gpcom}
  Lambert, D.~L., \& Slovak, M.~H.\ 1981, \pasp, 93, 477

\bibitem[Lasota et~al.\labelspace(1995)]{las95wzsge}
  Lasota, J.-P., Hameury, J.-M., \& Hur\'{e}, J.~M.\ 1995, \aap, 302, L29

\bibitem[Lasota et~al.\labelspace(1999)]{las99wzsgeIP}
  Lasota, J.-P., Kuulkers, E., \& Charles, P.\ 1999, \mnras, 305, 473

\bibitem[Leibowitz et~al.\labelspace(1987)]{lei87vyaqriauc}
  Leibowitz, E.~M., Laor, A., Mazeh, T., \& Brosch, N.\ 1987, \iaucirc, 4418

\bibitem[Leibowitz, Mazeh\labelspace(1981)]{lei81wzsge}
  Leibowitz, E.~M., \& Mazeh, T.\ 1981, \apj, 251, 214

\bibitem[Leibowitz et~al.\labelspace(1994)]{lei94hvvir}
  Leibowitz, E.~M., Mendelson, H., Bruch, A., Duerbeck, H.~W., Seitter, W.~C.,
  \& Richter, G.~A.\ 1994, \apj, 421, 771

\bibitem[Lemm et~al.\labelspace(1993)]{lem93tleo}
  Lemm, K., Patterson, J., Thomas, G., \& Skillman, D.~R.\ 1993, \pasp, 105,
  1120

\bibitem[Levy\labelspace(2000)]{lev00tvcrv}
  Levy, D.~H.\ 2000, \JAVSO, 28, 38

\bibitem[Levy et~al.\labelspace(1990a)]{lev90tvcrv}
  Levy, D.~H., Howell, S.~B., Kreidl, T.~J., Skiff, B.~A., \& Tombaugh, C.~W.\
  1990a, \pasp, 102, 1321

\bibitem[Levy et~al.\labelspace(1990b)]{lev90tvcrviauc}
  Levy, D.~H., Tombaugh, C.~W., \& Skiff, B.\ 1990b, \iaucirc, 4983

\bibitem[Liller\labelspace(1983)]{lil83vyaqr}
  Liller, M.~H.\ 1983, \ibvs, 2293

\bibitem[Liller\labelspace(1990a)]{lil90v4338sgriauc1}
  Liller, W.\ 1990a, \iaucirc, 4974

\bibitem[Liller\labelspace(1990b)]{lil90v4338sgriauc2}
  Liller, W.\ 1990b, \iaucirc, 4976

\bibitem[Liller et~al.\labelspace(1990)]{lil90vxforiauc}
  Liller, W., Phillips, M., Hamuy, M., Lamontagne, R., Baganoff, F., Maza, J.,
  \& Wischnjewsky, M.\ 1990, \iaucirc, 5127

\bibitem[Liu, Hu\labelspace(2000)]{liu00CVspec3}
  Liu, W., \& Hu, J.~Y.\ 2000, \apjs, 128, 387

\bibitem[Liu et~al.\labelspace(1999)]{liu99CVspec2}
  Liu, W., Hu, J.~Y., Li, Z.~Y., \& Cao, L.\ 1999, \apjs, 122, 257

\bibitem[Liu et~al.\labelspace(1998)]{liu98egcnc}
  Liu, W., Li, Z.~Y., \& Hu, J.~Y.\ 1998, \apss, 257, 183

\bibitem[Livio\labelspace(1992)]{liv92novaabsmag}
  Livio, M.\ 1992, \apj, 393, 516

\bibitem[Lubbock, McNaught\labelspace(1986)]{lub86vyaqriauc}
  Lubbock, S., \& McNaught, R.~H.\ 1986, \iaucirc, 4209

\bibitem[Lubow\labelspace(1991a)]{lub91SHa}
  Lubow, S.~H.\ 1991a, \apj, 381, 259

\bibitem[Lubow\labelspace(1991b)]{lub91SHb}
  Lubow, S.~H.\ 1991b, \apj, 381, 268

\bibitem[Lubow\labelspace(1992)]{lub92SH}
  Lubow, S.~H.\ 1992, \apj, 401, 317

\bibitem[McAdam et~al.\labelspace(1988)]{mca88pqandiauc}
  McAdam, D., Hurst, G.~M., Manning, B., Lubbock, S., \& Young, A.\ 1988,
  \iaucirc, 4620

\bibitem[McAdam et~al.\labelspace(1983)]{mca83vyaqriauc}
  McAdam, D., Huruhata, M., Bortle, J., Fujino, S., McNaught, R., Argyle,
  R.~W., \& Jones, D. H.~P.\ 1983, \iaucirc, 3896

\bibitem[Mackie\labelspace(1919)]{mac19wzsge}
  Mackie, J.~C.\ 1919, \an, 210, 79

\bibitem[McNaught\labelspace(1982)]{mcn82vyaqr}
  McNaught, R.~H.\ 1982, \ibvs, 2232

\bibitem[McNaught\labelspace(1985)]{mcn85rzleoiauc}
  McNaught, R.~H.\ 1985, \iaucirc, 4036

\bibitem[McNaught\labelspace(1986)]{mcn86egcnc}
  McNaught, R.~H.\ 1986, \ibvs, 2926

\bibitem[McNaught\labelspace(1990)]{mcn90v4338sgriauc}
  McNaught, R.~H.\ 1990, \iaucirc, 4978

\bibitem[McNaught et~al.\labelspace(1989)]{mcn89wxcetiauc}
  McNaught, R.~H., Hurst, G.~M., Pearce, A., Jones, A.~F., Bateson, F.~M.,
  Kilmartin, P.~M., \& Gilmore, A.~C.\ 1989, \iaucirc, 4792

\bibitem[McNaught, Wenzel\labelspace(1983)]{mcn83vyaqriauc}
  McNaught, R.~H., \& Wenzel, W.\ 1983, \iaucirc, 3759

\bibitem[Mantel et~al.\labelspace(1992)]{man92hvviriauc}
  Mantel, K.~H., Barwig, H., Patschke, M., \& Ritter, H.\ 1992, \iaucirc, 5517

\bibitem[Marsh\labelspace(1990)]{mar90htcas}
  Marsh, T.~R.\ 1990, \apj, 357, 621

\bibitem[Marsh\labelspace(1999)]{mar99gpcom}
  Marsh, T.~R.\ 1999, \mnras, 304, 443

\bibitem[Marsh et~al.\labelspace(1991)]{mar91gpcom}
  Marsh, T.~R., Horne, K., \& Rosen, S.\ 1991, \apj, 366, 535

\bibitem[Marsh et~al.\labelspace(1995)]{mar95gpcom}
  Marsh, T.~R., Wood, J.~H., Horne, K., \& Lambert, D.\ 1995, \mnras, 274, 452

\bibitem[Mason et~al.\labelspace(2000)]{mas00wzsge}
  Mason, E., Skidmore, W., Howell, S.~B., Ciardi, D.~R., Littlefair, S., \&
  Dhillon, V.~S.\ 2000, \mnras, 318, 440

\bibitem[Matsumoto et~al.\labelspace(1998a)]{mat98egcncqui}
  Matsumoto, K., Kato, T., Ayani, K., \& Kawabata, T.\ 1998a, \ibvs, 4613

\bibitem[Matsumoto et~al.\labelspace(1998b)]{mat98egcnc}
  Matsumoto, K., Nogami, D., Kato, T., \& Baba, H.\ 1998b, \pasj, 50, 405

\bibitem[Mattei et~al.\labelspace(2001a)]{mat01alcomiauc7669}
  Mattei, J., Abbott, P., Gunther, J., \& Schmeer, P.\ 2001a, \iaucirc, 7629

\bibitem[Mattei et~al.\labelspace(1985a)]{mat85rzleoiauc}
  Mattei, J., Ducoty, R., Stanton, R., \& Scovil, C.\ 1985a, \iaucirc, 4026

\bibitem[Mattei et~al.\labelspace(1987)]{mat87rzleoiauc}
  Mattei, J., Isles, J., \& Lubbock, S.\ 1987, \iaucirc, 4506

\bibitem[Mattei\labelspace(1980)]{mat80wzsge}
  Mattei, J.~A.\ 1980, \jrasc, 74, 53

\bibitem[Mattei\labelspace(1995)]{mat95v1028cyg}
  Mattei, J.~A.\ 1995, AAVSO Alert Notice, 211

\bibitem[Mattei et~al.\labelspace(1985b)]{mat85htcasiauc}
  Mattei, J.~A., Kinnunen, T., \& Hurst, G.\ 1985b, \iaucirc, 4027

\bibitem[Mattei et~al.\labelspace(1999)]{mat99qyperiauc}
  Mattei, J.~A., Poyner, G., Midlands, W., Simonsen, M., Muyllaert, E., Jones,
  C., McGee, H., Schmeer, P., \& Masi, G.\ 1999, \iaucirc, 7340

\bibitem[Mattei et~al.\labelspace(2001b)]{mat01wzsgeiauc7669}
  Mattei, J.~A., Poyner, G., Reszelski, M., McGee, H., Jones, C., Schmeeer, P.,
  Bouma, R.~J., Vohla, F., Comello, G., Bortle, J., Royer, R., \& West, J.~D.\
  2001b, \iaucirc, 7669

\bibitem[Mattei et~al.\labelspace(1995a)]{mat95alcomiauc}
  Mattei, J.~A., Szentasko, L., Schmeer, P., York, D., \& Cragg, T.\ 1995a,
  \iaucirc, 6155

\bibitem[Mattei et~al.\labelspace(1995b)]{mat95htcasiauc}
  Mattei, J.~A., York, D., Schmeer, P., Worraker, W., Stewart, R., Bortle, J.,
  \& Griese, J.\ 1995b, \iaucirc, 6264

\bibitem[Mayall\labelspace(1946)]{may46wzsge}
  Mayall, M.~W.\ 1946, Bull. Harv. Coll. Obs., 918, 3

\bibitem[Mayall\labelspace(1968)]{may68v1028cyg}
  Mayall, M.~W.\ 1968, \jrasc, 62, 141

\bibitem[Mayall\labelspace(1970)]{may70v1028cyg}
  Mayall, M.~W.\ 1970, \jrasc, 64, 205

\bibitem[Maza, Gonzalez\labelspace(1983)]{maz83gwlibiauc}
  Maza, J., \& Gonzalez, L.~E.\ 1983, \iaucirc, 3856

\bibitem[Meinunger\labelspace(1970)]{mei70v1289aql}
  Meinunger, L.\ 1970, \MitVS, 5, 126

\bibitem[Meinunger\labelspace(1976)]{mei76wxcet}
  Meinunger, L.\ 1976, \MitVS, 7, 192

\bibitem[Meinunger\labelspace(1977)]{mei77lsand}
  Meinunger, L.\ 1977, \ibvs, 1331

\bibitem[Meinunger\labelspace(1986)]{mei86uztri}
  Meinunger, L.\ 1986, \MitVS, 11, 1

\bibitem[Mendelson et~al.\labelspace(1992)]{men92hvviriauc}
  Mendelson, H., Leibowitz, E.~M., Brosch, N., \& Almoznino, E.\ 1992,
  \iaucirc, 5509

\bibitem[Mennicken, Vogt\labelspace(1993)]{men93wxcetaqericuvel}
  Mennicken, R., \& Vogt, N.\ 1993, \RMxAA, 26, 111

\bibitem[Mennickent\labelspace(1994)]{men94wxcet}
  Mennickent, R.\ 1994, \aap, 285, 979

\bibitem[Mennickent, Arenas\labelspace(1998)]{men98ektra}
  Mennickent, R.~E., \& Arenas, J.\ 1998, \pasj, 50, 333

\bibitem[Mennickent et~al.\labelspace(2001)]{men01j1050}
  Mennickent, R.~E., Diaz, M., Skidmore, W., \& Sterken, C.\ 2001, \aap,
  in press (astro-ph/0107207)

\bibitem[Mennickent et~al.\labelspace(1999)]{men99rzleo}
  Mennickent, R.~E., Sterken, C., Gieren, W., \& E., U.\ 1999, \aap, 352, 239

\bibitem[Mennickent, Tappert\labelspace(2001)]{men01rzleo}
  Mennickent, R.~E., \& Tappert, C.\ 2001, \aap, 372, 563

\bibitem[Meyer-Hofmeister et~al.\labelspace(1998)]{mey98wzsge}
  Meyer-Hofmeister, E., Meyer, F., \& Liu, B.~F.\ 1998, \aap, 339, 507

\bibitem[Mineshige et~al.\labelspace(1998)]{min98wzsge}
  Mineshige, S., Liu, B., Meyer, F., \& Meyer-Hofmeister, E.\ 1998, \pasj, 50,
  L5

\bibitem[Misselt, Shafter\labelspace(1995)]{mis95PGCV}
  Misselt, K.~A., \& Shafter, A.~W.\ 1995, \aj, 109, 1757

\bibitem[Mobberley et~al.\labelspace(1997)]{mob97uwper}
  Mobberley, M.~P., Barber, P.~M., \& Hurst, G.~M.\ 1997, \JBAA, 107, 65

\bibitem[Molnar, Kobulnicky\labelspace(1992)]{mol92SHexcess}
  Molnar, L.~A., \& Kobulnicky, H.~A.\ 1992, \apj, 392, 678

\bibitem[Moorhead\labelspace(1965)]{moo65alcom}
  Moorhead, J.~M.\ 1965, \pasp, 77, 468

\bibitem[Moriyama et~al.\labelspace(1991)]{mor91v1251cygiauc}
  Moriyama, M., Kato, T., Hurst, G.~M., \& Schmeer, P.\ 1991, \iaucirc, 5377

\bibitem[Mukai et~al.\labelspace(1990)]{muk90faintCV}
  Mukai, K., Mason, K.~O., Howell, S.~B., Allington-Smith, J., Callman, P.~J.,
  Charles, P.~A., Hassall, B. J.~M., Machin, G., Naylor, T., Smale, A.~P., \&
  van Paradijs, J.\ 1990, \mnras, 245, 385

\bibitem[Mukai et~al.\labelspace(1997)]{muk97htcas}
  Mukai, K., Wood, J.~H., Naylor, T., Schlegel, E.~M., \& Swank, J.~H.\ 1997,
  \apj, 475, 812

\bibitem[M\"{u}ndler\labelspace(1919)]{mun19rzleo}
  M\"{u}ndler, M.\ 1919, \an, 209, 65

\bibitem[Munari et~al.\labelspace(1990)]{mun90vyaqriauc}
  Munari, U., Bianchini, A., Claudi, R., Augusteijn, T., \& della Valle, M.\
  1990, \iaucirc, 5048

\bibitem[Munari, Zwitter\labelspace(1998)]{mun98CVspec5}
  Munari, U., \& Zwitter, T.\ 1998, \aaps, 128, 277

\bibitem[Narumi et~al.\labelspace(1989)]{nar89rzleoiauc}
  Narumi, H., Korth, S., Dyck, G., Iida, M., \& Hurst, G.\ 1989, \iaucirc, 4757

\bibitem[Nather et~al.\labelspace(1981)]{nat81gpcom}
  Nather, R.~E., Robinson, E.~L., \& Stover, R.~J.\ 1981, \apj, 244, 269

\bibitem[Nather, Stover\labelspace(1978)]{nat78wzsgeiauc3311}
  Nather, R.~E., \& Stover, R.\ 1978, \iaucirc, 3311

\bibitem[Naylor\labelspace(1989)]{nay89wzsgeIUEoutburst}
  Naylor, T.\ 1989, \mnras, 238, 587

\bibitem[Neustroev\labelspace(1998)]{neu98wzsge}
  Neustroev, V.~V.\ 1998, \ARep, 42, 748

\bibitem[Nogami et~al.\labelspace(1998)]{nog98swuma}
  Nogami, D., Baba, H., Kato, T., \& Nov\'{a}k, R.\ 1998, \pasj, 50, 297

\bibitem[Nogami et~al.\labelspace(2000)]{nog00kvdra}
  Nogami, D., Engels, D., G\"{a}nsicke, B.~T., Pavlenko, E.~P., Nov\'{a}k,
  R., \& Reinsch, K.\ 2000, \aap, 364, 701

\bibitem[Nogami, Kato\labelspace(1995)]{nog95alcomiauc}
  Nogami, D., \& Kato, T.\ 1995, \iaucirc, 6164

\bibitem[Nogami et~al.\labelspace(1997a)]{nog97alcom}
  Nogami, D., Kato, T., Baba, H., Matsumoto, K., Arimoto, J., Tanabe, K., \&
  Ishikawa, K.\ 1997a, \apj, 490, 840

\bibitem[Nogami et~al.\labelspace(2001)]{nog01dvuma}
  Nogami, D., Kato, T., Baba, H., Nov\'{a}k, R., Lockley, J.~J., \& Somers, M.\
  2001, \mnras, 322, 79

\bibitem[Nogami et~al.\labelspace(1996)]{nog96cthya}
  Nogami, D., Kato, T., \& Hirata, R.\ 1996, \pasj, 48, 607

\bibitem[Nogami et~al.\labelspace(1995)]{nog95v725aql}
  Nogami, D., Masuda, S., \& Kato, T.\ 1995, \ibvs, 4218

\bibitem[Nogami et~al.\labelspace(1997b)]{nog97sxlmi}
  Nogami, D., Masuda, S., \& Kato, T.\ 1997b, \pasp, 109, 1114

\bibitem[Nov\'{a}k et~al.\labelspace(2001)]{nov01v2176cyg}
  Nov\'{a}k, R., Vanmunster, T., Jensen, L.~T., \& Nogami, D.\ 2001, \ibvs,
  5108

\bibitem[O'Connell\labelspace(1936)]{oco36apcru}
  O'Connell, D. J.~K.\ 1936, \an, 259, 399

\bibitem[O'Donoghue\labelspace(1990)]{odo90wxcetiauc}
  O'Donoghue, D.\ 1990, \iaucirc, 5006

\bibitem[O'Donoghue et~al.\labelspace(1991)]{odo91wzsge}
  O'Donoghue, D., Chen, A., Marang, F., Mittaz, J. P.~D., Winkler, H., \&
  Warner, B.\ 1991, \mnras, 250, 363

\bibitem[Ohtani et~al.\labelspace(1992)]{Ouda}
  Ohtani, H., Uesugi, A., Tomita, Y., Yoshida, M., Kosugi, G., Noumaru, J.,
  Araya, S., \& Ohta, K.\ 1992, Memoirs of the Faculty of Science, Kyoto
  University, Series A of Physics, Astrophysics, Geophysics and Chemistry, 38,
  167

\bibitem[Oke, Wade\labelspace(1982)]{oke82CVspec}
  Oke, J.~B., \& Wade, R.~A.\ 1982, \aj, 87, 670

\bibitem[Olech\labelspace(1997)]{ole97v485cen}
  Olech, A.\ 1997, \AcA, 47, 281

\bibitem[Ortolani et~al.\labelspace(1980)]{ort80wzsge}
  Ortolani, S., Rafanelli, P., Rosino, L., \& Vittone, A.\ 1980, \aap, 87, 31

\bibitem[Osaki\labelspace(1989)]{osa89suuma}
  Osaki, Y.\ 1989, \pasj, 41, 1005

\bibitem[Osaki\labelspace(1995)]{osa95wzsge}
  Osaki, Y.\ 1995, \pasj, 47, 47

\bibitem[Osaki\labelspace(1996)]{osa96review}
  Osaki, Y.\ 1996, \pasp, 108, 39

\bibitem[Osaki et~al.\labelspace(1993)]{osa93unifiedmodel}
  Osaki, Y., Hirose, M., \& Ichikawa, S.\ 1993, \apss, 210, 359

\bibitem[Osaki et~al.\labelspace(2001)]{osa01egcnc}
  Osaki, Y., Meyer, F., \& Meyer-Hofmeister, E.\ 2001, \aap, 370, 488

\bibitem[Osaki et~al.\labelspace(1997)]{osa97egcnc}
  Osaki, Y., Shimizu, S., \& Tsugawa, M.\ 1997, \pasj, 49, 19

\bibitem[Overbeek, Pearce\labelspace(1989)]{ove89wxcetiauc}
  Overbeek, D., \& Pearce, A.\ 1989, \iaucirc, 4793

\bibitem[Papaloizou, Pringle\labelspace(1979)]{pap79SHmodel}
  Papaloizou, J., \& Pringle, J.~E.\ 1979, \mnras, 189, 293

\bibitem[Parsamian et~al.\labelspace(1983)]{par83nsv15820}
  Parsamian, E.~S., Ohanian, G.~B., Kazarian, E.~S., \& Yankovich, I.~I.\ 1983,
  \ATsir, 1269, 7

\bibitem[Patterson\labelspace(1979a)]{pat79htcas}
  Patterson, J.\ 1979a, \baas, 11, 664

\bibitem[Patterson\labelspace(1979b)]{pat79SH}
  Patterson, J.\ 1979b, \aj, 84, 804

\bibitem[Patterson\labelspace(1980)]{pat80wzsge}
  Patterson, J.\ 1980, \apj, 241, 235

\bibitem[Patterson\labelspace(1981)]{pat81DNOhtcas}
  Patterson, J.\ 1981, \apjs, 45, 517

\bibitem[Patterson\labelspace(1998)]{pat98evolution}
  Patterson, J.\ 1998, \pasp, 110, 1132

\bibitem[Patterson\labelspace(2001)]{pat01SH}
  Patterson, J.\ 2001, \pasp, 113, 736

\bibitem[Patterson et~al.\labelspace(1996)]{pat96alcom}
  Patterson, J., Augusteijn, T., Harvey, D.~A., Skillman, D.~R., Abbott, T.
  M.~C., \& Thorstensen, J.\ 1996, \pasp, 108, 748

\bibitem[Patterson et~al.\labelspace(1993)]{pat93vyaqr}
  Patterson, J., Bond, H.~E., Grauer, A.~D., Shafter, A.~W., \& Mattei, J.~A.\
  1993, \pasp, 105, 69

\bibitem[Patterson et~al.\labelspace(1995a)]{pat95alcomiauc}
  Patterson, J., Harvey, D., \& Deyoung, J.~A.\ 1995a, \iaucirc, 6157

\bibitem[Patterson et~al.\labelspace(1995b)]{pat95v1159ori}
  Patterson, J., Jablonski, F., Koen, C., O'Donoghue, D., \& Skillman, D.~R.\
  1995b, \pasp, 107, 1183

\bibitem[Patterson et~al.\labelspace(1998a)]{pat98egcnc}
  Patterson, J., Kemp, J., Skillman, D.~R., Harvey, D.~A., Shafter, A.~W.,
  Vanmunster, T., Jensen, L., Robert, F., Kiyota, S., Thorstensen, J.~R., \&
  Taylor, C.~J.\ 1998a, \pasp, 110, 1290

\bibitem[Patterson et~al.\labelspace(1981)]{pat81wzsge}
  Patterson, J., McGraw, J.~T., Coleman, L., \& Africano, J.~L.\ 1981, \apj,
  248, 1067

\bibitem[Patterson, McGraw\labelspace(1978)]{pat78wzsgeiauc3311}
  Patterson, J., \& McGraw, T.\ 1978, \iaucirc, 3311

\bibitem[Patterson et~al.\labelspace(1998b)]{pat98wzsge}
  Patterson, J., Richman, H., Kemp, J., \& Mukai, K.\ 1998b, \pasp, 110, 403

\bibitem[Patterson et~al.\labelspace(2000)]{pat00dvuma}
  Patterson, J., Vanmunster, T., Skillman, D.~R., Jensen, L., Stull, J.,
  Martin, B., Cook, L.~M., Kemp, J., \& Knigge, C.\ 2000, \pasp, 112, 1584

\bibitem[Pavlenko et~al.\labelspace(2000)]{pav00swuma}
  Pavlenko, E.~P., Shugarov, S.~Y., \& Katysheva, N.~A.\ 2000, \Ap, 43, 419

\bibitem[Pavlov, Shugarov\labelspace(1985)]{pav85dvdra}
  Pavlov, M.~V., \& Shugarov, S.~Y.\ 1985, \ATsir, 1373, 8

\bibitem[Pearce et~al.\labelspace(1989)]{pea89wxcetiauc}
  Pearce, A., Jones, A., \& McNaught, R.~H.\ 1989, \iaucirc, 4814

\bibitem[Pesch, Sanduleak\labelspace(1987)]{pes87eluma}
  Pesch, P., \& Sanduleak, N.\ 1987, \ibvs, 2989

\bibitem[Pesch, Sanduleak\labelspace(1988)]{pes88elumaCasesurvey}
  Pesch, P., \& Sanduleak, N.\ 1988, \apjs, 66, 297

\bibitem[Peters\labelspace(1865)]{pet1865tleo}
  Peters, C. H.~F.\ 1865, \an, 65, 55

\bibitem[Pfau\labelspace(1976)]{pfa76novaabsmag}
  Pfau, W.\ 1976, \aap, 50, 113

\bibitem[Politano et~al.\labelspace(1998)]{pol98TOAD}
  Politano, M., Howell, S.~B., \& Rappaport, S.\ 1998, in \ASPConf{137}{Wild
  Stars in the Old West}, ed. S. Howell, E. Kuulkers, \& C. Woodward (San
  Francisco: ASP), p.~207

\bibitem[Poyner\labelspace(1996)]{poy96alcom}
  Poyner, G.\ 1996, \JBAA, 106, 155

\bibitem[Pych, Olech\labelspace(1995a)]{pyc95alcomiauc}
  Pych, W., \& Olech, A.\ 1995a, \iaucirc, 6162

\bibitem[Pych, Olech\labelspace(1995b)]{pyc95alcom}
  Pych, W., \& Olech, A.\ 1995b, \AcA, 45, 385

\bibitem[Richer et~al.\labelspace(1973)]{ric73gpcom}
  Richer, H.~B., Auman, J.~R., Isherwood, B.~C., Steele, J.~P., \& Ulrych,
  T.~J.\ 1973, \apj, 180, 107

\bibitem[Richter\labelspace(1968)]{ric68v592her}
  Richter, G.~A.\ 1968, \ibvs, 293

\bibitem[Richter\labelspace(1970)]{ric70newvar}
  Richter, G.~A.\ 1970, \MitVS, 5, 99

\bibitem[Richter\labelspace(1979)]{ric79m33blueobjectaspsc}
  Richter, G.~A.\ 1979, \MitVS, 8, 119

\bibitem[Richter\labelspace(1983a)]{ric83vyaqraditional}
  Richter, G.~A.\ 1983a, \ibvs, 2332

\bibitem[Richter\labelspace(1983b)]{ric83aspsc}
  Richter, G.~A.\ 1983b, \ATsir, 1262, 7

\bibitem[Richter\labelspace(1983c)]{ric83vyaqr}
  Richter, G.~A.\ 1983c, \ibvs, 2267

\bibitem[Richter\labelspace(1985)]{ric85rzleo}
  Richter, G.~A.\ 1985, \ibvs, 2714

\bibitem[Richter\labelspace(1986a)]{ric86CVamplitudecyclelength}
  Richter, G.~A.\ 1986a, \an, 307, 221

\bibitem[Richter\labelspace(1986b)]{ric86v358lyr}
  Richter, G.~A.\ 1986b, \ibvs, 2971

\bibitem[Richter\labelspace(1987)]{ric87uwper}
  Richter, G.~A.\ 1987, \an, 308, 125

\bibitem[Richter\labelspace(1990a)]{ric90pqand}
  Richter, G.~A.\ 1990a, \ibvs, 3546

\bibitem[Richter\labelspace(1990b)]{ric90gd552}
  Richter, G.~A.\ 1990b, \MitVS, 12, 59

\bibitem[Richter\labelspace(1991a)]{ric91alcom}
  Richter, G.~A.\ 1991a, \MitVS, 12, 99

\bibitem[Richter\labelspace(1991b)]{ric91v592her}
  Richter, G.~A.\ 1991b, \ibvs, 3619

\bibitem[Richter\labelspace(1992)]{ric92wzsgedip}
  Richter, G.~A.\ 1992, in \ASPConf{29}{Vina del Mar Workshop on Cataclysmic
  Variable Stars}, ed. N. Vogt (San Francisco: ASP), p.~12

\bibitem[Richter, Borngen\labelspace(1989)]{ric89aspsc}
  Richter, G.~A., \& Borngen, F.\ 1989, \MitVS, 12, 1

\bibitem[Richter, Brauer\labelspace(1994)]{ric94iodel}
  Richter, G.~A., \& Brauer, H.~J.\ 1994, \ibvs, 4033

\bibitem[Richter, B\"{o}rngen\labelspace(1981)]{ric81aspsc}
  Richter, G.~A., \& B\"{o}rngen, F.\ 1981, \aplett, 21, 101

\bibitem[Richter, Greiner\labelspace(1995)]{ric95ircom}
  Richter, G.~A., \& Greiner, J.\ 1995, in Cataclysmic Variables, ed. A.
  Bianchini, M. Della~Valle, \& M. Orio (Kluwer Academic Publishers,
  Dordrecht), p.~177

\bibitem[Richter et~al.\labelspace(1996)]{ric96ircomiauc}
  Richter, G.~A., Kroll, P., Greiner, J., \& Bianchini, A.\ 1996, \iaucirc,
  6295

\bibitem[Richter et~al.\labelspace(1997)]{ric97ircom}
  Richter, G.~A., Kroll, P., Greiner, J., Wenzel, W., Luthardt, R., \&
  Schwartz, R.\ 1997, \aap, 325, 994

\bibitem[Ringwald et~al.\labelspace(1996)]{rin96oldnovaspec}
  Ringwald, F.~A., Naylor, T., \& Mukai, K.\ 1996, \mnras, 281, 192

\bibitem[Ritter, Kolb\labelspace(1998)]{RitterCV}
  Ritter, H., \& Kolb, U.\ 1998, \aaps, 129, 83

\bibitem[Ritter, Schroeder\labelspace(1979)]{rit79wzsge}
  Ritter, H., \& Schroeder, R.\ 1979, \aap, 76, 168

\bibitem[Roberts\labelspace(1914)]{rob14cpdra}
  Roberts, D.\ 1914, \an, 197, 57

\bibitem[Robertson, Honeycutt\labelspace(1996)]{rob96htcas}
  Robertson, J.~W., \& Honeycutt, R.~K.\ 1996, \aj, 112, 2248

\bibitem[Robertson et~al.\labelspace(2000)]{rob00oldnova}
  Robertson, J.~W., Honeycutt, R.~K., Hillwig, T., Jurcevic, J.~S., \& Henden,
  A.~A.\ 2000, \aj, 119, 1365

\bibitem[Robinson et~al.\labelspace(1978)]{rob78wzsge}
  Robinson, E.~L., Nather, R.~E., \& Patterson, J.\ 1978, \apj, 219, 168

\bibitem[Robinson et~al.\labelspace(1987)]{rob87swumaQPO}
  Robinson, E.~L., Shafter, A.~W., Hill, J.~A., Wood, M.~A., \& Mattei, J.~A.\
  1987, \apj, 313, 772

\bibitem[Robinson, Zhang\labelspace(1985)]{rob85htcas}
  Robinson, E.~L., \& Zhang, E.\ 1985, \baas, 17, 886

\bibitem[Robinson, Ashbrook\labelspace(1969)]{rob69tleo}
  Robinson, L.~J., \& Ashbrook, J.\ 1969, \skytel, 37, 128

\bibitem[Rogoziecki, Schwarzenberg-Czerny\labelspace(2001)]{rog01wxcet}
  Rogoziecki, P., \& Schwarzenberg-Czerny, A.\ 2001, \mnras, 323, 850

\bibitem[Romano\labelspace(1964)]{rom64bcuma}
  Romano, G.\ 1964, \MmSAI, 35, 101

\bibitem[Romano\labelspace(1977)]{rom77lsand}
  Romano, G.\ 1977, \aj, 82, 319

\bibitem[Romano, Minello\labelspace(1976)]{rom76DN}
  Romano, G., \& Minello, S.\ 1976, \ibvs, 1140

\bibitem[Rosen et~al.\labelspace(1994)]{ros94v426ophswumav348pup}
  Rosen, S.~R., Clayton, K.~L., Osborne, J.~P., \& McGale, P.~A.\ 1994, \mnras,
  269, 913

\bibitem[Rosino\labelspace(1961)]{ros61alcomiauc}
  Rosino, L.\ 1961, \iaucirc, 1782

\bibitem[Rosino, Candeo\labelspace(1989)]{ros89qyperiauc}
  Rosino, L., \& Candeo, G.\ 1989, \iaucirc, 4900

\bibitem[Rosino, Pigatto\labelspace(1972a)]{ros72xypsciauc1}
  Rosino, L., \& Pigatto, L.\ 1972a, \iaucirc, 2453

\bibitem[Rosino, Pigatto\labelspace(1972b)]{ros72xypsciauc2}
  Rosino, L., \& Pigatto, L.\ 1972b, \iaucirc, 2464

\bibitem[Ross\labelspace(1925)]{ros25vyaqr}
  Ross, F.~E.\ 1925, \aj, 36, 123

\bibitem[Royer\labelspace(1988)]{roy88pqandiauc}
  Royer, R.~E.\ 1988, \iaucirc, 4628

\bibitem[Ruiz et~al.\labelspace(2001)]{rui01ce315}
  Ruiz, M.~T., Rojo, P.~M., Garay, G., \& Maza, J.\ 2001, \apj, 552, 679

\bibitem[Samus, Kato\labelspace(2000)]{sam00dvdra}
  Samus, N.~N., \& Kato, T.\ 2000, \vsnetid{309}, \\
  http://www.kusastro.kyoto-u.ac.jp/vsnet/Mail/\\
  vsnet-id/msg00309.html

\bibitem[Sandage, Tammann\labelspace(1988)]{san88virgonovaabsmag}
  Sandage, A., \& Tammann, G.~A.\ 1988, \apj, 328, 1

\bibitem[Sandig\labelspace(1950)]{san50efpeg}
  Sandig, H.-U.\ 1950, \an, 279, 89

\bibitem[Schaefer, Hoffleit\labelspace(1994)]{sch94kyara}
  Schaefer, B.~E., \& Hoffleit, D.\ 1994, \ibvs, 4123

\bibitem[Schmeer\labelspace(2000)]{sch00htcam}
  Schmeer, P.\ 2000, \vsnetalert{4846}, \\
  http://www.kusastro.kyoto-u.ac.jp/vsnet/Mail/\\
  alert4000/msg00846.html

\bibitem[Schmeer, Duerbeck\labelspace(1999)]{sch99cigem}
  Schmeer, P., \& Duerbeck, H.~W.\ 1999, \ibvs, 4758

\bibitem[Schmeer et~al.\labelspace(1992)]{sch92hvviriauc}
  Schmeer, P., Hurst, G.~M., Kilmartin, P.~M., \& Gilmore, A.~C.\ 1992,
  \iaucirc, 5502

\bibitem[Schmidt\labelspace(1957)]{sch57novaabsmag}
  Schmidt, T.\ 1957, \ZA, 41, 182

\bibitem[Schneller\labelspace(1931)]{sch31hvvir}
  Schneller, H.\ 1931, \an, 243, 335

\bibitem[Schultheis et~al.\labelspace(2000)]{sch00ISOGAL}
  Schultheis, M., Ganesh, S., Glass, I.~S., Omont, A., Ortiz, R., Simon, G.,
  van Loon, J.~T., Alard, C., Blommaert, J. A. D.~L., Borsenberger, J., Fouqu,
  P., \& Habing, H.~J.\ 2000, \aap, 362, 215

\bibitem[Semeniuk et~al.\labelspace(1997a)]{sem97rzsge}
  Semeniuk, I., Nalezyty, M., Gembara, P., \& Kwast, T.\ 1997a, \AcA, 47, 299

\bibitem[Semeniuk et~al.\labelspace(1997b)]{sem97swuma}
  Semeniuk, I., Olech, A., Kwast, T., \& Nalezyty, M.\ 1997b, \AcA, 47, 201

\bibitem[Shafter\labelspace(1983)]{sha83swuma}
  Shafter, A.~W.\ 1983, \ibvs, 2354

\bibitem[Shafter, Campbell\labelspace(1990)]{sha90bccas}
  Shafter, A.~W., \& Campbell, R.\ 1990, \baas, 22, 1211

\bibitem[Shafter et~al.\labelspace(1987)]{sha87swuma}
  Shafter, A.~W., Hill, J.~A., Robinson, E.~L., Szkody, P., Thorstensen, J.~R.,
  \& Wood, M.~A.\ 1987, \apss, 130, 125

\bibitem[Shafter, Szkody\labelspace(1984)]{sha84tleo}
  Shafter, A.~W., \& Szkody, P.\ 1984, \apj, 276, 305

\bibitem[Shafter et~al.\labelspace(1986)]{sha86swumaXray}
  Shafter, A.~W., Szkody, P., \& Thorstensen, J.~R.\ 1986, \apj, 308, 765

\bibitem[Sharov\labelspace(1987)]{sha87aspsc}
  Sharov, A.~S.\ 1987, \PAZh, 13, 427

\bibitem[Sharov, Alksnis\labelspace(1989)]{sha89ptand}
  Sharov, A.~S., \& Alksnis, A.~K.\ 1989, \PAZh, 15, 885

\bibitem[Sharov, Karinova\labelspace(1978)]{sha78lsand}
  Sharov, A.~S., \& Karinova, D.~K.\ 1978, \ATsir, 998, 1

\bibitem[Sima, Hadrava\labelspace(1987)]{sim87wzsge}
  Sima, Z., \& Hadrava, P.\ 1987, \apss, 130, 151

\bibitem[Sion et~al.\labelspace(1995)]{sio95wzsgeHST}
  Sion, E.~M., Cheng, F.~H., Long, K.~S., Szkody, P., Gilliland, R.~L., Huang,
  M., \& Hubeny, I.\ 1995, \apj, 439, 957

\bibitem[Sion et~al.\labelspace(1990)]{sio90wzsge}
  Sion, E.~M., Leckenby, H.~J., \& Szkody, P.\ 1990, \apjl, 364, L41

\bibitem[Skidmore et~al.\labelspace(2000)]{ski00wzsge}
  Skidmore, W., Mason, E., Howell, S.~B., Ciardi, D.~R., Littlefair, S., \&
  Dhillon, V.~S.\ 2000, \mnras, 318, 429

\bibitem[Skidmore et~al.\labelspace(1999)]{ski99wzsge}
  Skidmore, W., Welsh, W.~F., Wood, J.~H., Catal\'{a}n, M.~S., \& Horne, K.\
  1999, \mnras, 310, 750

\bibitem[Skidmore et~al.\labelspace(1997)]{ski97wzsge}
  Skidmore, W., Welsh, W.~F., Wood, J.~H., \& Stiening, R.~F.\ 1997, \mnras,
  288, 189

\bibitem[Slevinsky et~al.\labelspace(1999)]{sle99wzsge}
  Slevinsky, R.~J., Stys, D., West, S., Sion, E.~M., \& Cheng, F.~H.\ 1999,
  \pasp, 111, 1292

\bibitem[Slevinsky et~al.\labelspace(2000)]{sle00wzsge}
  Slevinsky, R.~J., Stys, D., West, S., Sion, E.~M., \& Cheng, F.~H.\ 2000,
  \BaltA, 9, 247

\bibitem[Slovak et~al.\labelspace(1987)]{slo87tleoiauc}
  Slovak, M.~H., Nelson, M.~J., \& Shafter, A.~W.\ 1987, \iaucirc, 4314

\bibitem[Smak\labelspace(1975)]{sma75gpcom}
  Smak, J.\ 1975, \AcA, 25, 227

\bibitem[Smak\labelspace(1979)]{sma79wzsge}
  Smak, J.\ 1979, \AcA, 29, 325

\bibitem[Smak\labelspace(1983)]{sma83HeCV}
  Smak, J.\ 1983, \AcA, 33, 333

\bibitem[Smak\labelspace(2000)]{sma00DNabsmag}
  Smak, J.~I.\ 2000, \AcA, 50, 399

\bibitem[Spogli et~al.\labelspace(1998)]{spo98alcomv544herv660herv516cygdxand}
  Spogli, C., Fiorucci, M., \& Tosti, G.\ 1998, \aaps, 130, 485

\bibitem[Spruit, Rutten\labelspace(1998)]{spr98wzsge}
  Spruit, H.~C., \& Rutten, R. G.~M.\ 1998, \mnras, 299, 768

\bibitem[Steeghs et~al.\labelspace(2001)]{ste01wzsgeiauc7675}
  Steeghs, D., Marsh, T.~R., Kuulkers, E., \& Skidmore, W.\ 2001, \iaucirc,
  7675

\bibitem[Stellingwerf\labelspace(1978)]{PDM}
  Stellingwerf, R.~F.\ 1978, \apj, 224, 953

\bibitem[Stover\labelspace(1983)]{sto83gpcom}
  Stover, R.~J.\ 1983, \pasp, 95, 18

\bibitem[Strohmeier\labelspace(1964)]{str64wxcet}
  Strohmeier, W.\ 1964, \ibvs, 47

\bibitem[Sumner\labelspace(1997)]{sum97v4338sgr}
  Sumner, B.\ 1997, \vsnetchat{148}, \\
  http://www.kusastro.kyoto-u.ac.jp/vsnet/Mail/\\
  vsnet-chat/msg00148.html

\bibitem[Szkody\labelspace(1985)]{szk85CVmultiwavelength}
  Szkody, P.\ 1985, \aj, 90, 1837

\bibitem[Szkody\labelspace(1987)]{szk87shortPCV}
  Szkody, P.\ 1987, \apjs, 63, 685

\bibitem[Szkody\labelspace(1998)]{szk98multiwavelength}
  Szkody, P.\ 1998, in \ASPConf{137}{Wild Stars in the Old West}, ed. S.
  Howell, E. Kuulkers, \& C. Woodward (San Francisco: ASP), p.~18

\bibitem[Szkody et~al.\labelspace(2000a)]{szk00TOADs}
  Szkody, P., Desai, V., Burdullis, T., Hoard, D.~W., Fried, R., Garnavich, P.,
  \& G\"{a}nsicke, B.\ 2000a, \apj, 540, 983

\bibitem[Szkody et~al.\labelspace(2000b)]{szk00gwlib}
  Szkody, P., Desai, V., \& Hoard, D.~W.\ 2000b, \aj, 119, 365

\bibitem[Szkody et~al.\labelspace(1998)]{szk98alcom}
  Szkody, P., Hoard, D.~W., Sion, E.~M., Howell, S.~B., Cheng, F.~H., \&
  Sparks, W.~M.\ 1998, \apj, 497, 928

\bibitem[Szkody, Howell\labelspace(1992)]{szk92CVspec}
  Szkody, P., \& Howell, S.~B.\ 1992, \apjs, 78, 537

\bibitem[Szkody, Howell\labelspace(1993)]{szk93dvumaaypscv503cyg}
  Szkody, P., \& Howell, S.~B.\ 1993, \apj, 403, 743

\bibitem[Szkody et~al.\labelspace(1989)]{szk89faintCV2}
  Szkody, P., Howell, S.~B., Mateo, M., \& Kreidl, T.~J.\ 1989, \pasp, 101, 899

\bibitem[Szkody et~al.\labelspace(1992)]{szk92hvviriauc}
  Szkody, P., Ingram, D., Schmeer, P., Midtskogen, O., Dahle, H., \& Bortle,
  J.~E.\ 1992, \iaucirc, 5516

\bibitem[Szkody et~al.\labelspace(2001)]{szk01lspegtleoXray}
  Szkody, P., Nishikida, K., Long, K.~S., \& Fried, R.\ 2001, \aj, 121, 2761

\bibitem[Szkody et~al.\labelspace(1988)]{szk88swumaEXOSATIUE}
  Szkody, P., Osborne, J., \& Hassall, B. J.~M.\ 1988, \apj, 328, 243

\bibitem[Szkody et~al.\labelspace(1996)]{szk96alcomIUE}
  Szkody, P., Silber, A., Sion, E., Downes, R.~A., Howell, S.~B., Costa, E., \&
  Moreno, H.\ 1996, \aj, 111, 2379

\bibitem[Szkody et~al.\labelspace(1991)]{szk91interoutburst}
  Szkody, P., Stablein, C., Mattei, J.~A., \& Waagen, E.~O.\ 1991, \apjs, 76,
  359

\bibitem[Tappert, Mennickent\labelspace(2001)]{tap01kxaql}
  Tappert, C., \& Mennickent, R.~E.\ 2001, \ibvs, 5101

\bibitem[Targan\labelspace(1979a)]{tar79wzsge}
  Targan, D.\ 1979a, \ibvs, 1539

\bibitem[Targan\labelspace(1979b)]{tar79wzsgeiauc3320}
  Targan, D.\ 1979b, \iaucirc, 3320

\bibitem[Targan\labelspace(1979c)]{tar79wzsgeiauc3344}
  Targan, D.\ 1979c, \iaucirc, 3344

\bibitem[Tch\"{a}pe\labelspace(1963)]{tch63v1028cyg}
  Tch\"{a}pe, R.\ 1963, \MitVS, 2, 3

\bibitem[Thiele\labelspace(1916)]{thi16grori}
  Thiele, H.\ 1916, \an, 202, 213

\bibitem[Thorstensen et~al.\labelspace(1996)]{tho96Porb}
  Thorstensen, J.~R., Patterson, J.~O., Shambrook, A., \& Thomas, G.\ 1996,
  \pasp, 108, 73

\bibitem[Thorstensen, Taylor\labelspace(1997)]{tho97uvpervyaqrv1504cyg}
  Thorstensen, J.~R., \& Taylor, C.~J.\ 1997, \pasp, 109, 1359

\bibitem[Tovmassian et~al.\labelspace(1998)]{tov98htcam}
  Tovmassian, G.~H., Greiner, J., Kroll, P., Szkody, P., Mason, P.~A.,
  Zickgraf, F.-J., Krautter, J., Thiering, I., Serrano, A., Howell, S., \&
  Ciardi, D.~R.\ 1998, \aap, 335, 227

\bibitem[Tsesevich et~al.\labelspace(1979)]{tse79efpeg}
  Tsesevich, V.~P., Goranskij, V.~P., Samus', N.~N., \& Shugarov, S.~Y.\ 1979,
  \ATsir, 1043, 3

\bibitem[Tsugawa, Osaki\labelspace(1997)]{tsu97amcvn}
  Tsugawa, M., \& Osaki, Y.\ 1997, \pasj, 49, 75

\bibitem[Uemura\labelspace(2001)]{uem01cpdracampaigndn556}
  Uemura, M.\ 2001, \vsnetcampaigndn{556}, \\
  http://www.kusastro.kyoto-u.ac.jp/vsnet/Mail/\\
  vsnet-campaign-dn/msg00556.html

\bibitem[Uemura et~al.\labelspace(2001)]{uem01v725aql}
  Uemura, M., Kato, T., Pavlenko, E., Baklanov, A., \& J., P.\ 2001, \pasj, 53,
  539

\bibitem[Uemura et~al.\labelspace(2000)]{uem00bcumaalert4601}
  Uemura, M., Kato, T., Shugarov, S., Pavlenko, E., Schmeer, P., James, N.,
  Nov\'{a}k, R., \& Masi, G.\ 2000, \vsnetalert{4601}, \\
  http://www.kusastro.kyoto-u.ac.jp/vsnet/Mail/\\
  alert4000/msg00601.html

\bibitem[Ulla, Solheim\labelspace(1990)]{ull90amcvncrboov803cengpcom}
  Ulla, A.~M., \& Solheim, J.-E.\ 1990, \apss, 169, 189

\bibitem[Usher\labelspace(1981)]{ush81gocom}
  Usher, P.~D.\ 1981, \apjs, 46, 117

\bibitem[Usher et~al.\labelspace(1982)]{ush82dvuma}
  Usher, P.~D., Mattson, D., \& Warnock, A.~I.\ 1982, \apjs, 48, 51

\bibitem[Usher et~al.\labelspace(1983)]{ush83dvuma}
  Usher, P.~D., Warnock, A., \& Green, R.~F.\ 1983, \apj, 269, 73

\bibitem[van~den Bergh et~al.\labelspace(1973)]{vandenber73lsand}
  van~den Bergh, S., Herbst, E., \& Pritchet, C.\ 1973, \aj, 78, 275

\bibitem[van~den Bergh, Pritchet\labelspace(1986)]{vandenver86novaabsmag}
  van~den Bergh, S., \& Pritchet, C.~J.\ 1986, \pasp, 98, 110

\bibitem[van~der Woerd et~al.\labelspace(1988)]{vanderwoe88lateSH}
  van~der Woerd, H., van~der Klis, M., van Paradijs, J., Beuermann, K., \&
  Motch, C.\ 1988, \apj, 330, 911

\bibitem[van Paradijs et~al.\labelspace(1989)]{vanpar89wxcet}
  van Paradijs, J., van~der Klis, M., \& Pedersen, H.\ 1989, \aap, 225, L5

\bibitem[van Teeseling et~al.\labelspace(1999)]{vantee99v592her}
  van Teeseling, A., Hessman, F.~V., \& Romani, R.~W.\ 1999, \aap, 342, L45

\bibitem[van Zyl et~al.\labelspace(2000)]{vanzyl00gwlib}
  van Zyl, L., Warner, B., O'Donoghue, D., Sullivan, D., Pritchard, J., \&
  Kemp, J.\ 2000, \BaltA, 9, 231

\bibitem[van Zyl et~al.\labelspace(2001)]{vanzyl01gwlib}
  van Zyl, L., Warner, B., O'Donoghue, D., Sullivan, D., Pritchard, J., \&
  Kemp, J.\ 2001, \apj, 550, 231

\bibitem[Vanmunster\labelspace(1997)]{van97v2176cygiauc}
  Vanmunster, T.\ 1997, \iaucirc, 6740

\bibitem[Vogt\labelspace(1983)]{vog83lateSH}
  Vogt, N.\ 1983, \aap, 118, 95

\bibitem[Vogt, M.\labelspace(1982)]{vog82atlas}
  Vogt, N., \& M., B.~F.\ 1982, \aaps, 48, 383

\bibitem[Vogt, Semeniuk\labelspace(1980)]{vog80ektra}
  Vogt, N., \& Semeniuk, I.\ 1980, \aap, 89, 223

\bibitem[Voikhanskaya\labelspace(1983)]{voi83wzsge}
  Voikhanskaya, N.~F.\ 1983, \azh, 60, 938

\bibitem[Waagen et~al.\labelspace(1985)]{waa85htcasiauc}
  Waagen, E., Scovil, C., Sakuma, F., Borgman, M., Hurst, G., Heifner, M., \&
  Griese, J.\ 1985, \iaucirc, 4037

\bibitem[Waagen et~al.\labelspace(2000)]{waa00bcumaiauc}
  Waagen, E.~O., Muyllaert, E., Hanson, G., Schmeer, P., Kinnunen, T., \& Dyck,
  G.\ 2000, \iaucirc, 7393

\bibitem[Wade, Hamilton\labelspace(1988)]{wad88pqandiauc}
  Wade, R.~A., \& Hamilton, D.\ 1988, \iaucirc, 4629

\bibitem[Wagner et~al.\labelspace(1988)]{wag88CaseSurvey}
  Wagner, R.~M., Sion, E.~M., Liebert, J., \& Starrfield, S.~G.\ 1988, \apj,
  328, 213

\bibitem[Wagner et~al.\labelspace(1990)]{wag90v4338sgriauc}
  Wagner, R.~M., Starrfield, S.~G., \& Austin, S.\ 1990, \iaucirc, 5008

\bibitem[Walker, Bell\labelspace(1978)]{wal78wzsgeiauc3315}
  Walker, M., \& Bell, M.\ 1978, \iaucirc, 3315

\bibitem[Walker, Bell\labelspace(1980)]{wal80wzsgespec}
  Walker, M.~F., \& Bell, M.\ 1980, \apj, 237, 89

\bibitem[Warner\labelspace(1972)]{war72gpcom}
  Warner, B.\ 1972, \mnras, 159, 315

\bibitem[Warner\labelspace(1985)]{war85suuma}
  Warner, B.\ 1985, in Interacting Binaries, ed. P.~P. Eggelton, \& J.~E.
  Pringle (D. Reidel Publishing Company, Dordrecht), p.~367

\bibitem[Warner\labelspace(1987)]{war87CVabsmag}
  Warner, B.\ 1987, \mnras, 227, 23

\bibitem[Warner\labelspace(1995a)]{war95amcvn}
  Warner, B.\ 1995a, \apss, 225, 249

\bibitem[Warner\labelspace(1995b)]{war95suuma}
  Warner, B.\ 1995b, \apss, 226, 187

\bibitem[Warner, Nather\labelspace(1972)]{war72wzsge}
  Warner, B., \& Nather, E.~R.\ 1972, \mnras, 156, 297

\bibitem[Warner, Woudt\labelspace(2001)]{war01mmhyaalert5884}
  Warner, B., \& Woudt, P.\ 2001, \vsnetalert{5884}, \\
  http://www.kusastro.kyoto-u.ac.jp/vsnet/Mail/\\
  alert5000/msg00884.html

\bibitem[Watanabe\labelspace(1998)]{wat98htcam}
  Watanabe, T.\ 1998, \VSOLJBul, 32, 1

\bibitem[Watson et~al.\labelspace(1995)]{wat95j1255iauc}
  Watson, M.~G., McMahon, R.~G., \& Page, M.~J.\ 1995, \iaucirc, 6126

\bibitem[Watson et~al.\labelspace(1996)]{wat96j1255}
  Watson, M.~G., Marsh, T.~R., Fender, R.~P., Barstow, M.~A., Still, M., Page,
  M., Dhillon, V.~S., \& Beardmore, A.~P.\ 1996, \mnras, 281, 1016

\bibitem[Weber\labelspace(1966)]{web66v1251cyg}
  Weber, R.\ 1966, \ibvs, 123

\bibitem[Welsh, Wood\labelspace(1995)]{wel95htcas}
  Welsh, W., \& Wood, J.~H.\ 1995, in Flares and Flashes, ed. J. Greiner, H.~W.
  Duerbeck, \& R.~E. Gershberg (Springer-Verlag), p.~300

\bibitem[Welsh et~al.\labelspace(1997)]{wel97wzsgeUVoscillation}
  Welsh, W.~F., Skidmore, W., Wood, J.~H., Cheng, F.~H., \& Sion, E.~M.\ 1997,
  \mnras, 291, 57P

\bibitem[Welsh et~al.\labelspace(1996)]{wel96flickeringhtcas}
  Welsh, W.~F., Wood, J.~H., \& Horne, K.\ 1996, in \IAUColloq{158}{Cataclysmic
  Variables and Related Objects}, ed. A. Evans, \& J.~H. Wood (Kluwer Academic
  Publishers, Dordrecht), p.~29

\bibitem[Wenzel\labelspace(1965a)]{wen65alcom1}
  Wenzel, W.\ 1965a, \ibvs, 99

\bibitem[Wenzel\labelspace(1965b)]{wen65alcom2}
  Wenzel, W.\ 1965b, \ibvs, 110

\bibitem[Wenzel\labelspace(1983a)]{wen83tleo}
  Wenzel, W.\ 1983a, \ibvs, 2430

\bibitem[Wenzel\labelspace(1983b)]{wen83vyaqr}
  Wenzel, W.\ 1983b, \ibvs, 2261

\bibitem[Wenzel\labelspace(1984)]{wen84tleo}
  Wenzel, W.\ 1984, \MitVS, 10, 51

\bibitem[Wenzel\labelspace(1985)]{wen85htcas}
  Wenzel, W.\ 1985, \ibvs, 2832

\bibitem[Wenzel\labelspace(1987)]{wen87htcas}
  Wenzel, W.\ 1987, \an, 308, 75

\bibitem[Wenzel\labelspace(1990)]{wen90cigem}
  Wenzel, W.\ 1990, \ibvs, 3440

\bibitem[Wenzel\labelspace(1991a)]{wen91dvdra}
  Wenzel, W.\ 1991a, \ibvs, 3626

\bibitem[Wenzel\labelspace(1991b)]{wen91v1251cyg}
  Wenzel, W.\ 1991b, \ibvs, 3689

\bibitem[Wenzel, Fuhrmann\labelspace(1994)]{wen94rej1255iauc}
  Wenzel, W., \& Fuhrmann, B.\ 1994, \iaucirc, 6102

\bibitem[Wenzel et~al.\labelspace(1995)]{wen95ircom}
  Wenzel, W., Richter, G.~A., Luthardt, R., \& Schwartz, R.\ 1995, \ibvs, 4182

\bibitem[Wenzel, Wicklein\labelspace(1990)]{wen90nsv00895}
  Wenzel, W., \& Wicklein, A.\ 1990, \MitVS, 12, 76

\bibitem[Wheatley et~al.\labelspace(2000)]{whe00j1255}
  Wheatley, P.~J., Burleigh, M.~R., \& Watson, M.~G.\ 2000, \mnras, 317, 343

\bibitem[Wheatley et~al.\labelspace(2001)]{whe01wzsgeiauc7677}
  Wheatley, P.~J., Kuulkers, E., Drake, J.~J., Kaastra, J.~S., Mauche, C.~W.,
  Starrfield, S.~G., \& Wagner, R.~M.\ 2001, \iaucirc, 7677

\bibitem[Whitehurst\labelspace(1988)]{whi88tidal}
  Whitehurst, R.\ 1988, \mnras, 232, 35

\bibitem[Whyte, Eggleton\labelspace(1980)]{why80CVevolution}
  Whyte, C.~A., \& Eggleton, P.~P.\ 1980, \mnras, 190, 801

\bibitem[Wild\labelspace(1979)]{wil79lland}
  Wild, P.\ 1979, \iaucirc, 3412

\bibitem[Williams\labelspace(1983)]{wil83CVspec1}
  Williams, G.\ 1983, \apjs, 53, 523

\bibitem[Wlodarczyk\labelspace(1986)]{wlo86htcas}
  Wlodarczyk, K.\ 1986, \AcA, 36, 395

\bibitem[Wolf\labelspace(1904)]{wol04v1289aql}
  Wolf, M.\ 1904, \an, 165, 363

\bibitem[Wolf\labelspace(1919)]{wol19rzleo}
  Wolf, M.\ 1919, \an, 209, 85

\bibitem[Wolf, Wolf\labelspace(1905)]{wol05svari}
  Wolf, M., \& Wolf, G.\ 1905, \an, 169, 415

\bibitem[Wood et~al.\labelspace(1992)]{woo92htcas}
  Wood, J.~H., Horne, K., \& Vennes, S.\ 1992, \apj, 385, 294

\bibitem[Wood et~al.\labelspace(1995)]{woo95htcasXray}
  Wood, J.~H., Naylor, T., Hassall, B. J.~M., \& F., R.~T.\ 1995, \mnras, 273,
  772

\bibitem[Woudt, Warner\labelspace(2001)]{wou01v359cenxzeriyytel}
  Woudt, P.~A., \& Warner, B.\ 2001, \mnras, in press (astro-ph/0107505)

\bibitem[Wyckoff, Wehinger\labelspace(1978)]{wyc78oldnovaID}
  Wyckoff, S., \& Wehinger, P.~A.\ 1978, \pasp, 90, 557

\bibitem[Young et~al.\labelspace(1981)]{you81htcas}
  Young, P., Schneider, D.~P., \& Shectman, S.~A.\ 1981, \apj, 245, 1035

\bibitem[Zhang et~al.\labelspace(1986)]{zha86htcas}
  Zhang, E.~H., Robinson, E.~L., \& Nather, R.~E.\ 1986, \apj, 305, 740

\bibitem[Zwicky\labelspace(1965)]{zwi65alcom}
  Zwicky, F.\ 1965, \iaucirc, 1902

\bibitem[Zwitter, Munari\labelspace(1994)]{zwi94CVspec1}
  Zwitter, T., \& Munari, U.\ 1994, \aaps, 107, 503

\end{thebibliography}
\end{document}